\documentclass[12pt]{article}
\usepackage{amsfonts,amsmath}

\usepackage{amssymb}
\usepackage[hypertex] {hyperref}
\makeatletter
\@addtoreset{equation}{section}
\makeatother
\def\theequation{\thesection.\arabic{equation}}
\newcommand{\Ycoh}{{\Y'}}
\newcommand{\Ydelnol}{{\Y''}}
\newcommand{\Imm}{\mathrm{Im}}
\newcommand{\ath}{\mathbf{a}}
  \newcommand{\ld}{{\mathbf{m}}}
\newcommand{\la}{{\mathbf{n}}}

\newcommand{\TTT}{\mathcal{T}}

\newcommand{\del}{{\vartriangle}}
\newcommand{\bel}{\overline{\vartriangle}}

\newcommand{\oSW}{{\,\otimes_{\rule{0pt}{4pt}\atop \ls\!\!  SW}}\!}
\newcommand{\ddn}{{N}}
\newcommand{\ddd}{{D}}
\newcommand{\aaa}{{a}}

\newcommand{\sh}{{ \mathbb{S}  }^{sh}   }
\newcommand{\Sm}{{ \mathcal{S}  }    }
\newcommand{\shm}{{ \mathbb{S}}^{Msh}   }
\newcommand{\shi}{{\rm S} }

\newcommand{\MM}{{  \mathcal{{M}}_M^{Mnk}} }
\newcommand{\hei}{{ \tt{h}} }
\newcommand{\len}{{   \tt{l}} }

 \newcommand{\bhi}{{ \bar{\phi}}}

\newcommand{\ie}{{\it i.e.,} }

\newcommand{\dis}{\displaystyle}
\newcommand{\Y}{{\mathbf{Y}} }
\newcommand{\Yod}{{\mathbf{Y}''\left\{{\Y_0,\Y_\gd }\right\}} }

\newcommand{\Yoda}{{\mathbf{Y}'}  }
\newcommand{\Yodm}{{\mathbf{Y}'}}
\newcommand{\dr}{{\rm d}}
\newcommand{\Yn}{{\mathbb{Y}} }

\newcommand{\YY}{ \mathbf{{Y'}}}
\newcommand{\YYY}{ \mathbf{{\overline{Y}'}}}

\newcommand{\ppi}{{\mathbf{p}} }
\newcommand{\rhs}{{\it r.h.s.} }
\newcommand{\lhs}{{\it l.h.s.} }
\newcommand{\etc}{{\it etc}}

\newcommand{\ga}{\alpha}
\newcommand{\pb}{{\gb^\prime}}
\newcommand{\pga}{{\gga^\prime}}
\newcommand{\pa}{{\ga^\prime}}

\newcommand{\gp}{\rho}

\newcommand{\gb}{\beta}

\newcommand{\gga}{\gamma}

\newcommand{\gx}{\xi}

\newcommand{\Hh}{{\cal H} {}}
\newcommand{\Hhs}{{\cal Y} {}}
\newcommand{\Hr} {\mathbb{H}}
 
\newcommand{\hh}{ {\mathfrak{h}} }

\newcommand{\be}{\begin{equation}}
\newcommand{\ee}{\end{equation}}
\newcommand{\bee}{\begin{eqnarray}}
\newcommand{\beee}{\begin{array}}
\newcommand{\eee}{\end{eqnarray}}
\newcommand{\eeee}{\end{array}}

\newcommand{\gn}{\nu}
\newcommand{\ddm}{M}
\newcommand{\gch}{\rho}

\newcommand{\dda}{A}
\newcommand{\ddb}{B}
\newcommand{\kk}{\xi}

\newcommand{\M}{{\cal M}}
\newcommand{\gd}{\delta}
\newcommand{\gl}{\lambda}

\newcommand{\gvep}{\varepsilon}
\newcommand{\gs}{\sigma}
\newcommand{\bz}{\bar z}
\newcommand{\go}{\omega}

\newcommand{\gta}{\tau}
\newcommand{\by}{{\bar y}}
\newcommand{\q}{\,,\qquad}
\newcommand{\nn}{\nonumber}
\newcommand{\half}{\frac{1}{2}}
\newcommand{\ptl}{\partial}

\newcommand{\p}{\partial}

\newcommand{\K}{{\mathbf{r }}}
\newcommand{\KM} { \mathbf{r}}

\newcommand{\A}{{\cal A}}
\newcommand{\AAA}{{ \mathrm{ A}}}
\newcommand{\B}{{\cal B}}
\newcommand{\C}{{\cal C}}

\newcommand{\V}{{\cal V}}

\newcommand{\E}{{\cal E}}
\newcommand{\F}{{\cal F}}

\renewcommand{\S}{{\cal S}}
\newcommand{\f}{\frac}

\newcommand{\ls}{\!\!\!\!\!\!}

\newsavebox{\ver}
\newsavebox{\verp}
\newsavebox{\gorp}
\newsavebox{\toch}

\begin{document}
\sbox{\gorp}{\line(1,0){5}} \sbox{\ver }{\line(0,1){3}}
\sbox{\verp}{\line(0,1){5}} \sbox{\toch}{\circle*{1}}

\begin{flushright}
  FIAN/TD/23--13\\
 \end{flushright}\vspace{2cm}

\begin{center}
{\large\bf Higher-Rank Fields and Currents} \vglue 0.6  true cm \vskip1cm

O.A.~Gelfond$^1$ and M.A.~Vasiliev$^2$ \vglue 0.3  true cm

${}^1$Institute of System Research of Russian Academy of Sciences,\\
Nakhimovsky prospect 36-1, 117218, Moscow, Russia
and \\ I.E.Tamm Department of Theoretical Physics, Lebedev Physical Institute,\\
Leninsky prospect 53, 119991, Moscow, Russia
\vglue 0.3  true cm

 ${}^2$I.E.Tamm Department of Theoretical Physics, Lebedev Physical Institute,\\
Leninsky prospect 53, 119991, Moscow, Russia
 \vskip1.5cm
\date{date}
\end{center}

\begin{abstract}

 $Sp(2M)$  invariant field equations in the space
 $\M_M$ with symmetric matrix coordinates are classified. Analogous results
 are obtained for Minkowski-like subspaces of $\M_M$
which include usual $4d$ Minkowski space as a particular case. The
constructed equations are associated with the tensor products of the Fock
(singleton) representation of $Sp(2M)$ of any rank $\K$.
 The infinite set of higher-spin conserved currents  multilinear in rank-one fields
 in $\M_M$ is found. The associated conserved charges are supported by
$(\K  M-\f{\K (\K -1)}{2})-$dimensional differential forms in $\M_M$, that are closed
by virtue of the rank-$2\K$ field equations. The cohomology groups $H^p(\sigma^\K_-)$
with all $p$ and $\K$,  which determine the form of appropriate gauge
fields and their field equations, are found both for $\M_M$ and for its
Minkowski-like  subspace.

 \end{abstract}

\newpage
\tableofcontents
\newpage

\section{Introduction}\label{Generalities}

As originally observed by Fronsdal \cite{Fr1}, infinite towers of massless
fields in four dimensions admit an extension of the conformal algebra
$\mathfrak{su}(2,2)$ to $\mathfrak{sp}(4)$ allowing a description in the generalized matrix
space $\M_4$  with symmetric matrix coordinates $X^{AB}=X^{BA}$ $(A,B=1,
\ldots,4)$. This observation was
farther elaborated in \cite{BL,BLS,BHS}. The rank-one
$Sp(8)$-invariant unfolded equation describing massless fields of all spins
is \cite{BHS}
 \be
 \label{dydy}
\left( \kk^{AB} \f{\p}{\p X^{AB}} + \gs_-^{1 }\right ) \C(Y|X)
=0\q \gs_-^{1 }=\kk^{AB}\f{\p^2}{\p Y^A \p Y^B} \,,
\ee
where $Y^A$ are
auxiliary commuting variables that will be referred to as {\it twistor
variables}. To simplify formulae we  use  notation $\kk^{AB }$ for
anticommuting differentials $dX ^{AB }$.

 Note that Eq.(\ref{dydy}) admits an interesting interpretation in terms
of world-like particle models of \cite{BL,BLS}, providing also
 a field-theoretical realization of the observation of Fronsdal
 \cite{Fr1}
 that the infinite tower of all $4d$ massless fields
 enjoys $ {Sp}(8)$ symmetry.
More generally $\M_M$ is the space with local coordinates $X^{AB}$ which are symmetric $M\times M$ matrices.
$Sp(2M)$ invariant unfolded field equations corresponding
to rank-$\K$ tensor products of the Fock (singleton) representation of
$Sp(2M)$ were introduced in \cite{tens2} where
these equations were argued to describe  ``branes''
of different dimensions in the $Sp(2M)$ invariant generalized space-time.

 In \cite{tens2}, the case of rank-two   equations
\be \label{dydy2} \left( \xi ^{AB} \f{\p}{\p
X^{AB}} +\gs_-^2 \right ) \C(Y|X) =0\,\q \ee where \be\label{sigma-2}
\gs_-^2= \xi ^{AB}\f{\p^2}{\p Y_1^A \p Y_1^B}+\xi ^{AB}\f{\p^2}{\p Y_2^A \p
Y_2^B} \ee
 was considered in detail. In
particular, all rank-two  dynamical (primary) fields and  field equations
were found and it was shown that dynamical equations for most of the rank-two
fields  have
 the form of conservation conditions for conserved currents found
in \cite{cur}, which give rise to the full set of bilinear conserved charges
in the rank-one  theory.

Rank-$\K$ unfolded equations are \be \label{dydyr} \left( \xi ^{AB} \f{\p}{\p
X^{AB}} + \gs_-^\K \right ) \C(Y|X) =0\,\q \ee where \be\label{sigmar}
\gs_-^\K=
\xi ^{AB}\f{\p^2}{\p Y_k^A \p Y_j^B}\, \gd_{kj} \,    \,\quad ( k,j=1,\ldots,\K)\,. \ee

In this paper the analysis of \cite{cur} is extended to the fields and equations
of arbitrary rank.
Namely we find all dynamical fields, which are primary fields from the
conformal field theory perspective, along with the explicit form of their
field equations. It is shown that, similarly to the rank-two case, some of
these fields give rise to differential forms that are closed by virtue of
their field equations, thus generating conserved currents.

The $sp(2M)$ invariant field equations are appropriate for  description
of infinite towers of massless fields that appear in higher-spin  theories.
For $M=2,4,8,16$, the $Sp(2M)$ invariant field equations describe towers of
conformal massless fields in usual Minkowski spaces of dimensions $d=3,4,6$ and $10$,
respectively  \cite{BLS,Mar,Bandos:2005mb}. Pattern of  $Sp(2M)$ invariant field equations
for other values $M$ so far has not been analysed including the case of
$M=32$ which is most interesting in the context of $M$ theory.

The finite subsets of relativistic fields in
$4d$  Minkowski space are most conveniently described by
the unfolded   equations of motion for massless fields of all spins \cite{Ann,BHS},
 \be \label{minun}
\gx^{\ga \pb}\left(\f{\p}{\p x^{\ga\pb}} +i\,\f{\p^2}{\p y^\ga\p
\by{}^{\pb}}\right)\C(y,\by|x)=0\,.
\ee
Here $y^\ga$ and $\by^\pb$ are
auxiliary commuting complex conjugated two-component spinor variables
($\ga,\gb = 1,2$; $\pa,\pb =  {1},  {2}$), $x ^{\ga\pb}$ are Minkowski
coordinates in two-component spinor notations, and  $ \gx^{\ga \pa }=dx^{\ga
\pa }$ are anticommuting differentials. Equations (\ref{minun}) decompose into
an infinite set of subsystems for fields of different helicities $h$
\be
\C(\mu y,\mu^{-1}\by|x) = \mu^{2h} \C(y,\by|x)\,.
\ee

 The space with coordinates $x^{\ga\pb}$  with $\ga=1,\ldots,K$
and $\pa=1,\ldots,K$ for any $ K$ we
call generalized Minkowski space $\M^{Mnk}_{2K}$. The rank-$\KM$
generalization  of (\ref{minun})   is
 \be \label{minunK} \gx^{\ga \pb}\left(\f{\p}{\p x^{\ga\pb}}
+i\, \frac{\ptl^2}{\ptl y^\ga_k\ptl {\by}^\pb_j}\eta^{kj}
\right)\C(y,\by|x)=0\,\quad
\ee
for any Hermitian form $\eta^{kj}$\,,\,
$k,j=1,\ldots,\KM$\,,\, $ \ga,\pa=1,\ldots,K$.
Though interpretation of these equations for higher $K$ from the perspective
of usual Minkowski space embedded into  $\M^{Mnk}_{2K}$ demands more
detailed analysis which is beyond the scope of this paper, we briefly comment on
the cases of  $K=2,4$ and $8$
being reductions of the $Sp(2M)$ systems with $M=4,8$ and $16$, respectively.\footnote{
We are grateful to the referee for raising this question.}

The case of $K=2$ gives the genuine $4d$ Minkowski space.

The case of $K=4$
results from the reduction of the $Sp(16)$ invariant system. The latter was shown in
\cite{Mar,Bandos:2005mb} to describe conformal massless fields in the $6d$ Minkowski
space carrying spinning degrees of freedom valued in $SU(2)$. As emphasized in
\cite{Bandos:2005mb} this implies that the original $Sp(16)$ invariant system describes
an infinite tower of $6d$ massless fields of all spins such that the multiplicity of
a spin $s$ is $2s+1$ coinciding  with the dimension of the spin-$s$ representation of the
spinning $SU(2)$. Coordinates $x^{\ga\pb}$ can be interpreted as the part of the coordinates
$X^{AB}$ that are invariant under the action of one of the  generators $\Hh$ in the spinning
$\mathfrak{su}(2)$, that acts on the primed and unprimed indices by the conjugated phase transformations.
As a result, the irreducible subsystems in $\M^{Mnk}_{8}$ are characterized  by
$\mathfrak{sl}(2)$ helicities $h$. Each of these systems contains an infinite tower of
  conformal fields in which every spin $s\geq |h|$ appears once ($s$ is related to the
length of the rectangular Young diagram associated with a  $6d$  conformal field in Minkowski
space).

The case of $K=8$
results from the reduction of the $Sp(32)$ invariant system. The latter
 describes a tower of conformal massless self-dual
fields of all spins in the $10d$ Minkowski space \cite{Bandos:2005mb}
such that each spin appears once. Now the reduction of the $Sp(32)$ invariant system
to $\M^{Mnk}_{2K}$ has different interpretation. Since  indices of
the coordinates $X^{AB}$ of  $\M_{16}$ are associated with
$10d$ chiral spinors, the coordinates $x^{\ga\pb}$ cannot be obtained by a $10d$ Lorentz
covariant projection from $X^{AB}$. In other words, the helicity-like operator
distinguishing between primed and unprimed indices cannot be $10d$ Lorentz
covariant. Hence, the  $\M^{Mnk}_{16}$ setup must break manifest $10d$ Lorentz
covariance describing the system in a  Minkowski space of some lower dimension $d<10$.
The simplest option is to identify the helicity-like generator with one of the $10d$ rotations,
considering the system in  eight dimensions. An interesting alternative possibility to be explored
is  to
embed a $9d$ Minkowski space in $\M^{Mnk}_{16}$ using   another helicity operator that would partially act in the spinning space.  In both of these cases
the $10d$ conformal invariance will be a kind of hidden and the same time the self-dual
$10d$ fields will be traded for lower-dimensional fields not restricted by the self-duality
conditions.
We hope to elaborate details of this analysis elsewhere.

In this paper we analyse
the pattern of Minkowski-like equations with general $\KM$ and $K$ from the
perspective of the generalized space  $\M^{Mnk}_{2K}$ with no reference to the
underlying physical space.
However,  since in the particular case of $K=2$ the space $\M^{Mnk}_{4}$ is
  the  usual $4d$
  Minkowsky space, this allows us to derive all conformal primary
currents in the four-dimensional Minkowski space that are built from $4d$
massless fields of all spins. (These results have been already announced and
used in \cite{{Gelfond:2013xt}} for the analysis of the operator algebra and
correlators of conserved currents in four dimensions.)

We expect that results of the present paper may have applications in the
context of $AdS/CFT$ holography \cite{JM,GKP,Wit1} and especially,
higher-spin holography (see, e.g.,
\cite{Klebanov:2002ja,Sezgin:2003pt,Giombi:2009wh,Maldacena:2011jn,
Maldacena:2012sf,Giombi:2012ms,Gaberdiel:2012uj,Vasiliev:2012vf,Alkalaev:2012rg,
Didenko:2012vh,Metsaev:2013wza} and references therein) because, as
emphasized in \cite{BRST2010}, the duality between fields in higher
dimensions and currents in lower dimensions to large extent amounts in the
language of this paper to the duality between lower-rank fields in $\M_M$
with higher $M$ and higher-rank fields in $\M_M$ with lower $M$.

Another interesting application is to the analysis of multiparticle amplitudes
(see e.g. \cite{Elvang:2013cua} and references therein). The key fact here is
that the product of $r$ rank-one solutions
\be
\C(y_i,\bar y_i|x)=
\C(y_1,\bar y_1|x) \C(y_2,\bar y_2|x) \ldots \C(y_\KM,\bar y_\KM|x)
\ee
gives a solution to the rank-$\KM$ equation.
In these terms
an $\KM$-particle amplitude represents a solution to the rank-$\KM$ system.
From this
perspective the constructed conserved multiparticle charges may be of most
interest since amplitudes supported by such charges can be represented as
multiple integrals independent of local variations of the integration cycle.
In this language the nontrivial dynamics of a model in question should be
hidden in the so-called parameters which in our formalism are arbitrary
functions of certain twistor variables, which parameterize different amplitudes.
We hope to come back to a more detailed analysis of this issue in a future publication.

The analysis of the pattern of dynamical fields and their field equations is
performed in the $\sigma_-$-cohomology language which is analogous (and in
many cases equivalent) to the search of singular vectors of conformal
modules. Specifically, zero-form fields and their field equations are
classified by $H^0(\sigma_-)$ and  $H^1(\sigma_-)$, respectively. In this
paper, we find all cohomology groups $H^p(\sigma_-)$ with $p\geq 0$.
 These
results determine the form of appropriate gauge fields and their field
equations both for $\M_M$ and for its Minkowski-like  subspace $\MM  $.

The rest of the paper is organized as follows. Section \ref{Conventions}
contains our Young diagram conventions and some their properties.
In particular, the structure of differential forms in $\M_M$
is analyzed here. Section \ref{spinv} presents the full
list of dynamical fields and field equations of any rank-$\K$ in $\M_M$.
Details of derivation of the equations of motion  as well as the form of
multilinear conserved  currents  in $\M_M$ are also given here. Section \ref{Mres} contains
  summary of our results for conformal fields and their field equations both in the usual
  four-dimensional Minkowski space and in the generalized Minkowski space  $\MM  $.
Section \ref{Details} contains the construction of {\it Homotopy equation}  which determines the $\sigma_-$-cohomology
 in $\M_{ M}$.
{The main tool in the analysis of   Homotopy equation is the  South-West principle} allowing to
minimize the positive semi-definite homotopy operators.
 Details of the analysis
of the  $\sigma_-$-cohomology in $\MM  $ are sketched in Section  \ref{CohomMin}.
 In Section \ref{Conclusions} some perspectives are briefly
discussed. Appendix A
contains
 details of the analysis of the  $\sigma_-$-cohomology
 in $\M_{ M}$.

\section{Young diagrams}
\label{Conventions}

A tensor $A_{B_1^1\ldots\,B_{l_1}^1\,,\ldots,B_1^m \ldots\,B_{l_m}^m\,}$
($l_1\ge...\ge l_m$) obeys symmetry properties of the Young diagram (YD)
${\Y}(l_1,...,l_m)$ with manifest symmetrization provided that $A$ is
symmetric with respect to permutations of
$l_k $ indices $ B_1^k,\ldots,B_{l_k}^k $ of any $k-$th row, while  the
 symmetrization over $l_k+1$ indices $ B_1^k,\ldots,B_{l_k}^k,\,B_j^p $
     yields zero  for any $p>k$.
Such tensors are conventionally   denoted  $A_{B^1( l_1)\,,\ldots,B^m(
l_m)\,}$.

Analogously, a tensor $A_{B_1^1,\ldots,
B_1^{h_1}\,;\ldots;B_p^1,\ldots,B_p^{h_p} \,}$ ($h_1\ge...\ge h_p$) obeys the
symmetry properties of the YD ${\Y}[h_1,...,h_p]$
 with manifest antisymmetrization provided that it is antisymmetric
with respect to permutations of $h_k $ indices $ B_k^1,\ldots,B_k^{h_k}  $ of
any $k-$th column while the antisymmetrization over $h_k+1$ indices $
B_k^1,\ldots,B_k^{h_k},\,B_q^j  $ yields zero for any $q>k$. Such  tensors are
conventionally  denoted  $A_{B_1[ h_1]\,,\ldots,B_p[ h_p]\,}$.

A height of the $k-$th column  of ${\Y} $ is denoted  $ \hei_k({\Y} ) $ or
simply $h_k $, while a length
 of the $k-$th row of  ${ \Y} $ is denoted
$\len_k({\Y})$ or $l_k$. Let us stress that components of a tensor with
symmetry properties of any Young diagram with manifest symmetrization
 are   linear combinations of the components of a tensor
with symmetry properties of  the same Young diagram with manifest
antisymmetrization and vise versa.  {\it Weight} $|\Y|$
 equals to the number of cells of a YD ${\Y}[h_1,...,h_p]$, \ie
 \be
 |{\Y}|:=h_1+...+h_p=l_1+...+l_m\,.
 \ee

Any Young diagram can be represented as the unification of its elementary
cells $\mathcal{S}(i\,,j)$ on the intersection of its $i-th$ row  and $j-th$
column, \ie $\dis{\Y=\bigcup_{\mathcal{S}(i,j)\in \Y} \mathcal{S}(i,j)}$. For
any cell $\mathcal{S}(i\,,j)$ and parameter $a\in \mathbb{R}$ we introduce
the  \emph{characteristic function}
\be\label{chicaz1S}
\chi^a(\mathcal{S}(i\,,j))=  (j-i+a)\,. \ee
For any set of cells $\mathcal{A}$ we introduce its characteristic function\footnote{We are grateful to Andrey Mironov
for bringing to our attention the paper \cite{mir} where the function $n_\gl=2\chi^0(\mathcal{\gl})$
characterized YD $\gl$  was used.}
\be\label{chicaz} \chi^a(\mathcal{A}):=
 \sum_{\mathcal{S}(i\,,j)\in \mathcal{A}}\chi^a(\mathcal{S}(i\,,j))=
  \sum_{\mathcal{S}(i\,,j)\in \mathcal{A}}(j-i+a)\,.
\ee
Considering a diagram  $ \Y[h_1,\ldots ,h_k]$ as the unification of either
its columns $\mathbf{H}_j(h_j)$ or  rows  $\mathbf{L}_j(l_j)$ we obtain
 \be
\label{sumH} \chi^a(\Y)
=\sum_{\mathbf{H}_j\subseteq \Y}\,\, \sum_{\mathcal{S}(i\,,j)\in
\mathbf{H}_j}(j-i +a )
 =-\half  \sum_{j} h_j(h_j -2j+1-2a)\, \ee  and
 \be\label{sumL}
\chi^a(\Y)=\sum_{ \mathbf{L}_i\subseteq \Y}\,\, \sum_{\mathcal{S}(i\,,j)\in
\mathbf{L}_i}(j-i+a ) =\half  \sum_{i  }
 l_i(l_i -2 i+1 +2a)\,.\ee
 Hence, the right-hand sides of
(\ref{sumH}) and (\ref{sumL}) are equal for any YD $\Y$.

Cells of a Young diagram can be ordered  as follows:
for any two cells $\S_1$ and $\S_2$, $\S_1\prec  \S_2$ if $\chi^a (\S_1)<
\chi^a (\S_2)$ and $\S_1\preceq  \S_2$ if $\chi^a (\S_1)\le
\chi^a (\S_2)$. (Note that this definition is insensitive to $a$.)
Definition (\ref{chicaz}) implies that, with respect to this ordering, for
any two cells the smaller is situated {\it South-West}  to the larger.

If $\S^i_1\prec \S^i_2$  ($\S^i_1\preceq \S^i_2$) for $i=1,\ldots,N$ then
$\A_1\prec \A_2$  ($\A_1\preceq \A_2$) where $\A_{1,2}=
 \bigcup_{i\le N} \,\mathcal{S}_{1,2}^i $. Thus the partial ordering can be
 introduced  in particular for Young  diagrams containing    equal numbers of
 cells.

The symmetrized tensor product
 of a number of rank-two symmetric tensors $\delta_{ij}$  described by
the YD $\Y[1,1]$
decomposes into a linear combination of tensors with the symmetry
properties \be\label{prodsym2}
 \begin{picture}( 50,40)(  -40,23)%
\put(-127,40){\small$\Y_\gd[d_1,d_1,d_2,d_2,\ldots,d_m,d_m ] =$}%
 \begin{picture}(150,40)(-22,-25)%
\put(43,27){\scriptsize $d_m$} \put(13, -03){\scriptsize $d_2$}
\put(0,-08){\scriptsize $d_1$}
\multiput(0,00)(5,0){02}{\usebox{\gorp}}%
\multiput(0,00)(5,0){03}{\usebox{\verp}}%
\multiput(0,05)(5,0){04}{\usebox{\gorp}}%
\multiput(0,05)(5,0){05}{\usebox{\verp}}%
\multiput(0,10)(5,0){6}{\usebox{\gorp}}%
\multiput(0,10)(5,0){7}{\usebox{\verp}}%
\multiput(0,15)(5,0){6}{\usebox{\gorp}}%
\multiput(0,15)(5,0){7}{\usebox{\verp}}%
\multiput(0,20)(5,0){6}{\usebox{\gorp}}%
\multiput(0,20)(5,0){7}{\usebox{\verp}}%
\multiput(0,25)(5,0){8}{\usebox{\gorp}}%
\multiput(0,25)(5,0){9}{\usebox{\verp}}%
\multiput(0,30)(5,0){9}{\usebox{\verp}}%
\multiput(0,30)(5,0){8}{\usebox{\gorp}}%
\multiput(0,35)(5,0){10}{\usebox{\gorp}}%
\multiput(0,35)(5,0){11}{\usebox{\verp}}%
\multiput(0,40)(5,0){10}{\usebox{\gorp}}%
\multiput(0,40)(5,0){11}{\usebox{\verp}}%
\multiput(0,45)(5,0){10}{\usebox{\gorp}}%
 \end{picture}
\end{picture}\qquad\qquad\,.
\ee This is because  antisymmetrization of indices $i_1, i_2,\ldots,  i_d$,
say, in $\delta_{i_1j_1} \ldots \delta_{i_d j_d}$   implies
antisymmetrization of indices $j_1, j_2,\ldots,  j_d$. Components of the
symmetrized tensor product of a number of $\Y[1,1]$
\be\label{Krond}\Y_\gd[d_1,d_1,
\ldots,d_m,d_m]\in \left(\otimes_{sym} \Y[1,1]\right) \, \ee
 will be referred to as Kronecker diagrams.

On the other hand, as shown below,
the antisymmetrized tensor product of tensors with the symmetry properties of
 $\Y[1,1]$
 contains various tensors with the symmetry properties of {\it almost symmetric} Young
diagrams $\Y_{A}$ defined as follows. For any  Young diagram invariant under
reflection with respect to the diagonal, one cell should be added to each row
that intersects the diagonal. Pictorially, shading the added cells, \bee
\label{exalsim}
 \begin{picture}(40,45)(00,-8)
 \multiput(0,35)(1 ,-1 ){20}{\usebox{\toch}}%
\multiput(0,-5)(5,0){01}{\usebox{\gorp}}
\multiput(0,00)(5,0){01}{\usebox{\gorp}}
\multiput(0,05)(5,0){02}{\usebox{\gorp}}
\multiput(0,10)(5,0){03}{\usebox{\gorp}}
\multiput(0,15)(5,0){03}{\usebox{\gorp}}
\multiput(0,35)(5,0){08}{\usebox{\gorp}}
\multiput(0,30)(5,0){08}{\usebox{\gorp}}
\multiput(0,25)(5,0){06}{\usebox{\gorp}}
\multiput(0,20)(5,0){05}{\usebox{\gorp}}
 \multiput(0,30)(5,0){09}{\usebox{\verp}}
\multiput(0,25)(5,0){07}{\usebox{\verp}}
\multiput(0,20)(5,0){06}{\usebox{\verp}}
\multiput(0,15)(5,0){04}{\usebox{\verp}}
\multiput(0,10)(5,0){04}{\usebox{\verp}}
\multiput(0,05)(5,0){03}{\usebox{\verp}}
\multiput(0,00)(5,0){02}{\usebox{\verp}}
\multiput(0,-5)(5,0){02}{\usebox{\verp}}
\put(-6,-16 ){\small   Symmetric $\Y_{S}$}
\end{picture}
\qquad \qquad\Rightarrow\qquad\qquad
 \begin{picture}(40,40)(00,-8)
 \multiput(0,35)(1 ,-1 ){20}{\usebox{\toch}}
 \multiput(0,-5)(5,0){01}{\usebox{\gorp}}
\multiput(0,00)(5,0){01}{\usebox{\gorp}}
\multiput(0,05)(5,0){02}{\usebox{\gorp}}
\multiput(0,10)(5,0){03}{\usebox{\gorp}}
\multiput(0,15)(5,0){03}{\usebox{\gorp}}
\multiput(0,35)(5,0){09}{\usebox{\gorp}}
\multiput(0,30)(5,0){09}{\usebox{\gorp}}
\multiput(0,25)(5,0){07}{\usebox{\gorp}}
\multiput(0,20)(5,0){06}{\usebox{\gorp}}
\multiput(25,20)(1,0){05}{\usebox{\verp}}
\multiput(30,25)(1,0){05}{\usebox{\verp}}
\multiput(40,30)(1,0){05}{\usebox{\verp}}
 \multiput(0,30)(5,0){10}{\usebox{\verp}}
\multiput(0,25)(5,0){08}{\usebox{\verp}}
\multiput(0,20)(5,0){07}{\usebox{\verp}}
\multiput(0,15)(5,0){04}{\usebox{\verp}}
\multiput(0,10)(5,0){04}{\usebox{\verp}}
\multiput(0,05)(5,0){03}{\usebox{\verp}}
\multiput(0,00)(5,0){02}{\usebox{\verp}}
\multiput(0,-5)(5,0){02}{\usebox{\verp}}
\put(-8,-16 ){\small  Almost symmetric $\Y_{A}$}
 \end{picture}.
\eee
The simplest one   is the almost symmetric hook of height $h$
\be\label{hhook}\Y[h,\underbrace{1,\ldots,1}_{h }]=\quad \left\{\rule{0pt}{25pt}\right.\!
 \overbrace{{\begin{picture}(45,25)(0,-18) \put(-17,-23){\scriptsize{$h$} }
 \put( 25,14){\scriptsize{$h+1$} }
\put(00,00){\line(1,0){45}}%
\put(00,05){\line(1,0){45}}%
\multiput(00,00)(5,0){10}{\usebox{\verp}}%
\begin{picture}(05,05)(0,35)%
\put(00,00){\line(0,1){35}}%
\put(05,00){\line(0,1){35}}%
\multiput(00,00)(0,5){8}{\usebox{\gorp}}%
\end{picture}\end{picture}
}  }.
 \ee
For instance,
     the full list of  almost symmetric
 Young diagrams  belonging to    $\dis{\otimes{}_{asym}^n \Y[1,1]}$ with $ {n\le5}$  is  \\%
$\bullet$\quad n=1:\quad $\Y(2)$ 
\begin{picture}(15,15)
{
\put(00,00){\line(1,0){10}}%
\put(00,05){\line(1,0){10}}%
\multiput(00,00)(5,0){3}{\usebox{\verp}}%
}
\end{picture}\,\,\,;\\
$\bullet $\quad n=2:\quad  $\Y(3,1)$ 
\begin{picture}(15,15)(0,-5)
{
\put(00,00){\line(1,0){15}}%
\put(00,05){\line(1,0){15}}%
\multiput(00,00)(5,0){4}{\usebox{\verp}}%
\begin{picture}(05,05)(0,05)%
\put(00,00){\line(0,1){05}}%
\put(05,00){\line(0,1){05}}%
\put(00,00){\line(1,0){5}}%
\end{picture}
}
\end{picture}\,\,\,;  \\
$\bullet $\quad n=3:\quad
 $\Y(4,1,1)=$
\begin{picture}(20,15)(0,-5)
{
\put(00,00){\line(1,0){20}}%
\put(00,05){\line(1,0){20}}%
\multiput(00,00)(5,0){5}{\usebox{\verp}}%
\begin{picture}(05,05)(0,10)%
\put(00,00){\line(0,1){10}}%
\put(05,00){\line(0,1){10}}%
\multiput(00,00)(0,5){2}{\usebox{\gorp}}%
\end{picture}
}
\end{picture}\,
and $\Y(3,3)=$
\begin{picture}(15,15)(0,-5)
{
\put(00,00){\line(1,0){15}}%
\put(00,05){\line(1,0){15}}%
\multiput(00,00)(5,0){4}{\usebox{\verp}}%
\begin{picture}(05,05)(0,05)%
\put(00,00){\line(1,0){15}}%
\multiput(00,00)(5,0){4}{\usebox{\verp}}%
\end{picture}
}
\end{picture}\,\,;
\\
$\bullet $\quad n=4:\quad $\Y(5,1,1,1)=$
\begin{picture}(25,10)(0,-5)
{
\put(00,00){\line(1,0){25}}%
\put(00,05){\line(1,0){25}}%
\multiput(00,00)(5,0){6}{\usebox{\verp}}%
\begin{picture}(05,05)(0,15)%
\put(00,00){\line(0,1){15}}%
\put(05,00){\line(0,1){15}}%
\multiput(00,00)(0,5){3}{\usebox{\gorp}}%
\end{picture}
}
\end{picture}\,
and $\Y(4,3,1)=$
\begin{picture}(20,10)(0,-0)
{
\put(00,05){\line(1,0){20}}%
\put(00,10){\line(1,0){20}}%
\multiput(00,05)(5,0){5}{\usebox{\verp}}%
\put(00,00){\line(1,0){15}}%
\multiput(00,00)(5,0){4}{\usebox{\verp}}%
\begin{picture}(05,10)(0,05)%
\put(00,00){\line(0,1){05}}%
\put(05,00){\line(0,1){05}}%
\multiput(00,00)(0,5){1}{\usebox{\gorp}}%
\end{picture}
}
\end{picture}\,\,;\\
$\bullet $\quad n=5:\quad  $\Y(6,1,1,1,1)=$ %
\begin{picture}(30,25)(0,-5)
{\put(00,00){\line(1,0){30}}%
\put(00,05){\line(1,0){30}}%
\multiput(00,00)(5,0){7}{\usebox{\verp}}%
\begin{picture}(10,35)(0,20)%
\put(00,00){\line(0,1){20}}%
\put(05,00){\line(0,1){20}}%
\multiput(00,00)(0,5){4}{\usebox{\gorp}}%
\end{picture}
}
\end{picture}\,
, $\Y(5,3,1,1)=$
\begin{picture}(25,05)(0,-5)
{
\put(00,00){\line(1,0){25}}%
\put(00,05){\line(1,0){25}}%
\multiput(00,00)(5,0){6}{\usebox{\verp}}%
\begin{picture}(05,05)(0,15)%
\put(05,10){\line(1,0){10}}%
\multiput(10,10)(5,0){2}{\usebox{\verp}}%
\put(00,00){\line(0,1){15}}%
\put(05,00){\line(0,1){15}}%
\multiput(00,00)(0,5){3}{\usebox{\gorp}}%
\end{picture}
}
\end{picture}\,
and $\Y(4,4,2)=$
\begin{picture}(20,05)(0,-0)
{
\put(00,05){\line(1,0){20}}%
\put(00,10){\line(1,0){20}}%
\multiput(00,05)(5,0){5}{\usebox{\verp}}%
\put(00,00){\line(1,0){20}}%
\multiput(00,00)(5,0){5}{\usebox{\verp}}%
\begin{picture}(05,05)(0,05)%
\put(00,00){\line(1,0){10}}%
\multiput(00,00)(5,0){3}{\usebox{\verp}}%
\end{picture}
}
\end{picture}\,.\\\\

Any almost symmetric diagram $\Y_{A}$ (\ref{exalsim}) admits a
 {\it nested hook}  realization, namely
$\Y_{A}$
is a unification of a
set of {\it shifted almost symmetric hooks} $\Hhs( k,a_k)$
\be\label{nestedhooks}  \Y_A = \Yn^{nest} \{a_1,\ldots a_\la\}=\bigcup_{1\le
k\le \la}\Hhs( k,a_k) \q a_j\ge a_{j+1}+1\,,  \quad   a_{\la }
\ge 1\,, \quad
 \ee
where
$\Hhs( k,a_k)$ is the almost symmetric hook 
(\ref{hhook}) with $h=a_k$
shifted by $k-1$ cells down along the diagonal.
More precisely,
\be\label{shiftedhook} \Hhs( k,a_k)= \bigcup_{j=1}^{a_k } \left(\S(k,k+j)
\bigcup
 \S(j+k-1,k)\,\right)\,.
\ee
  For example, $\Hhs( 1,h )=\Y[h ,\underbrace{1,\ldots,1}_{h }]$ (\ref{hhook}).

Pictorially, in the almost symmetric diagram shown below the bolded almost
symmetric hook is $\Hhs( 2,5)$ while  the other
two are $\Hhs( 1,8)$ and $\Hhs( 3,3)$
\sbox{\gorp}{\linethickness{.250mm}\line(1,0){10}}
\sbox{\verp}{\linethickness{.250mm}\line(0,1){10}} \bee\label{nested411}
    \begin{picture}(80,90)(-80,-20)
\multiput(0,-10)(10,0){01}{\usebox{\gorp}}
\multiput(0,00)(10,0){01}{\usebox{\gorp}}
\multiput(0,10)(10,0){02}{\usebox{\gorp}}
\multiput(0,20)(10,0){03}{\usebox{\gorp}}
\multiput(0,30)(10,0){03}{\usebox{\gorp}}
\multiput(0,70)(10,0){09}{\usebox{\gorp}}
\multiput(0,60)(10,0){09}{\usebox{\gorp}}
\multiput(0,50)(10,0){07}{\usebox{\gorp}}
\multiput(0,40)(10,0){06}{\usebox{\gorp}}
\multiput(0,60)(10,0){10}{\usebox{\verp}}
\multiput(0,50)(10,0){08}{\usebox{\verp}}
\multiput(0,40)(10,0){07}{\usebox{\verp}}
\multiput(0,30)(10,0){04}{\usebox{\verp}}
\multiput(0,20)(10,0){04}{\usebox{\verp}}
\multiput(0,10)(10,0){03}{\usebox{\verp}}
\multiput(0,00)(10,0){02}{\usebox{\verp}}
\multiput(0,-10)(10,0){02}{\usebox{\verp}}
{\linethickness{.70mm}  %
%
\put(16.5,50) {  \line(1,0){50}}
\put(6.5,60) {  \line(1,0){60}}
\put(66.5,50) {  \line(0,1){10}}
 \put(6.5,60) {  \line(0,-1){50}}
 \put(16.5,50) {  \line(0,-1){40}}
 \put(6.5,10) {  \line(1,0){10}}
} \put(-240, 20){$ \Yn^{nest}\{8,5,3 \}=\Hhs( 1,8)\cup\Hhs( 2,5)\cup\Hhs( 3,3)
=\qquad\qquad\,\,\,\,\,.$}
  \end{picture}
\eee
Note that by definition   \be\label{nesthn}
\hei_{ \la+1}(\Yn^{nest}\{a_1,\ldots a_\la\})=\la:=\sharp(\Yn^{nest}\{a_1,\ldots a_\la\})\,.\ee
For instance, the height of $4$-th column of $ \Yn^{nest}\{8,5,3 \} $ (\ref{nested411})
is $3$.

The  nested   realization  (\ref{nestedhooks}) of  almost symmetric diagrams $\Y_{A}$ (\ref{exalsim})
yields that  all of them obey
\be\label{almostchi}
 \chi^q(\Y_A)
  =  \left ( q+\half \right ) \big|\Y_A\big|\,.  \ee
Indeed, by definition (\ref{chicaz1S})
any pair of cells symmetric with respect to the diagonal does not contribute to
$\chi^0$,
hence for an  almost symmetric shifted hook $ \Hhs( k,h_k ) $ (\ref{shiftedhook})
 $$
 \chi^0(\Hhs( k,h_k ))= h_k=\half \big| \Hhs( k,h_k )\big|\,.
 $$ Summation over all shifted
 almost symmetric hooks gives (\ref{almostchi}).

It is  convenient to introduce {\it block hooks} of the form
$\Yn{}^{nest} \{  \underbrace{m,m-1, \ldots ,m-n }_{n+1}\}$ with arbitrary integer $m>n$.
Pictorially,
\bee\label{nested122+}
  \begin{picture}(77,100)(-20,-10)
\multiput(-20,-10)( 0,10){11}{\usebox{\gorp}}%
\multiput(-10,-10)( 0,10){11}{\usebox{\gorp}}
\multiput(- 0,-10)( 0,10){11}{\usebox{\gorp}}
\put(-20,-10){\linethickness{.250mm}\line(0,1){100}}%
\put(-10,-10){\linethickness{.250mm}\line(0,1){100}}%
\put(-0,-10){\linethickness{.250mm}\line(0,1){100}}%
\put( 10,-10){\linethickness{.250mm}\line(0,1){100}}%
\multiput(0,70)(10,0){10}{\usebox{\verp}}
\multiput(0,80)(10,0){10}{\usebox{\verp}}
\multiput(0,60)(10,0){10}{\usebox{\verp}}
\put(0,90) {\linethickness{.250mm}\line(1,0){90}}
\put(0,80){\linethickness{.250mm}\line(1,0){90}}
\put(0,70) {\linethickness{.250mm}\line(1,0){90}}
\put(0,60){\linethickness{.250mm}\line(1,0){90}}
{   %
 \put(-120 , 20){ $
 \Yn{}^{nest} \{ {10,9,8}   \,\}
= $}
}
  \end{picture}\quad
   \begin{picture}(80,100)(-160, 30)
\multiput(-20,50)( 0,10){5}{\usebox{\gorp}}%
\multiput(-10,50)( 0,10){5}{\usebox{\gorp}}
\put(-20,50){\linethickness{.250mm}\line(0,1){40}}%
\put(-10,50){\linethickness{.250mm}\line(0,1){40}}%
\put(-0,50){\linethickness{.250mm}\line(0,1){40}}%
\multiput(0,70)(10,0){4}{\usebox{\verp}}
\multiput(0,80)(10,0){4}{\usebox{\verp}}
\put(0,90) {\linethickness{.250mm}\line(1,0){30}}
\put(0,80){\linethickness{.250mm}\line(1,0){30}}
\put(0,70) {\linethickness{.250mm}\line(1,0){30}}
\put(-100 , 60){ $
 \Yn{}^{nest} \{ {4,3}   \,\}
= $}
   \end{picture}
\eee
A general almost symmetric diagram
  can be treated as {\it  nested
block hooks} consisting of $\ppi$ block hooks
\bee\label{nested'blocs}
 \Y_A=\Yn{}^{nest} \{\underbrace{\ath_1  ,\ath_1-1 ,\ldots ,\ath_1-n_1+1}_{n_1}\,,\,\ldots\,,\,
 \underbrace{\ath_\ppi\,,\ath_\ppi-1 \,,\ldots ,\,\ath_\ppi-n_\ppi+1}_{n_\ppi}  \,\}\,
 \eee under conditions\bee \nn n_1+\ldots+n_\ppi=\la\q
 \ath_i-n_i>\ath_{i+1}>0 \quad\mbox{for }i+1\le \ppi\,\q \ath_\ppi\ge n_\ppi,\qquad\eee
where $n_i$ is the number of hooks in the $i^{th}$ block and $\la$ is the total
  number  of  hooks in  $ \Y_A $.

Let  $\gx^{AB}$ be anticommuting differentials
  \be \label{ksigen}
 \kk^{AB }=\kk^{BA }\q
\kk^{AB}\kk^{C \ddd }=-\kk^{C \ddd}\kk^{AB }\,.
 \ee
The coefficients of differential forms
 \be \label{2antisym} Q_{\dda_1 \ddb_1; \ldots
;\dda_n \ddb_n }\xi ^{\dda_1 \ddb_1} {}\ldots{} \xi ^{\dda_n \ddb_n}
\ee
belong to the space of antisymmetric tensor products of $Y[1,1]$.
To show that they are described by almost symmetric
diagrams we observe that anticommutativity
of the differentials implies that the
projection of $ \gx^{AB}\gx^{CD}$ to the window diagram is zero \be
\label{wind} \gx^{AB}\gx^{CD} \Big |_{\begin{picture}(13,12)(0,0)
{
\put(00,10){\line(1,0){10}} \put(00,05){\line(1,0){10}}
\put(00,00){\line(1,0){10}} \put(00,00){\line(0,1){10}}
\put(05,00){\line(0,1){10}} \put(10,00){\line(0,1){10}} }
\end{picture}}=0\,.\ee
From here it follows that the product of $n$ variables $\gx$ antisymmetrized over
$n$ indices  has the symmetry  of the hook $\Y[
n,\underbrace{1,\ldots,1}_{ n}]$. Indeed, antisymmetrization
 over   $n$ indices of $n$ variables $\gx$ implies symmetrization over the other
 $n$ indices. There are two   diagrams containing a symmetrization over $n$ indices
 \sbox{\gorp}{\line(1,0){05}}
\sbox{\verp}{\line(0,1){05}} \be
\begin{picture}(30,15)(0,-20)
 {\put(-10,-15){\scriptsize$n$}%
 \put( 10, 8){\scriptsize$n+1$}%
 \put(00,00){\line(1,0){30}}%
\put(00,05){\line(1,0){30}}%
\multiput(00,00)(5,0){7}{\usebox{\verp}}%
\put(00,-20){\line(0,1){20}}%
\put(05,-20){\line(0,1){20}}%
\multiput(00,-20)(0,5){4}{\usebox{\gorp}}%
}\end{picture}\,\q\qquad
\begin{picture}(30,25)(0,-20)
{\put(-10,-15){\scriptsize$n$}%
 \put( 10, 8){\scriptsize$n $}%
 \put(00,00){\line(1,0){25}}%
\put(00,05){\line(1,0){25}}%
\multiput(00,00)(5,0){6}{\usebox{\verp}}%
\put(05,-5){\line(1,0){05}}%
\put(10,-5){\line(0,1){05}}%
\put(00,-20){\line(0,1){20}}%
\put(05,-20){\line(0,1){20}}%
\multiput(00,-20)(0,5){4}{\usebox{\gorp}}%
}
\end{picture}\,.
 \ee
However the one containing window is zero by  (\ref{wind}).
Due to antisymmetrization in every column, this implies
that any Young diagram associated with a differential form has the nested
hook structure, \ie is almost symmetric.

\section{$Sp(2M)$ invariant space}
\label{spinv}
\subsection{Fields and equations}
\subsubsection{Lower-rank examples}
\label{lowerspinex}
From  the rank-one $Sp(2M)$-invariant unfolded equation (\ref{dydy})  it follows that most of the component fields in the expansion
\be \C(Y|X) = \sum_{n=0}^\infty Y^{A_1} \ldots Y^{A_n} C_{A_1\ldots
A_n}(X) \ee are reconstructed in terms of ($X$-derivatives of) the primary
fields that  satisfy \be\label{din1}
  \gs_-^{1 }  C(Y|X) =0\,.
\ee In the rank-one case, the primary  fields are \cite{BHS}
\be\label{field1} C(X):= \C(0|X)\q C_{A }(X)\,Y^{A }. \ee The symmetry
properties of $C$ and $C_A$ are represented by the Young diagrams $\bullet$
(empty diagram) and $\begin{picture}(05,13)(05,-01)
{
\put(05,05){\line(1,0){05}}%
\put(05,0){\line(1,0){05}}%
\put(05,0){\line(0,1){05}}%
\put(10,0.0){\line(0,1){05}} }
\end{picture}$\,,
respectively.

The following equations hold as a consequence of  unfolded equations (\ref{dydy}):
\bee\label{fieldeq1F}  \left( \frac{\ptl}{\ptl X^{ A C}}
\frac{\ptl}{\ptl X^{ B  D}} -\frac{\ptl}{\ptl X^{ B  C}} \frac{\ptl}{\ptl X^{
A D }} \right)
\C (Y|X) &=&0\q\\\nn
\left(
 \frac{\ptl}{\ptl X^{ B  D }}\frac{\ptl}{\ptl Y^{ A  }}-
\frac{\ptl}{\ptl X^{ A D }} \frac{\ptl}{\ptl Y^{B }}\right)\C (Y|X)&=&0\,
\eee and, in particular, \bee\label{fieldeq1}  \left( \frac{\ptl}{\ptl X^{ A
C}} \frac{\ptl}{\ptl X^{ B  D}} -\frac{\ptl}{\ptl X^{ B  C}} \frac{\ptl}{\ptl
X^{ A D }} \right) C(X) &=&0\q\\ \label{fieldeq2}
 \frac{\ptl}{\ptl X^{ B  D }}C_{A}(X)-
\frac{\ptl}{\ptl X^{ A D }} C_{B}(X)&=&0\,. \eee
The symmetry properties of
the left-hand sides of equations (\ref{fieldeq1}) and (\ref{fieldeq2})
are represented by the Young diagrams $\begin{picture}(13,12)(0,0)
{
\put(00,10){\line(1,0){10}} \put(00,05){\line(1,0){10}}
\put(00,00){\line(1,0){10}} \put(00,00){\line(0,1){10}}
\put(05,00){\line(0,1){10}} \put(10,00){\line(0,1){10}} }
\end{picture}$ and
$\begin{picture}(13,12)(0,0)
{
\put(00,00){\line(1,0){05}} \put(00,05){\line(1,0){10}}
\put(00,10){\line(1,0){10}} \put(00,00){\line(0,1){10}}
\put(05,00){\line(0,1){10}} \put(10,05){\line(0,1){05}} }
\end{picture}$\,, respectively.
In the language of $\sigma_-$--cohomology \cite{SVsc} (see also
\cite{Bekaert:2005vh}) convenient for the analysis of the pattern of
zero-form higher-rank fields, primary fields  and their field equations are
represented by the cohomology groups $H^{0 }(\gs_-^{1})$ and $H^{1}(\gs_-^{1}
)$, respectively. Hence,  the structure of rank-one fields and field
equations is represented by the following diagrams
 \bee\label{cor1}
\begin{tabular}{|c|c|c|}
\hline
\rule[-3pt]{0pt}{23pt}
$H^0(\gs_-^{1 }):\Y_0{}$& $\bullet$ &
\begin{picture}(05,13)(05,00)
{
\put(05,05){\line(1,0){05}}%
\put(05,0){\line(1,0){05}}%
\put(05,0){\line(0,1){05}}%
\put(10,0.0){\line(0,1){05}}
}
\end{picture}\\
\hline
\rule[-3pt]{0pt}{23pt}
$H^1(\gs_-^{1 }):\Y_1{}$&
\begin{picture}(13,12)(0,0)
{
\put(00,10){\line(1,0){10}}
\put(00,05){\line(1,0){10}}
\put(00,00){\line(1,0){10}}
\put(00,00){\line(0,1){10}}
\put(05,00){\line(0,1){10}}
\put(10,00){\line(0,1){10}}
}
\end{picture}&
\begin{picture}(13,12)(0,0)
{
\put(00,00){\line(1,0){05}}
\put(00,05){\line(1,0){10}}
\put(00,10){\line(1,0){10}}
\put(00,00){\line(0,1){10}}
\put(05,00){\line(0,1){10}}
\put(10,05){\line(0,1){05}}
}
\end{picture}\\
\hline
\end{tabular}\,.
\eee

Note that Eqs.~(\ref{fieldeq1}), (\ref{fieldeq2}) are the only independent
equations obeyed by the primary (=dynamical) fields as a consequence of
(\ref{dydy}). Such equations will be referred to as dynamical.

Rank-two unfolded equations (\ref{dydy2}) are most conveniently analyzed
in terms of variables \be\label{twistorz}
  {2} z^A= Y^A_1+i {Y}_2^A\,,\qquad
  {2}{\bz}^A= Y^A_1-i {Y}_2^A\,
\ee
with $\gs_-^2$ (\ref{dydy2}) taking the form
\be \label{unfz}
\gs_-^2= \, \xi ^{AB}\,\f{\p^2}{\p z^A \p \bz^B}\,.
 \ee

The following homogeneous differential equations hold as a consequence of
(\ref{unfz}) \bee\label{fieldeq2F}\ls\ls \gvep^{B_1,B_2,B_3 } \frac{\ptl^3}{\ptl
X^{ B_1 D_1} \ptl X^{ B_2 D_2}\ptl
  X^{ B_3 D_3}}\C (z,\bz|X) =0\,,&&
\gvep^{B_1,B_2,B_3 } \frac{\ptl^3}{\ptl X^{ B_1 D }\ptl z^{ B_2}\ptl \bz^{
B_3}}
  \C (z,\bz|X) =0\q\\\nn
\gvep^{B_1,B_2,B_3 } \frac{\ptl^3}{\ptl X^{ B_1 D_1} \ptl X^{ B_2 D_2}\ptl
z^{ B_3}}
 \C (z,\bz|X) =0\,,&&
\gvep^{B_1,B_2,B_3 } \frac{\ptl^3}{\ptl X^{ B_1 D_1} \ptl X^{ B_2 D_2}\ptl
\bz^{ B_3}}
 \C (z,\bz|X) =0\q\\ \nn
\gvep^{A ,B  }\gvep^{C ,D  } \frac{\ptl^3}{\ptl X^{ B  C}\ptl z^{A}\ptl z^{
D}} \C (z,\bz|X)=0\,,&&\gvep^{A ,B  }\gvep^{C ,D  } \frac{\ptl^3}{\ptl X^{ B
C}\ptl \bz^{A}\ptl \bz^{ D}} \C (z,\bz|X)=0 \,, \qquad\eee
 where arbitrary  rank-$m$ totally antisymmetric tensors  $\gvep^{A_1 ,\ldots, A_m} $ are
 introduced to impose appropriate antisymmetrizations.

As shown in \cite{tens2},   the  rank-two primary fields are
\bee\label{field2}\!\!\!& C_{A_1\ldots A_m }(X)\,z^{A_1 }\!\ldots  z^{A_m
}\!\!,\quad
 \overline{C}_{A_1\ldots A_m }(X) \bz^{A_1 }\!\ldots \bz^{A_m }
 \!\!,\quad
C_{A,B}~(X) z^{A}\bar{z}^{B} \quad ( C_{A,B}=-C_{ B,A}  )\,.\qquad& \eee
Dynamical   equations  for the primary fields are \cite{tens2}
 \bee\label{fieldeq22} \gvep^{B_1,B_2,B_3 }
\frac{\ptl}{\ptl X^{ B_1 D_1}} \frac{\ptl}{\ptl X^{ B_2 D_2}}
\frac{\ptl}{\ptl X^{ B_3 D_3}}C (X) =0\,,&& \gvep^{B_1,B_2,B_3 }
\frac{\ptl}{\ptl X^{ B_1 D }}
 C_{B_2,B_3 } (X) =0\q\\ \nn
\gvep^{B_1,B_2,B_3 } \frac{\ptl}{\ptl X^{ B_1 D_1}} \frac{\ptl}{\ptl X^{ B_2
D_2}} C_{B_3}(X) =0\,,&& \gvep^{B_1,B_2,B_3 } \frac{\ptl}{\ptl X^{ B_1 D_1}}
\frac{\ptl}{\ptl X^{ B_2 D_2}} \overline{C}_{B_3}(X) =0\q\\ \nn \gvep^{E ,B
}\gvep^{C ,D  } \frac{\ptl}{\ptl X^{ B  C}}\overline{C}_{E\,D\,A(m-2)}(X)=0\,,&&
\gvep^{E ,B  }\gvep^{C ,D  } \frac{\ptl}{\ptl X^{ B
C}}C_{E\,D\,A(m-2)}(X)=0 \,. \qquad\eee Hence, the list of the  Young
diagrams associated with
    $H^{0 }(\gs_-^{2})$ and  $H^{1}(\gs_-^{ 2} )$ is
  \bee\label{cor2} \begin{tabular}{|c| c| c| c|c|}
\hline \rule[-3pt]{0pt}{23pt} $H^0(\gs_-^{2}):\Y_0{}$& $\bullet$ &
\begin{picture}(05,13)(05,00)
{
\put(05,05){\line(1,0){05}}%
\put(05,0){\line(1,0){05}}%
\put(05,0){\line(0,1){05}}%
\put(10,0.0){\line(0,1){05}} }
\end{picture}&
\begin{picture}(05,13)(05,00)
{
\put(05,05){\line(1,0){05}}%
\put(05,10){\line(1,0){05}}%
\put(05,0){\line(1,0){05}}%
\put(05,0){\line(0,1){10}}%
\put(10,0.0){\line(0,1){10}} }
\end{picture}&
\begin{picture}(40,10)
{
\put(00,00){\line(1,0){40}}%
\put(00,05){\line(1,0){40}}%
\put(00,00){\line(0,1){05}}%
\put(05,00.0){\line(0,1){05}} \put(10,00.0){\line(0,1){05}}
\put(15,00.0){\line(0,1){05}} \put(20,00.0){\line(0,1){05}}
\put(25,00.0){\line(0,1){05}} \put(30,00.0){\line(0,1){05}}
\put(35,00.0){\line(0,1){05}} \put(40,00.0){\line(0,1){05}} }
\put(16,6.2){\scriptsize  $2+{n}$}
\end{picture}\\
\hline \rule[-3pt]{0pt}{23pt} $H^1(\gs_-^{2}):\Y_1{}$&
\begin{picture}(10,20)
{
\put(00,05){\line(1,0){10}}%
\put(00,10){\line(1,0){10}}%
\put(00,15){\line(1,0){10}}%
\put(00,00){\line(1,0){10}}%
\put(00,00){\line(0,1){15}}%
\put(05,00.0){\line(0,1){15}} \put(10,00.0){\line(0,1){15}} }
\end{picture}&
\begin{picture}(10,20)
{
\put(00,15){\line(1,0){10}}%
\put(00,10){\line(1,0){10}}%
\put(00,05){\line(1,0){10}}%
\put(00,00){\line(1,0){05}}%
\put(00,00){\line(0,1){15}}%
\put(05,00.0){\line(0,1){15}} \put(10,05.0){\line(0,1){10}} }
\end{picture}&
\begin{picture}(10,20)
{
\put(00,15){\line(1,0){10}}%
\put(00,10){\line(1,0){10}}%
\put(00,05){\line(1,0){05}}%
\put(00,00){\line(1,0){05}}%
\put(00,00){\line(0,1){15}}%
\put(05,00){\line(0,1){15}} \put(10,10){\line(0,1){05}} }
\end{picture}&
\begin{picture}(40,20)
 \put(16,11.2){\scriptsize  $2+n $}
{
\put(00,05){\line(1,0){40}}%
\put(00,10){\line(1,0){40}}%
\put(00,00){\line(1,0){10}}%
\put(00,00){\line(0,1){10}}%
\put(05,00.0){\line(0,1){10}} \put(10,00.0){\line(0,1){10}}
\put(15,05.0){\line(0,1){05}} \put(20,05.0){\line(0,1){05}}
\put(25,05.0){\line(0,1){05}} \put(30,05.0){\line(0,1){05}}
\put(35,05.0){\line(0,1){05}} \put(40,05.0){\line(0,1){05}} }
\end{picture}   \\
\hline
\end{tabular}\,\,.
\eee
Fields (\ref{field2}) and equations  (\ref{fieldeq22})
 are in one-to-one correspondence with
 elements of $H^{0 }(\gs_-^{2})$ and $H^{1 }(\gs_-^{2})$,
respectively. The latter are tensorial spaces  with respect to indices
$A,B\ldots=1,\ldots M$, characterized by the Young diagrams of
(\ref{cor2}).

 One observes that in the examples of ranks one and two, dynamical fields
associated with $H^{0 }(\gs_-^{\K})$ are such that the total number of
indices in the first two columns of the respective Young diagrams does not
exceed $\K$.

Another property illustrated by these examples  is that all columns starting
from the third one of the Young diagrams $\Y_0{} $ and $\Y_1{} $ associated
with $H^{0 }(\gs_-^{\K})$ and $H^{1 }(\gs_-^{\K})$
   are  equal, while the first two columns  are such that together  they form a
rectangular two-column block of height $\K+1$, \ie \be \label{2du}
\hei_1(\Y_0{})+\hei_2(\Y_1{})=\hei_2(\Y_0{})+\hei_1(\Y_1{}) =\K+1\,, \quad
\hei_k(\Y_0{})=\hei_k(\Y_1{}) \,\quad  \forall k\ge3, \ee where $\hei_k(\Y_i)$
is the height of the $k^{th}$ column of $\Y_i$.

We say that a pair of Young diagrams are {\it rank-$\K$ two-column  dual} if
they obey  (\ref{2du}). The  rank-one and two examples   suggest that
dynamical fields and   their field equations are described by {rank-$\K$
two-column  dual} Young diagrams. As we explain now this is indeed  true for
 fields of any rank.

\subsubsection{Any rank}
\label{Any rank}

Rank-$\K$ unfolded equations (\ref{dydyr}) are 
  $\mathfrak{sp}(2M)$  symmetric
 by the general argument of \cite{tens2}.
Indeed, it is well known (for more details see e.g. \cite{BHS}) that any
 system of equations of the form
\be \label{genequ} (\dr +\go )C(X) = 0 \,,\qquad  \dr:=d X^\A\frac{\p}{\p X^\A}\,,
\ee where $C(X) $ is some  set of $p$-forms taking values in a
$\mathfrak{g}$-module $\V$ and the one-form $\go (X)=d X^\A\go_\A (X)$ is some
fixed connection of $\mathfrak{g}$ obeying the flatness condition \be
\label{R0} \dr\go +\half [\go \,,\wedge  \go ]=0\, \ee ($[\,,]$ denotes the Lie
product in $\mathfrak{g}$),
 is invariant under the global symmetry  $\mathfrak{g}$.

Eq.~(\ref{dydyr}) is a  particular case of Eq.~(\ref{genequ}) with
$\mathfrak{g}=\mathfrak{sp}(2M)$. The generators of $\mathfrak{sp}(2M)$ can
be realized as bilinears built from oscillators $Y_i^A$ and
$W{}^i_{A}=\f{\p}{\p Y_i^A}$,
\bee  \label{glM} [W{}^i_{A}
\,,Y_j^B]=\delta^i_{j}\delta_A{}^B \,,\qquad [Y_i^A \,,Y_j^B]=0\,,\qquad
[W{}^i_{A} \,,W{}^i_{B}]=0\,,\qquad i,j = 1\ldots \K\,,\quad\\\label{glM2}
T^{\dda\ddb}   = Y_i^{\dda} Y_j^{\ddb}\gd^{ij}\q T_{\dda \ddb}   =
W{}^i_{\dda} W{}^j_{\ddb}\gd_{ij}\q
  T^\dda_\ddb=\half \{Y_j^{\dda}\,,  W{}^j_{\ddb}\}  \q
 \eee  \bee
 \label{TT}[T_{AB}\,,   T^{CD}]&=&
  \gd_A^C T_B^D+\gd_A^D T_B^C
+\gd_B^C T_A^D+\gd_B^D T_A^C\,\q\\ \nn [T_{A}^{B}\,,   T^{CD}]&=&
  \gd_A^C T^B{}^D+\gd_A^D T^B{}^C\q
  [T_{A}^{B}\,,   T_{CD}]=
  -\gd_C^B T_A{}_D-\gd_D^B T_A{}_C
 .\quad\eee
Note that the oscillator representation provides a standard tool for the study
of representations of $\mathfrak{sp}(2M)$
\cite{Gunaydin:1998km,Gunaydin:1998bt}
 (and references therein).

Rank-$\K$ equations (\ref{dydyr}) have the form (\ref{genequ}). The
operator $\sigma_-^\K$ (\ref{sigmar}) obeys \be
\nn
(\sigma_-^\K)^2 =0\,, \qquad \{ \dr , \sigma_-^\K \} =0\,, \ee which implies
that the corresponding connection is flat.

System (\ref{dydyr}) as well as  generators (\ref{glM}) are invariant
under the action of $\mathfrak{o}(\K)$ on the {\it color  indices}
$i,j,\ldots=1,\ldots \K$, that leaves invariant $\delta_{ij}$. Generators of
$\mathfrak{o}(\K)$
 \be \label{or}
 \gta_{m}{}_{k}=t_m{}_k-t_k{}_m\q t_m{}_j= \gd_j{}_i t_m{}^i\q
 t_m{}^i=\half \{Y_m^{\dda}\,,W{}^i_{\dda}\}\,,
\ee obey the standard commutation relations \be
 \label{tt}\rule{0pt}{14pt}
 [\gta_m{}_p ,\gta_q{}{}_j ]=
 \gd_{ p}{}_{q}  \gta_m{}{}_j+  \gd_m{}_j \gta_p{}_q
-\gd_{ m}{}_{q}  \gta_p{}{}_j-  \gd_p{}_j \gta_m{}_q
 \,.
\ee Being mutually commuting, $\mathfrak{o}(\K)$ and $\mathfrak{sp}(2M)$ form
a Howe-dual pair  \cite{Howe}.

As in the lower-rank cases, rank-$r$ primary fields obey the condition
\be \label{trr} 
\gs_-^\K \,C(Y|X)=0\,. \ee
 In terms  of the expansion
\be   \label{field}
 C(Y|X)=\sum_n
 C_{A_1; \ldots; A_{n}}^{i_1; \ldots; i_{n}}(X)
Y_{i_1}^{A_1}\cdots Y_{i_n}^{A_{n}}
\ee Eq.~(\ref{trr})
implies
tracelessness of the component fields with respect to color indices \be
\label{APH0} \gd_{i_1 i_2} {C}^{\,i_1;\,\dots\,;\,i_n} _{A_1;\,\dots;A_n}(X)
=0\,\quad \forall n\ge2\,. \ee

Since  $Y_i^A$  commute,  the tensors
 $ C_{A_1;\ldots;A_{n}}^{\,i_1; \ldots;\, i_{n}}(X)$
 are symmetric with respect to   permutation of any pair of
upper and lower indices. Hence
 $ C_{A_1;\ldots;A_{n}}^{\,i_1; \ldots;\, i_{n}}(X)$
can be decomposed into a direct sum of tensors forming irreducible
representations of  $\mathfrak{o}(\K)$ as well as
 of  $\mathfrak{gl}_M$ with the same symmetry properties described by
 some  YD $\Y(l_1,...,l_m)$ ($l_1+...+l_m=n$).
Abusing terminology, a $\mathfrak{gl}(M)$   YD associated with a traceless tensor  with respect to color indices will be referred to as  traceless YD.

Recall that if a traceless tensor with respect to color indices taking $\K$ values
 has symmetry of $\Y[h_1,...,h_m]$, it can be nonzero only if
 \be \label{konstein} \hei_1(\Y)+\hei_2(\Y) \le \K\,.
\ee
Equivalently,  for a traceless YD $\Y_0{}(n_1,\ldots,n_m,\underbrace{1,\ldots,1}_q)$ ($n_m\ge2$)
     \be\label{2}
 2m+q\le\K\,.
\ee

 As shown in Section \ref{CohomMin}, for any rank-$\K $    primary field  $C
\in H^0(\gs_-^\K)$
 associated with a YD $\Y_0{}(n_1,\ldots,n_m,\underbrace{1,\ldots,1}_q)$
that obeys (\ref{2}),
the left-hand side  of the dynamical equation has the symmetry properties of
the rank-$\K$  two-column  dual  YD $\Y_1{}(n_1,\ldots,n_m,
\underbrace{2,\ldots,2}_{\K+1-2m-q}, \underbrace{1,\ldots,1}_q)$ (\ref{2du}).

  \sbox{\gorp}{\linethickness{.15mm}\line(1,0){05}}
  \sbox{\verp}{\linethickness{.15mm}\line(0,1){05}}

Pictorially, \bee\label{pict2dual}
\begin{picture}(150,125)
\put(07,49){\small$q$}%
\put(10,35){\line(1,0){9}}%
\put(10,65){\line(1,0){9}}%
\put(11,45){\vector(0,-1){10}}%
\put(11,55){\vector(0,01){10}}%
\multiput(22,35)(0,5){06}{\multiput(00,00)(5,0){01}{\usebox{\gorp}}}%
\multiput(22,35)(0,5){06}{\multiput(00,00)(5,0){02}{\usebox{\verp}}}%
\put(15,15){$n_m\ge 2$}
\begin{picture}(150,75)(0,-65)%
\put( 4,20){\small$  m $}%
\put(10,00){\line(1,0){9}}%
\put(10,45){\line(1,0){9}}%
\put(11,18){\vector(0,-1){18}}%
\put(11,30){\vector(0,01){15}}%
\begin{picture}(120,55)(-22,0)%
\put(92,41){\small $n_1$} \put(26,-02){\small $n_m$} \put(23,55){$\Y_0{}$}
\multiput(0,00)(5,0){04}{\usebox{\gorp}}%
\multiput(0,00)(5,0){05}{\usebox{\verp}}%
\multiput(0,05)(5,0){07}{\usebox{\gorp}}%
\multiput(0,05)(5,0){08}{\usebox{\verp}}%
\multiput(0,10)(5,0){10}{\usebox{\gorp}}%
\multiput(0,10)(5,0){11}{\usebox{\verp}}%
\multiput(0,15)(5,0){10}{\usebox{\gorp}}%
\multiput(0,35)(5,0){15}{\usebox{\gorp}}%
\multiput(0,35)(5,0){16}{\usebox{\verp}}%
\multiput(0,40)(5,0){17}{\usebox{\gorp}}%
\multiput(0,40)(5,0){18}{\usebox{\verp}}%
\multiput(0,45)(5,0){17}{\usebox{\gorp}}%
 \multiput(0,0)(0,5){08}{\usebox{\ver}}
\multiput(3,30)(4,0){15}{\usebox{\toch}}%
\multiput(3,25)(4,0){14}{\usebox{\toch}}%
\multiput(3,20)(4,0){12}{\usebox{\toch}}%
\end{picture}
\end{picture}
\end{picture}\qquad\qquad
 \begin{picture}(150,120)%
\put(07,13){\small$q$}%
\put(-150, -10.0){  {Rank-$\K$ two-column  duality.} }%
\put(-50,45.0){\small$\K-2m-q+1$}%
\put(10,00){\line(1,0){9}}
\put(10,30){\line(1,0){9}}%
\put(10,65){\line(1,0){9}}%
\put(11,10){\vector(0,-1){10}}
\put(11,20){\vector(0,01){10}}%
\put(11,42.5){\vector(0,-1){12.50}}%
\put(11,52.5){\vector(0,01){12.50}}%
\multiput(22,30)(0,5){07}{\multiput(00,00)(5,0){02}{\usebox{\gorp}}}%
\multiput(22,30)(0,5){07}{\multiput(00,00)(5,0){03}{\usebox{\verp}}}%
\multiput(22,00)(0,5){06}{\multiput(00,00)(5,0){01}{\usebox{\gorp}}}%
\multiput(22,00)(0,5){06}{\multiput(00,00)(5,0){02}{\usebox{\verp}}}%
\begin{picture}(120,75)(0,-65)
\put(07,20){\small$m$}%
\put(10,00){\line(1,0){9}}
\put(10,45){\line(1,0){9}}%
\put(11,18){\vector(0,-1){18}}
\put(11,30){\vector(0,01){15}}%
\begin{picture}(120,55)(-22,0)%
\put(23,55){$\Y_1{}$} \put(92,41){\small $n_1$} \put(26,-02){\small $n_m$}
\multiput(0,00)(5,0){04}{\usebox{\gorp}}%
\multiput(0,00)(5,0){05}{\usebox{\verp}}%
\multiput(0,05)(5,0){07}{\usebox{\gorp}}%
\multiput(0,05)(5,0){08}{\usebox{\verp}}%
\multiput(0,10)(5,0){10}{\usebox{\gorp}}%
\multiput(0,10)(5,0){11}{\usebox{\verp}}%
\multiput(0,15)(5,0){10}{\usebox{\gorp}}%
\multiput(0,35)(5,0){15}{\usebox{\gorp}}%
\multiput(0,35)(5,0){16}{\usebox{\verp}}%
\multiput(0,40)(5,0){17}{\usebox{\gorp}}%
\multiput(0,40)(5,0){18}{\usebox{\verp}}%
\multiput(0,45)(5,0){17}{\usebox{\gorp}}%
 \multiput(0,0)(0,5){08}{\usebox{\ver}}
\multiput(3,30)(4,0){15}{\usebox{\toch}}%
\multiput(3,25)(4,0){14}{\usebox{\toch}}%
\multiput(3,20)(4,0){12}{\usebox{\toch}}%
\end{picture}
\end{picture}
\end{picture}
\eee
\\ \\
$\Y_0{}$ and $\Y_1{}$ are $\mathfrak{gl}_{M}$--Young diagrams with respect to
indices
$A,B=1,\ldots, M$.
\,\,Note that for $M<\KM$ some of them may be zero which would mean either that
the corresponding primary field is absent ($\Y_0{}=0$) or that it does not
obey any dynamical equations ($\Y_1{}=0$), \ie the system is off-shell.

Dynamical equations are most conveniently described in terms of Young
diagrams with manifest antisymmetrization. For nonnegative integers $h_1\ge
h_2$,
   $h_1+h_2\le \K$
 consider a tensor
\be \label{DUALT}
\E_{\,\,i_1[h_1]\,,\quad\quad\,\,\,i_2[h_2]\,,\quad\quad\,\,\,i_3[h_3],\ldots,\,\,i_k[h_n]}
^{\dda_1[\K-h_2+1]\,,\,\dda_2[\K-h_1+1]\,,\,\dda_3[h_3],\ldots,\dda_k[h_n]}
\ee with the symmetry  $\Y_0{}[h_1,h_2,h_3,\ldots,h_n]$ in the lower indices
and
 $\Y_1{}[\K-h_2+1,\K-h_1+1,h_3,\ldots,h_n]$
in the upper ones. So defined  $\Y_0{}$  and $\Y_1{}$ obey  two-column duality
condition (\ref{2du}). Let $\E$, which  plays a role of projector to the
respective tensor representations, be traceless with respect to the color
indices. It is not difficult to see that, for any $\E$   (\ref{DUALT}), the
following equations hold as a consequence of  (\ref{dydyr})
 \bee \label{conseqr}
\widehat{\E}_{\Y_0{}}\C(Y|X)=0 \q\eee where \bee \label{conseqrop}
\widehat{\E}_{\Y_0{}}=
\E_{\,\,i_1[h_1]\,,\quad\quad\,\,\,i_2[h_2]\,,\quad\quad
\,\,i_3[h_3],\dots,\,\,i_n[h_n]}
^{\dda_1[\K-h_2+1]\,,\,\dda_2[\K-h_1+1]\,,\,\dda_3[h_3],\dots,\dda_n[h_n]}
\qquad\qquad\qquad\\ \nn \underbrace{ \f{\ptl}{\ptl
Y_{i_1^{{1}}}^{\dda_1^{{1}}}} \ldots \f{\ptl}{\ptl
Y_{i_1^{{h_1}}}^{\dda_1^{{h_1}}}}}_{h_1} \,\,
 \ldots
\underbrace{ \f{\ptl}{\ptl Y_{i_n^{{1}}}^{\dda_n^{{1}}}} \ldots \f{\ptl}{
\ptl Y_{i_{n \phantom{{{}_{1_{1}}}}}^{h_n}}^{\dda_n^{h_n}}}}_{h_n}
 \,\,
 \underbrace{\f{\ptl}{\ptl X^{\dda_1^{h_1+1}}{} ^{ \,\dda_2^{{h_2+1}} } }
\cdots \f{\ptl }{\rule{0pt}{16pt}\ptl X_{\phantom{{{}_{1
}}}}^{\dda_{1}^{\K-h_2+1}}{} ^{ \,\dda_2^{{\K-h_1+1}} } }
 }_{\K+1-h_1-h_2}%
  \,. \quad\qquad
\eee

Indeed, by virtue of (\ref{dydyr}), Eq.~(\ref{conseqr}) is equivalent to \bee
\label{conseq2prove}
\E_{\,\,i_1[h_1]\,,\quad\quad\,\,\,i_2[h_2]\,,\quad\quad
\,\,i_3[h_3],\dots,\,\,i_n[h_n]}
^{\dda_1[\K-h_2+1]\,,\,\dda_2[\K-h_1+1]\,,\,\dda_3[h_3],\dots,\dda_n[h_n]}
&&\\ \nn
 \gd_{  j_1^1  j_2^1 }
\cdots \gd_{j_1^{N} j_2^{N}}\, \f{\ptl^2}{\ptl Y_{j_1^1}^{\dda_1^{h_1+1}}\ptl
Y_{j_2^1}^{ \,\dda_2^{{h_2+1}} } } \cdots \f{\ptl^2}{\ptl
Y_{j_1^{N}}^{\dda_1^{{\K-h_1+1}}}\ptl Y_{j_2^{N}}^{ \,\dda_2^{{{\K-h_2+1}}} } } &&\\ \nn
\f{\ptl}{\ptl Y_{i_1^{{1}}}^{\dda_1^{{1}}}} \ldots \f{\ptl}{\ptl
Y_{i_1^{{h_1}}}^{\dda_1^{{h_1}}}} \,\, \f{\ptl}{\ptl
Y_{i_2^{{1}}}^{\dda_2^{{1}}}} \ldots \f{\ptl}{\ptl
Y_{i_2^{h_2}}^{\dda_2^{h_2}}} \,\,\ldots\ldots \f{\ptl}{\ptl
Y_{i_n^{{1}}}^{\dda_n^{{1}}}} \ldots
\f{\ptl}{\ptl Y_{i_n^{h_n}}^{\dda_n^{h_n}}}
\C(Y|X)&=& 0  \q
 \eee
where $N=\K+1-h_1-h_2.$ Since derivatives $\f{\ptl}{\ptl Y_{j}^{\ddb }}$
commute, while the indices in columns of the both of the Young diagrams,
$\Y[h_1,h_2,h_3 ,\ldots,h_ò]$ and its two-column dual
$\Y[\K-h_2+1,\K-h_1+1,h_3 ,\ldots,h_ò ]$, are antisymmetrized, the indices
$j_1^{1}\ldots j_1^{N}$ are antisymmetrized with the indices $i_1[  h_1 ]$,
while the indices $j_2^{1}\ldots j_2^{N}$ are antisymmetrized with the
indices $i_2[ h_2]$.
Such antisymmetrizations yields  zero by virtue of the following\\

\smallskip
{\it Lemma 1}\label{Lemma 1}\\
Let a tensor $\F_{ \,i[m]\,, j[n]} $
 be traceless and antisymmetric both in indices $i$ and in $j$
 taking $\K$ values
 ($\F_{ \,i[m]\,, j[n]}$  is not demanded to have properties of a Young diagram).
 Consider the tensor
 \be
 \mathcal{G}_{ \,i[m+p]\,, j[n+p]} =\F_{ \big[i[m]\,, \big[j[n]}
 \delta_{i_{m+1}j_{n+1}}\ldots \delta_{i_{m+p}j_{n+p}\big]_j \big]_i}
 \ee
 resulting from the total antisymmetrization of indices $i$ and $j$.
 Then
 \be
 \label{lem}
 \mathcal{G}_{ \,i[m+p]\,, j[n+p]}=0 \qquad \mbox{at}\quad m+n> \K-p\,.
 \ee

 The proof follows from the determinant formula applied to the double dual
 tensor $\tilde {\mathcal{G}}$
 \be
 \tilde{ \mathcal{G}}^{k[\K-m-p]\,, l[\K-n-p]}:=\epsilon^{k[\K-m-p] \,i[m+p]}
 \epsilon^{l[\K-n-p] \,j[n+p]} \mathcal{G}_{ \,i[m+p]\,, j[n+p]}\,.
 \ee
 Indeed,
 \be
 \tilde{ \mathcal{G}}^{k[\K-m-p]\,, l[\K-n-p]}:=\epsilon^{k[\K-m-p] \,i[m]}{}_{\K[p]}
 \epsilon^{l[\K-n-p] \,j[n]\,\K[p]} {\F}_{ \,i[m]\,, j[n]}\,.
 \ee
Expressing the product of two totally antisymmetric symbols in terms of
Kronecker symbols one observes that for $ m+n> \K-p$ at least one pair of
indices of the traceless tensor ${\F}_{ \,i[m]\,, j[n]}$ is contracted hence
giving zero. $\Box$

\medskip

Note that Eq.~(\ref{lem}) at $p=0$ yields Eq.~(\ref{konstein}).

Since the differential operator $\widehat{\E}_{\Y_0{}}$ (\ref{conseqrop}) is
homogeneous, the primary fields also satisfy (\ref{conseqr}). The parameter
(\ref{DUALT}) is designed in such a way that Eq.~(\ref{conseqr}) is
nontrivial only for primaries $C (Y|X)$ with the symmetry properties of
$\Y_0{}[h_1, \ldots,h_n ]$. The nontrivial part of the story is to prove
that the presented list of dynamical equations is complete. This follows from
the analysis of the cohomology group $H^1(\gs_-^\K)$ in Section
\ref{Details}.

As anticipated, there is precise matching between the primaries and field
equations. This means that the respective subspaces of  $H^0(\gs_-^\K)$ and
$H^1(\gs_-^\K)$ form isomorphic $\mathfrak{o}(\K)$-modules. Here one should
not be confused by the fact that the two-column dual diagram $\Y_1{}$ does
not respect condition (\ref{konstein}). The point is that, as an
$\mathfrak{o}(\K)$--tensor, $\Y_1{}$ is not traceless, containing explicitly
a number of  $\mathfrak{o}(\K)$ metric tensors $\delta_{ij}$ which add
additional $\mathfrak{o}(\K)$ indices to $\Y_1{}$ compared to $\Y_0{}$.

The particular case of (\ref{conseqr})  which plays the  key role in the
construction of conserved currents is that with $\K=2\kappa$ and a traceless
$\mathfrak{o}(2\kappa)$-diagram $\Y_0=\Y^{2\kappa_{cur}}_{0\,} $  of the form
\be\label{eqdiag}\Y^{2\kappa_{cur}}_{0\,} [h_1,h_2,
h_3, h_4,\ldots] =\Y^{2\kappa_{cur}}_{0\,} [\kappa,\kappa,
h_3, h_4,\ldots]  .\ee Then operator $\widehat{\E}_{\Y^{2\kappa_{cur}}_{0\,}}$
 (\ref{conseqrop}) is of the first order in $\f{\ptl}{\ptl X}$. Hence,
Eq.~(\ref{conseqr}) acquires  the form of  {\it   conservation condition}
    \bee \label{conseqroptok}
 \E_{\,\,i_1[\kappa]\,,\quad\quad\,\,\,i_2[\kappa]\,,\,
\,\,i_3[h_3],\dots,\,\,i_n[h_n]}
^{\dda_1[\kappa+1]\,,\,\dda_2[\kappa+1]\,,\,\dda_3[h_3],\dots,\dda_n[h_n]}\underbrace{
\f{\ptl}{\ptl Y_{i_1^{{1}}}^{\dda_1^{{1}}}} \ldots \f{\ptl}{\ptl
Y_{i_1^{{\kappa}}}^{\dda_1^{{\kappa}}}}}_{\kappa} \,\,
 \underbrace{ \f{\ptl}{\ptl
Y_{i_2^{{1}}}^{\dda_2^{{1}}}} \ldots \f{\ptl}{\ptl
Y_{i_2^{{\kappa}}}^{\dda_2^{{\kappa}}}}}_{\kappa} \,\,
  \f{\ptl}{\ptl X^{\dda_1^{\kappa+1}}{} ^{ \,\dda_2^{{\kappa+1}} } }
\qquad\\ \nn \underbrace{ \f{\ptl}{\ptl Y_{i_3^{{1}}}^{\dda_3^{{1}}}} \ldots
\f{\ptl}{ \ptl Y_{i_{3 \phantom{{{}_{1_{1}}}}}^{h_3}}^{\dda_3^{h_3}}}}_{h_3}
 \,\,
 \ldots
\underbrace{ \f{\ptl}{\ptl Y_{i_n^{{1}}}^{\dda_n^{{1}}}} \ldots \f{\ptl}{
\ptl Y_{i_{n \phantom{{{}_{1_{1}}}}}^{h_n}}^{\dda_n^{h_n}}}}_{h_n}
 \quad
\C(Y|X)=0\,.\eee

For instance, in the  particular case of two-column diagrams this yields
a straightforward generalization of the conservation condition   of
\cite{cur}
 \bee \label{conserv}
 \E_{\,\,i_1[\kappa]\,,\quad\quad\,\,\,i_2[\kappa] ]}
^{\dda_1[\kappa+1]\,,\,\dda_2[\kappa+1]\, }\underbrace{ \f{\ptl}{\ptl
Y_{i_1^{{1}}}^{\dda_1^{{1}}}} \ldots \f{\ptl}{\ptl
Y_{i_1^{{\kappa}}}^{\dda_1^{{\kappa}}}}}_{\kappa} \,\,
 \underbrace{ \f{\ptl}{\ptl
Y_{i_2^{{1}}}^{\dda_2^{{1}}}} \ldots \f{\ptl}{\ptl
Y_{i_2^{{\kappa}}}^{\dda_2^{{\kappa}}}}}_{\kappa} \,\, \f{\ptl}{\ptl
X^{\dda_1^{\kappa+1}}{} ^{ \,\dda_2^{{\kappa+1}} } } \C(Y|X)=0\,. \eee

 \subsection{Higher $\sigma_-^\K$- cohomology in $\M_M$}
    \label{allsigcohomology}
    As shown in \cite{SVsc} (for more detail see \cite{tens2,Bekaert:2005vh}),
when the fields $\C(Y|X)$ satisfying (\ref{dydyr}) are $p$-forms, dynamical fields
and their field equations are associated
with $H^{p} (\gs_- )$ and  $H^{p+1} (\gs_- )$, respectively. Recall that
     $H^p(\sigma_-^\K )= $
$ {\ker}(\sigma_-^\K )/ {\Imm } (\sigma_-^\K )\big |_p$, where $\big |_p$
denotes the restriction to  $p$-forms. Hence, we are interested in
 $\gs_-^\K$-closed $p$-forms  $P (\gx,Y|X)$,
  \be\label{F}
  \gs_-^\K  P(\gx,Y|X)=0\q \ee that are not   $\gs_-^\K$-exact.

  In  the expansion
\be   \label{fieldp}
 P(\gx,Y|X)=\sum_n
  C ^{\,i_1; \ldots; i_{n}}_{A_1; \ldots; A_{n}\, |\, C_1; \ldots; C_{2p}}(X)
Y_{i_1}^{A_1}\cdots Y_{i_n}^{A_{n}}
\,\gx^{C_1C_2}\cdots \gx^{C_{2p-1}C_{2p}} \,, \ee
  the coefficients
 $C ^{\,i_1; \ldots; i_{n}}_{A_1; \ldots; A_{n}\, |\, C_1; \ldots; C_{2p}}(X)$
 form almost symmetric diagrams (\ref{exalsim})
 with respect the indices  ${C_1; \ldots; C_{2p}}$ contracted with
 the differentials $\xi^{AB}$ and
 are symmetric with respect to   permutation of any pair of
upper and lower indices $A_k, i_k$. Hence, with respect to the latter indices,
$C ^{\,i_1; \ldots; i_{n}}_{A_1; \ldots; A_{n}\, |\, C_1; \ldots; C_{2p}}$
can be decomposed into a direct sum of tensors forming irreducible
representations of  $\mathfrak{o}(\K)$ as well as of  $\mathfrak{gl}_M$ with the same symmetry properties described by
 some  YD $\YY $.

Differently from the zero-form case, the coefficients
$C_{A_1; \ldots; A_{n} }^{i_1; \ldots; i_{n} }{}_{| C_1; \ldots; C_{2p}}(X)$ in
 Eq.~(\ref{fieldp}) are not necessarily traceless  with respect to color
indices. In addition to the indices of a   $\mathfrak{o}(\K)$-traceless YD,
the diagram  $\YY $   can
contain indices  carried by  products of  $\delta^{ij}$.
The respective $\mathfrak{gl}(M)$ tensor   contains    products of
  the  ``tracefull" combinations
  \be \label{Kron}
 U^{AB}= \delta^{ij} Y^A_i Y^B_j\,,
  \ee
that, as  mentioned in Section
\ref{Conventions}, are described by the  Kronecker diagrams (\ref{prodsym2}).

  As a result,  $P(\gx,Y|X)$ satisfying  (\ref{prodsym2})
 has symmetry properties that described by direct sum of the following tensor products of YD
        \be\label{TrlessKronAL } \Y_0 [h_1, h_2, \ldots]\otimes\Y_A[\hh_1,\ldots,\hh_m]\otimes
        \Y_\delta[d_1,\ldots,d_{2n}] \ee for different $m,n$ such that $|\Y_A|=2p$.

  Analysis of the higher cohomology with
$p>1$ is also interesting in many respects. For instance, it  characterizes
further relations on the left-hand sides of field equations as well as
gauge fields in higher-spin gauge theories, described by higher differential forms.
In this section we present the final results for  $H^p (\sigma_-^\K)$ leaving details
of their derivation to Section \ref{Details} and  Appendix A.

 $H^p(\gs^\K_-)$ is realized by a space of homogeneous polynomials $P(\xi, Y)$
of degree $p$ in $\xi$,   forming an
$\mathfrak{o}(\K)$-module with respect to the color indices $i$ carried by
$Y^A_i$ and a $\mathfrak{gl}_M$-module with respect to the spinor indices
$A,B,\ldots$ carried by $Y^A_i$ and  $\xi^{AB}$.   To avoid the restriction
that the maximal height of $\mathfrak{gl}_M$-Young diagrams is $M$,
we will assume that $M $ is large enough. In the
final result one should simply take into account that such diagrams are zero.
On the other hand, the rank $\K$ is not supposed to be large because nontrivial
cohomology is just associated with special values of $\K$.

\subsubsection{Main results}

It turns out that the cohomology groups $H^p(\gs^\K_-)$ are characterized by
various $\mathfrak{o}(\K)$-modules and $\mathfrak{gl}_M$-modules formed by
the differentials $\xi^{AB}$. The former can be described by a traceless
 $\mathfrak{o}(\K)$-diagram  $ \Y_0$ while the latter by an almost symmetric
 $\mathfrak{gl}_M$-diagram $\Y_{A}$. Given $ \Y_0$ and  $\Y_{A}$, there
exists   some tensor with the symmetry  of a  Kronecker  YD $\Y_\gd$ (\ref{prodsym2}),
composed of the Kronecker symbols $\delta^{ij}$, and some    component
$\Ycoh \in\Y_0\otimes\Y_\gd\otimes \Y_{A}$ that represents
 $H^p(\gs^\K_-)$.

To identify $\Ycoh $
it is convenient to introduce the  {\it infinite  almost symmetric matrix}
 $\shi=  \shi(\K|\Y_0)$, that has the form 
  \bee \label{nestSinf}
\shi   =\left(
\beee{c c c c c c c c c c  }
   {\del}_1^0&{\del}_1^0&{\del}_{2}&{\del}_3&{\del}_4&{\del}_5&{\del}_6&{\del}_7&\ldots  \\
   {\del}_2&{\del}_1&{\del}_1  &{\del}_2&{\del}_3  &{\del}_4&{\del}_5&{\del}_6  &\ldots        \\
   {\del}_3&{\del}_2&{\del}_1^0&{\del}_1^0&{\del}_2  &{\del}_3&{\del}_4&{\del}_5& \ldots \\
   {\del}_4&{\del}_3&{\del}_2  &{\del}_1&{\del}_1&{\del}_2&{\del}_3  &{\del}_4  & \ldots    \\
   {\del}_5&{\del}_4&{\del}_3  & {\del}_2&{\del}_1^0&{\del}_1^0&{\del}_2  &{\del}_3 &\ldots    \\
   {\del}_6&{\del}_5&{\del}_4  & {\del}_3&{\del}_2  &{\del}_1&{\del}_1&{\del}_2 &\ldots \\
   {\del}_7&{\del}_6&{\del}_5  &{\del}_4&{\del}_3  & {\del}_2&{\del}_1^0&{\del}_1^0 & \ldots \\
   \vdots  & \vdots & \vdots & \vdots & \vdots  & \vdots  &  \vdots& \vdots & \ddots     \\
  \eeee\right)\,,
\eee
  where \be\label{deltfCar}   \del_k=\hei_k(\Y_0) -\hei_{k+1}(\Y_0)\qquad \ee
 are the Cartan weights of $\Y_0$ and
 \be\label{deltf0} \del^0_1=\K-\hei_1(\Y_0)-\hei_2(\Y_0)\,.\qquad
 \ee
 Note that
here all  zero-height columns of $\Y_0$ are in the game, \ie      $\Y_0$ is treated as a generalized Young
diagram with an infinite number of columns.

Elements of matrix (\ref{nestSinf}) obey
\bee
 \label{shiftinfin}\shi_{n\,\,(n+j)}&=&\shi_{(n+j-1) \,\,n }\quad \forall\,\,
n\ge1,\,\,\,\,j\ge1 \, ,\\\nn
 \shi_{n\,\,m}&=& \del_{m-n  }\qquad\quad \forall\,\,   m-n\ge2\q     \\ \nn
  \shi_{(2k+1)\,\,(2k+1)}  &=&\del^0_1\,\q \quad  \\ \nn
 \shi_{(2k)\,\,(2k) } &=&  \del_{1}\,.\,
\eee

For any
almost symmetric $\Y_{A}$    we introduce the  {\it   shift matrix} \,\,$\sh=\sh(\K|\Y_A,\Y_0)$ resulting from the
  intersection of the {infinite  almost symmetric matrix}
 $  \shi(\K|\Y_0)$ (\ref{nestSinf})  with  $\Y_A$. In other words,
   elements of  the  shift  matrix $\sh$ are \be
   \label{shiftS}\sh_{n\,\,m}=
 \shi _{n\,\,m}  \theta\big(\hei_m({\Y_A})-n\big)
\q\ee
 where $\theta\big( n\big)=0 $ if $n<0$, $\theta\big( n\big)=1 $ if $n\ge 0$.
  For example, for $\Y_A=\Yn^{nest}\{8,5,3 \}$ (\ref{nested411}), the shift matrix \,$ \sh(\K|\Yn^{nest}\{8,5,3 \},\Y_0)$
   is
  \bee\label{nestS1} \sh  =
  \left(
\beee{c c c c c c c c c c  }
   {\del}_1^0&{\del}_1^0&{\del}_{2}&{\del}_3&{\del}_4&{\del}_5&{\del}_6&{\del}_7&{\del}_8  \\
   {\del}_2&{\del}_1&{\del}_1&{\del}_2&{\del}_3  &{\del}_4&{\del}_5&  0     & 0        \\
   {\del}_3&{\del}_2&{\del}_1^0&{\del}_1^0&{\del}_2  &{\del}_3&  0     &  0     & 0    \\
   {\del}_4&{\del}_3&{\del}_2&  0     &  0       &  0     &  0     &  0     &  0      \\
   {\del}_5&{\del}_4&{\del}_3&  0     &  0       &  0     &  0     &  0     &  0       \\
   {\del}_6&{\del}_5&  0     &  0     &  0       &  0     &  0     &  0     &  0    \\
   {\del}_7&  0     &  0     &  0     &  0       &  0     &  0     &  0     &  0     \\
   {\del}_8&  0     &  0     &  0     &  0       &  0     &  0     &  0     &  0         \\
  \eeee\right).
\eee

 The central result is

 {\it Theorem} \label{ Theorem}

 The $\mathfrak{gl}(M)$--Young diagrams associated
with $H^{p}(\gs_-^{\K} )$ are \be\label{otvetec}\Ycoh=\Ycoh(p|\K| \Y_A,\Y_0)\ee where
$\Y_A$ is any almost symmetric
diagram, $ \big | Y_A \big |=2p $, $\Y_0$ is any traceless diagram,
  \be\label{otvetec1}
 \hei_j(\Ycoh)=\Hh_j+ \hh_j\q \Hh_j=h_j+
  \sum _i    \sh_{i\,j}  \,,\quad\hh_j=\hei_j(\Y_A)\,,\quad h_j=\hei_j(\Y_0)\,,
\ee
   and $\sh_{i\,j}$ are elements of the shift  matrix $\sh(\K|\Y_A\,,\,\Y_0)$ (\ref{shiftS}).

The proof of  {\it Theorem} is given in Appendix using  the homotopy
trick   explained in Section \ref{Details}.

As shown in Appendix, the  {Kronecker YD}      $\Y_\gd[d_1,d_1,\ldots,d_{\ld},d_{\ld}]$ can be expressed via
 $\Hh_{j}$ (\ref{otvetec1}) as follows
 \be \label{Ydi}
  \,   d_{k}= \Hh_{2k}+\Hh_{2k-1}-\K\,\qquad \forall \,\, k\le  {\ld}
  =\left[\half(\la+1)\right]\,\q\la= \sharp(\Y_A) \,. \ee
($[a]\leq a$ denotes the integer part of $a$.)
For example, for   the nested hook \,$\Y_A$ (\ref{nested411}) $\ld=2$ and
$  d_1=\K-h_6-h_9$, $ d_2 =  \K-h_1-h_4\,.$

By   Eqs.~(\ref{otvetec1}), (\ref{Ydi})  the nonzero
heights of  the  Kronecker YD $\Y_\gd $
are
\be\label{resKron}
\hei_{2j-1}(\Y_\gd)=\hei_{2j}(\Y_\gd)= h_{2j-1} +\sum _i   \sh_{i\,(2j-1)}
+h_{2j } +\sum _i  \sh_{i\,(2j )}\,-\K\q j\le\ld.\ee

For the  case of  one-forms  described by the simplest nested hook
 $\Y_A=\Y[1,1]$  this gives  $ \sh_{1\,1}=\sh_{1\,2}=\K-h_1 -h_2 =d_1. $
Hence for  $H^{1}(\gs_-^{\K} )$
  Eq.~(\ref{otvetec}) yields $\hei_1(\Ycoh)=\K+1 -h_2  $,  $\hei_2(\Ycoh)=\K+1 -h_1 $.
In accordance with Section \ref{Any rank} and  definition (\ref{2du}),  $\Ycoh$ is  two-column dual to $ \Y_0$.

\subsection{Multilinear currents in $\M_M$}

The construction of conserved currents in terms of closed forms  proposed in
\cite{cur,tens2} for rank-one fields admits a higher-rank generalization.

 That dynamical degrees of freedom associated with the
rank-one  equations  (\ref{dydy}) live on a $M$-dimensional
surface $S\subset \M_M\otimes \mathbb{R}^M$ 
suggests that conserved charges associated with these equations have to be
built in terms of $M$-forms that are closed
 as a consequence of  rank-two field equations  (\ref{dydy2}).
 As shown in \cite{gelcur}, the following $M-$form in $\M_M\otimes
\mathbb{R}^{2M}$ \be\label{Warpi}     \Big( \, \gx{}^{AB}\f{\p}{\p z ^B}
-    d\, \bz {}^A
\Big)^{\,M}\,\,J^2(z,\,\bz\,|X)\Big|_{ \,z =0}\,\qquad
\ee is closed  provided that   $J^2(z,\,\bz\,|X)$   solves
 (\ref{dydy2}) with  $z,\bz   $   (\ref{twistorz}). Usual conserved currents
 $$
J^2_\eta(z,\,\bz\,|X)=\eta\delta^{kj} \C_k(z+\bz|X)\C_j(i(\bz- z)|X)\,
$$
 are built in terms of
bilinears in  solutions $\C_{j}$  of (\ref{dydy}) with
parameters $\eta$ that commute with $\gs_-^2$ (\ref{sigma-2})
\cite{tens2,BRST2010,{Gelfond:2013xt}}.

Since modules of solutions of the rank-$\K$ equations in $\M_M\otimes
\mathbb{R}^{\K M}$ are functions of $\K M$ variables $Y_i^A$ one might
guess that in the rank-$\K$ case the dimension of a ``local Cauchy bundle''
\cite{Mar} on which  initial data should be given to determine a solution
everywhere in $\M_M$ is $\K M$. One can see however that the following
straightforward generalization of (\ref{Warpi}) to conserved currents in
$\M_M\otimes \mathbb{R}^{2\K  M}$ with arbitrary $\K$
 \bee\label{Warpir}
\prod _{j=1}^\K 
  \Big(  \kk{}^{A^j  \,B}\f{\p}{\p z_j^B} -    d\, \bz_j{}^{A^j }
\Big)^M   J^{2\K}_\eta (z,\,\bz\,|X)\Big|_{  z  =0}\q \eee 
 where 
   $ Y_j=z_j+\bz_j \,,$ $   Y_{j+\K}=i(\bz_j-z_j)
 $\,, 
  $\gvep_{A_1\ldots A_M}$ is the rank-$M$ Levi-Civita symbol, and   \be J_\eta^{2\K} (z,\,\bz\,|X)=
  \eta\delta^{kj}
   \C{}^\K_k(z+\bz|X)\C{}^\K_j(i(\bz- z)|X)\ee is bilinear in rank-$\K$ fields $\C_{i}^\K$, does
  not work. The reason is that, although these forms are closed by
virtue of the rank-$2\K $ unfolded equations, their pullback to $\M_M$ is
zero for $\K>1$ thus obstructing the construction of conserved charges in the
form of integrals over $\M_M$.

Indeed, up to a numerical factor, the
pullback of the form (\ref{Warpir}) to $\M_M$  is \be\label{Warpireps}
\left\{\prod _{j=1}^\K \gvep_{A^j_1,\ldots, A^j_M} \xi{}^{A^j_1\,{B^j_1}}
 \ldots
   \xi^{A^j_M\,{B^j_M}}\right\}\left(\f{\p}{\p  z_j^{B^j_1}}\ldots\f{\p}{\p  z_j^{B^j_M}}
 J^{2\K} (z, \bz |X)\right)\Big|_{z= \bz =0}\,.
\ee Any Young diagram associated with the combination of differentials $\xi^{AB}$
can only contain rows of lengths $l\leq M+1$ since otherwise it
would contain  symmetrization of a pair of differentials. On the
other hand, because of $\gvep$-symbols, it contains $\K$ columns of the
maximal height $M$ associated with the indices $A$ in (\ref{Warpireps}).
Anticommutativity of the differentials implies that the part of the diagram
associated with the indices $B$ should contain $\K$ total symmetrizations.
However, since the first $\K$ columns are occupied by the indices $A$ and
the total number of symmetrized indices cannot exceed $M+1$, this
is not possible for $\K>1$ as at least two symmetrized indices $B$ would
belong to the same column, which implies antisymmetrization. Thus
(\ref{Warpireps}) is zero for $\K>1$.

 The reason why the naive construction does not work is that
the $\mathfrak{o}( \K)$, that acts on the primary (dynamical) rank-$\K$
fields, relates different solutions to each other not affecting evolution of
any given solution. This suggests that the dimension of the true local Cauchy
bundle is $\K M-\half \K(\K-1)$. Correspondingly, the respective currents
should be represented by closed $\K( M-\half (\K-1))$-forms.

 As   shown in Section \ref{Conventions},  products of
 anticommuting variables $\gx^{AB}$ have  symmetries of  almost symmetric
 Young diagrams.
The product of $\K$ totally antisymmetric tensors  with $1\leq\K<M$ admits a
unique extension to such a tensor  with $(2\K M-{\K (\K -1)} )$ cells described by
the YD
  ${\Y}\big[\underbrace{M ,\dots,M}_\K,\underbrace{\K,\dots,\K}_{M+1-\K}\big]$. 
  Pictorially,
\sbox{\gorp}{\line(1,0){5}} \sbox{\verp}{\line(0,1){5}} \be \label{krukM_M+1}
\begin{picture}(40,38)(0,-3)
\put(-162,15){  ${\Y}\big[\underbrace{M ,\dots,M}_\K,\underbrace{\K,\dots,\K}_{M+1-\K}\big]\,=$}
\multiput(0,00)(5,0){03}{\usebox{\gorp}}%
\multiput(0,05)(5,0){03}{\usebox{\gorp}}%
\multiput(0,10)(5,0){03}{\usebox{\gorp}}%
\multiput(0,15)(5,0){03}{\usebox{\gorp}}%
\multiput(0,35)(5,0){08}{\usebox{\gorp}}%
\multiput(0,30)(5,0){08}{\usebox{\gorp}}%
\multiput(0,25)(5,0){08}{\usebox{\gorp}}%
\multiput(0,20)(5,0){08}{\usebox{\gorp}}%
\put(5,37){\scriptsize $M+1$} \put(-12,15){\scriptsize $M$}
\put(42,25){\scriptsize $\K$}
\multiput(0,30)(5,0){09}{\usebox{\verp}}%
\multiput(0,25)(5,0){09}{\usebox{\verp}}%
\multiput(0,20)(5,0){09}{\usebox{\verp}}%
\multiput(0,15)(5,0){04}{\usebox{\verp}}%
\multiput(0,10)(5,0){04}{\usebox{\verp}}%
\multiput(0,05)(5,0){04}{\usebox{\verp}}%
\multiput(0,00)(5,0){04}{\usebox{\verp}}%
\put(7,-5){\scriptsize $\K$}
\end{picture} \,.
\ee

 Setting $N=M+1-\K$ consider the   differential form \bee\label{formaMr}\ls
\Xi^{  D_1[M],\ldots,D_\K[M],D_{\K+1}[\K],\ldots,D_{M+1}[\K]}
&=&
\xi ^{\ddd_1^1 \ddd_2^1}                            
{} \xi ^{\ddd_1^2 \ddd_3^1} {} \ldots {} \xi ^{\ddd_{1}^{\K -1}
\ddd_{\K}^{1}} {} \xi ^{\ddd_{1}^{\K} \dda_{1}^{1}} {} \ldots {}
\xi ^{\ddd_{1}^{M} \dda_{N}^{1}} \ldots\\\nn
  & &\ls \!\!\!
\xi ^{\ddd_n^n \ddd_{n+1}^n}                        
{} \xi ^{\ddd_n^{n+1}  \ddd_{n+2}^n} {} \ldots {} \xi ^{\ddd_{n}^{\K -1}
\ddd_{\K}^{n}} {} \xi ^{\ddd_{n}^{\K} \dda_{1}^{n}} {} \ldots {} \xi
^{\ddd_{n}^{M} \dda_{N}^{n}}\ldots \\ \nn & &\ls \!\!\!
\xi ^{\ddd_{\K}^{\K} \dda_{1}^{\K}}                     
{} \xi ^{\ddd_{\K}^{\K +1} \dda_{2}^{\K}} {} \ldots {} \xi ^{\ddd_{\K}^{M}
\dda_{N}^{\K}} \,,
 \eee   which by construction  has  symmetry  of
 ${\Y}\big[\underbrace{M ,\dots,M}_\K,\underbrace{\K,\dots,\K}_{N}\big]$ (\ref{krukM_M+1}).
Contracting indices of the first $\K$ columns with the totally
antisymmetric symbols $\epsilon_{A_1  \ldots A_\K}$ yields a tensor $\B$
of the symmetry of $\Y[\underbrace{\K,\ldots
\K}_{N}]$ \be   \label{tok_r D} \nn \B^{A_{ 1}[\K],\ldots,D_{N}[\K]}
= {\epsilon_{\ddd_1^1 \ldots \ddd^M_1}} \ldots
{\epsilon_{\ddd^1_\K  \ldots \ddd^M_\K}}
\Xi^{  D_1[M],\ldots,D_\K[M],A_{ 1}[\K],\ldots,A_{N}[\K]}    \,.
 \ee

Let a tensor $ \F_{\,\,i_1[\K]\,, \,i_2[\K]\,,\,
 \dots,\,\,i_N[\K]}$ be  traceless with respect to the color indices and have
the symmetry  of    $\Y{}[\underbrace{\K,\ldots \K}_{N} ].$ Then the
      $(\K M-\f{\K (\K -1)}{2})$-form
   \bee   \label{tok_r}
\Omega_{2\K}(J )=
 \B^{\dda_{1}[\K]\,,\dda_{ 2}[\K]\,,\ldots,\dda_{N}[\K]}
 \F_{\,\,i_1[\K]\,, \,i_2[\K]\,,\,
 \dots,\,\,i_N[\K]}
\\ \nn
\underbrace{ \f{\ptl}{\ptl Y_{i_1^{{1}}}^{\dda_1^{{1}}}} \ldots \f{\ptl}{\ptl
Y_{i_1^{{\K}}}^{\dda_1^{{\K}}}}}_{\K} \,\,
  \ldots
\underbrace{ \f{\ptl}{\ptl Y_{i_N^{{1}}}^{\dda_N^{{1}}}} \ldots \f{\ptl}{
\ptl Y_{i_{N \phantom{{{}_{1_{1}}}}}^{\K}}^{\dda_N^{\K}}}}_{\K}
 \,\,
  J^{2\K}(Y| X)\big|_{Y=0}
  \,  \eee
  is closed provided that a rank-$2\K $ field $J^{2\K}$  (current)
  obeys the rank-$ 2\K $
    {\it current equation}
\be \label{dydy2r}   \left (\f{\p}{\p X^{ A B}}+ \f{\p^2}{\p Y_k^A \p
Y_j^B}\, \gd_{kj}\, \right ) J^{2\K}(Y| X) =0\,\ \quad ( k,j=1,\ldots,2\K)
\,, \ee that   coincides with unfolded equations (\ref{dydyr}) with
$\K$ replaced by $2\K$.

Indeed,   consider   a $(\K M-\f{\K (\K -1)}{2})$-form \bee
\label{AlSyfo}\nn \Omega_{2\K}  (\AAA ) & =&
\Xi^{  D_1[M],\ldots,D_\K[M],B_{ 1}[\K],\ldots,B_{N}[\K]}
 {\AAA}_{\,\ddd_1[M ]\,,\,,\ldots, \ddd_\K[M],
 \,B_1 [\K ]\,,\,B_N[\K] }(X), \eee
  where ${\AAA}$ has symmetry of almost symmetric
  YD
 (\ref{krukM_M+1}).
Then  the only non-zero component of the differential form $d  \Omega_{2\K}(\AAA )=\gx^{AB}\f{\p}{\p X^{ A B}}
\Omega_{2\K}(\AAA ) $  is represented by  the  almost symmetric Young
diagram
\bee
 \label{dkrukM_M+12}
 \Y[\underbrace{M ,\ldots,M }_\K ,\K+1,\K+1,\underbrace{\K
,\ldots,\K}_{N-2}] \,=\qquad\qquad
\begin{picture}(40,30)(0,15)%
\multiput(0,00)(5,0){03}{\usebox{\gorp}}%
\multiput(0,05)(5,0){03}{\usebox{\gorp}}%
\multiput(0,10)(5,0){03}{\usebox{\gorp}}%
\multiput(0,15)(5,0){05}{\usebox{\gorp}}%
\multiput(0,35)(5,0){08}{\usebox{\gorp}}%
\multiput(0,30)(5,0){08}{\usebox{\gorp}}%
\multiput(0,25)(5,0){08}{\usebox{\gorp}}%
\multiput(0,20)(5,0){08}{\usebox{\gorp}}%
\put(5,37){\scriptsize $M+1$} \put(-12,15){\scriptsize $M$}
\put(42,25){\scriptsize $\K$ } 
\multiput(0,30)(5,0){09}{\usebox{\verp}}%
\multiput(0,25)(5,0){09}{\usebox{\verp}}%
\multiput(0,20)(5,0){09}{\usebox{\verp}}%
\multiput(0,15)(5,0){06}{\usebox{\verp}}%
\multiput(0,10)(5,0){04}{\usebox{\verp}}%
\multiput(0,05)(5,0){04}{\usebox{\verp}}%
\multiput(0,00)(5,0){04}{\usebox{\verp}}%
\put(7,-5){\scriptsize $\K$ }
\end{picture}
\eee
because  the first column of $\Y[\underbrace{M ,\ldots,M }_\K ,\underbrace{\K ,\ldots,\K}_{N}]$
 (\ref{krukM_M+1}) is of the maximal height and cannot be enlarged further.

However, Eq.~(\ref{conseqroptok}) associated with the traceless
$\mathfrak{o}(2\K)$-diagram $\Y^{2\K_{cur}}_{0}= \Y^{2\K_{cur}}_{0}[\K,\K, \ldots]$
 (\ref{eqdiag}), which is
the  first-order differential equation with respect to $X$ derivatives,
just implies that the projection to this diagram is zero. Hence,
the form
 $\Omega_{2\K}(J )$ (\ref{tok_r}) is closed on shell.

 In particular, closed form (\ref{tok_r}) with
 \be\label{T_r} J^{2\K}(Y|X)\sim \C_{ 1}(Y_{ 1}|X) \ldots \C_{ {2\K}}(Y_{2 \K}|X)
 \ee
 where $\C_{j }(Y_{j}|X)$ solve rank-one equations
(\ref{dydyr}), generates $2\K$-linear conserved currents.
As in the case of bilinear currents   \cite{{Gelfond:2013xt}}, the
first-order differential operators $\A^{1 }(Y|X)$ and  $\A^{ 2}(Y|X)$
\be\label{param12} \A_j^1 {}^B(Y_j|X)=2   X^{AB}\f{\p}{\p Y_j^A}-Y_j^B\,\q\,
\A_j^2 {}_C(Y_j|X)= \f{\p}{\p Y_j^C}\,\q  j=1,\ldots,2\K\q
\ee as well as
any polynomial $\eta(\A ) $, obey
\be \label{dgydgyh1} \left[ \left
(\f{\p}{\p X^{ A B}}+ \f{\p^2}{\p Y_k^A \p Y_j^B}\, \gd_{kj}\, \right
),\eta(\A ) \right ] =0\,\ \quad ( k,j=1,\ldots,2\K) \,.
\ee
 Therefore,  for any    $\eta^{j_1,\ldots,j_{2\K}}(\A ) $,
\be\label{Jeta} J_\eta(Y|X) =
 \eta^{j_1,\ldots,j_{2\K}} (\A )  \C_{j_1}(Y_{ 1}|X) \ldots \C_{j_{2\K}}(Y_{2 \K}|X), \ee
also obeys  current equation (\ref{dydy2r}), giving rise to  the
${2\K}$-linear charges $Q^{2\K}_\eta$ where $\eta^{j_1,\ldots,j_{2\K}} (\A )$
are  parameters of the symmetries generated by these multilinear charges,
which are analogous to the multiparticle symmetries considered in
\cite{Vasiliev:2012tv}.

\section{Minkowski-like reduction}
\label{Mres}
\subsection{Fields and equations   in $\MM $}
\label{MH1res} \sbox{\gorp}{\line(1,0){5}} \sbox{\ver }{\line(0,1){3}}
\sbox{\verp}{\line(0,1){5}} \sbox{\toch}{\circle*{1}}

The unfolded form of the  equations of motion
for massless fields of all spins in  $4d$  Minkowski space (\ref{minun}) is a subsystem of (\ref{dydyr})
at $M=4$,  $\K=2$ and  $\dda=(\ga,\,\,\pa)$, $\ddb=(\gb,\,\,\pb)$ \etc, \ie
\be\label{MinXY} Y^\dda=(y^\ga,\,\,\by^\pb)\q
X^{\dda\ddb}=(x^{\ga\pb},\,\,x^{\ga\gb},\,\,\overline{x}^{\pa\pb}).\ee

More generally,  consider $X^{\dda\ddb}$ (\ref{MinXY}) with $\ga=1,\ldots,K$
and $\pa=1,\ldots,K$ for any $ K$. The rank-$\KM$
generalization  (\ref{minunK}) of (\ref{minun}),   is a subsystem of (\ref{dydyr}) with
$M=2K$,    We set $\eta^{kj}=\gd^{kj}$\, in (\ref{minunK}),
setting
\be \label{Msigmar} \gs_-^\KM{}^{Mnk}= i\, \gx^{\ga \pb} \frac{\p
}{\p  y^\ga_k }  \frac{\p }{\p  {\by}^\pb{}^k}\,\q
{\by}^\pb{}^k={\by}^\pb_j\gd^{kj}. \ee Zero-form primary fields in $\MM $ \be
\label{fieldM}
  C(y,\by|x)=\sum_{p,q}
C_{\ga_1;\ldots;\ga_{p}\,;\pa_1;\ldots;\pa_{q}}
 ^{\,i_1; \ldots;\, i_{p}\,;\,j_1; \ldots;\, j_{q}}(x)
y_{i_1}^{\ga_1}\ldots y_{i_p}^{\ga_{p}}\,\,\by_{j_1}^{\pa_1}\ldots
\by_{j_{q}}^{\pa_{q}}\, \ee
 satisfy the {\it mutual tracelessness condition}  with respect to color indices
\be\label{trrM}
 \gs_-^\KM{}^{Mnk}   C(y,\by|x) =0\,,
\ee which implies \be \label{tr0M} \gd_{mn}
C_{\,\ga_1;\ga_2;\ldots;\ga_{p}\,;\pa_1;\pa_2;\ldots;\pa_{q}}
 ^{\,m\,;\,i_2; \ldots;\,\, i_{p}\,;\,\,n\,\,\,;\,\,j_{\,2}\,; \ldots\,;\, j_{\,q}}
=0.\ee In these terms, the algebra $\mathfrak{o}(\K)$ extends to
$\mathfrak{u}(\K)$ acting on the conjugated representations carried by lower
and upper color indices, while $\mathfrak{sp}(2M)$ reduces to
$\mathfrak{u}(K,K)$ that acts on spinor indices and commutes with the
  $\mathfrak{u}(\K)$.  Note that $\mathfrak{u}(\K)$ and $\mathfrak{u}(K,K)$
contain the common central element.

Since the variables $y_i^\ga$ and  $\by_j^\pa$ are commuting, the tensors
 $ C_{\ga_1;\ldots;\ga_{p}\,;\pa_1;\ldots;\pa_{q}}
 ^{\,i_1; \ldots;\, i_{p}\,;\,j_1; \ldots;\, j_{q}}(x)$
are symmetric with respect to  permutation of any pair of\, upper and lower
indices (for Greek indices of the same type). Hence every such a tensor can be decomposed into a direct sum of
tensors described by irreducible representations of  $\mathfrak{u}(\K) $ as
well as   $\mathfrak{gl}_K\oplus \mathfrak{gl}_K$. The irreducible tensors
have the
 symmetries described by a pair of YD $\Y (n_1,\ldots,n_m)$ and
  $\overline{\Y}(\overline{n}_1,\ldots,\overline{n}_{\bar m})$
  with $n_1+...+n_m=p$, $\bar{n}_1+...+\bar{n}_{\bar m}=q$.

Because of the symmetry with respect to exchange of pairs of indices,
Young diagrams for color indices $i(j)$ and spinor-like indices $\ga (\pa)$
must have the same shape. Hence, primary fields in $\MM $ are \bee\label{leadprimM}
  &&C^{\,\,i^1(n_1)\,,    \ldots,\,\,\, \,i^m(n_m)\,;}
_{\ga^1( {n}_1 )\,,\,  \ldots,\, \ga^m(n_m)\,;}
{\,\, }^{\,j^1(\bar{n}_1)\,,   \ldots,\,\, \,\, \,j^{\bar m}(\bar{n}_{\bar m})}
_{\pa_1(\bar{n}_1)\,, \, \ldots,\pa_{\bar m}(\bar{n}_{\bar m})} (x)\qquad\\ \nn&&
\times y_{i_1^1}^{\ga_1^1 }\ldots  y_{i^1_{n_1}} ^{\ga^1_{n_1}} \ldots
y_{i_1^m}^{\ga_1^m }\ldots  y_{i^m_{n_m}} ^{\ga^m_{n_m}}\,\,
\by_{i_1^1}^{\pa_1^1 }\ldots  \by_{i^1_{\bar{n}_1}} ^{\pa^1_{\bar{n}_1}}
\ldots \by_{i_1^{\bar m}}^{\pa_1^{\bar m} }\ldots  \by_{i^{\bar m}_{\bar{n}_{\bar m}}}
^{\pa^{\bar m}_{\bar{n}_{\bar m}}}\,\,
 \eee
 with tensors $C$
that obey the mutual tracelessness condition (\ref{tr0M})
  and have definite symmetry properties in $y$ and $\by$
described by a pair of  Young diagrams
 \bee  \label{halfdimK}
\begin{picture}(50,60)(0,-10)%
\put(0,20){$\Y_0{}\,=$}
\end{picture}
\begin{picture}(150,65)(0,-10)
\put( 5,23){\small$  m$}%
\put(10,00){\line(1,0){9}}
\put(10,50){\line(1,0){9}}%
\put(11,18){\vector(0,-1){18}}
\put(11,30){\vector(0,01){20}}%
\begin{picture}(120,50)(-22,0)%
\put(92, 45){\small $n_1\,$ } \put(19, -1){\small $n_m> 0\, $ }
\multiput(0,00)(0,7){3}{\usebox{\ver}}
\multiput(0,00)(5,0){01}{\usebox{\gorp}}%
\multiput(0,00)(5,0){02}{\usebox{\verp}}%
\multiput(0,05)(5,0){04}{\usebox{\gorp}}%
\multiput(0,05)(5,0){05}{\usebox{\verp}}%
\multiput(0,10)(5,0){10}{\usebox{\gorp}}%
\multiput(0,10)(5,0){11}{\usebox{\verp}}%
\multiput(0,15)(5,0){10}{\usebox{\gorp}}%
\multiput(0,35)(5,0){15}{\usebox{\gorp}}%
\multiput(0,35)(5,0){16}{\usebox{\verp}}%
\multiput(0,40)(5,0){15}{\usebox{\gorp}}%
\multiput(0,40)(5,0){16}{\usebox{\verp}}%
\multiput(0,45)(5,0){17}{\usebox{\gorp}}%
\multiput(0,45)(5,0){18}{\usebox{\verp}}%
\multiput(0,50)(5,0){17}{\usebox{\gorp}}%
\multiput(3,30)(4,0){14}{\usebox{\toch}}%
\multiput(3,25)(4,0){13}{\usebox{\toch}}%
\multiput(3,20)(4,0){12}{\usebox{\toch}}%
\end{picture}
\end{picture}\,,
\begin{picture}(50,20)(0,-10)%
\put(-2,20){$  \overline{\Y }_0  \,=$}\quad
\end{picture}
\begin{picture}(150,65)(0,-10)
\put( 4,23){\small${\bar m}$}%
\put(127, 45){\small $\bar{n}_1\,   $ }
\put(10,10){\line(1,0){9}}
\put(10,50){\line(1,0){9}}%
\put(11,18){\vector(0,-1){8}}
\put(11,30){\vector(0,01){20}}%
\begin{picture}(120,50)(-22,0)%
\put( 40, 10 ){\small $ \overline{{n}}_{{\bar m}}> 0\, $}
\multiput(0,10)(0,6){3}{\usebox{\ver}}
\multiput(0,10)(5,0){6}{\usebox{\gorp}}%
\multiput(0,10)(5,0){7}{\usebox{\verp}}%
\multiput(0,15)(5,0){6}{\usebox{\gorp}}%
\multiput(0,35)(5,0){15}{\usebox{\gorp}}%
\multiput(0,35)(5,0){16}{\usebox{\verp}}%
\multiput(0,40)(5,0){17}{\usebox{\gorp}}%
\multiput(0,40)(5,0){18}{\usebox{\verp}}%
\multiput(0,45)(5,0){20}{\usebox{\gorp}}%
\multiput(0,45)(5,0){21}{\usebox{\verp}}%
\multiput(0,50)(5,0){20}{\usebox{\gorp}}%
\multiput(3,30)(4,0){15}{\usebox{\toch}}%
\multiput(3,25)(4,0){14}{\usebox{\toch}}%
\multiput(3,20)(4,0){12}{\usebox{\toch}}%
\end{picture}
\end{picture}\ls .
\eee
As a consequence of (\ref{tr0M}) and  {\it Lemma 1} on p.\pageref{Lemma 1}  at $p=0$,
the latter have to obey the condition
\be\label{trlessmnk} \KM \ge {\bar m}+m\ge0\,.
\ee

In the Minkowski case, a pair of Young diagrams ${\Y}_0$ and $\overline{\Y}_0$ (\ref{halfdimK}) will be called
{\it rank-$\KM$ two-column dual} to the   pair $
{\Y}_1(n_1,\ldots,n_m,\underbrace{1, \ldots,1}_q)$ and $
{\overline{\Y}}_1(\overline{n}_1,\ldots,\overline{n}_{\bar m},\underbrace{1, \ldots,1}_q)$
\bee  \label{halfdimKdual}
\begin{picture}(50,60)(0,-40)%
\put(0,20){$ \Y_1{} \,=$}
\end{picture}
\begin{picture}(150,95)(0,-40)
\put( 5,23){\small$  m$}%
\put(10,00){\line(1,0){9}}
\put(10,50){\line(1,0){9}}%
\put(11,18){\vector(0,-1){18}}
\put(11,30){\vector(0,01){20}}%
\put(10,-30){\line(1,0){9}}%
\put(11,-20){\vector(0,-1){10}}
\put(11,-10){\vector(0,01){10}}%
\put( 6,-16){\small$q$}%
\begin{picture}(120,50)(-22,0)%
\put(92, 45){\small $n_1\,  $ }
\multiput(0,-30)(0,5){6}{\usebox{\verp}}
\multiput(0,-30)(0,5){6}{\usebox{\gorp}}%
\multiput(5,-30)(0,5){6}{\usebox{\verp}}%
  \put(18, -1){\small $n_m \,  $ }
\multiput(0,00)(0,7){3}{\usebox{\ver}}
\multiput(0,00)(5,0){01}{\usebox{\gorp}}%
\multiput(0,00)(5,0){02}{\usebox{\verp}}%
\multiput(0,05)(5,0){04}{\usebox{\gorp}}%
\multiput(0,05)(5,0){05}{\usebox{\verp}}%
\multiput(0,10)(5,0){10}{\usebox{\gorp}}%
\multiput(0,10)(5,0){11}{\usebox{\verp}}%
\multiput(0,15)(5,0){10}{\usebox{\gorp}}%
\multiput(0,35)(5,0){15}{\usebox{\gorp}}%
\multiput(0,35)(5,0){16}{\usebox{\verp}}%
\multiput(0,40)(5,0){15}{\usebox{\gorp}}%
\multiput(0,40)(5,0){16}{\usebox{\verp}}%
\multiput(0,45)(5,0){17}{\usebox{\gorp}}%
\multiput(0,45)(5,0){18}{\usebox{\verp}}%
\multiput(0,50)(5,0){17}{\usebox{\gorp}}%
\multiput(3,30)(4,0){14}{\usebox{\toch}}%
\multiput(3,25)(4,0){13}{\usebox{\toch}}%
\multiput(3,20)(4,0){12}{\usebox{\toch}}%
\put( 18, -36){ $\qquad  $    \quad   $q=\KM+1-{\bar m}-m\,  $}
\end{picture}
\end{picture}
\begin{picture}(50,20)(0,-40)%
\put(-2,20){$   \overline{\Y}_1\,=$}\quad
\end{picture}
\begin{picture}(150,95)(0,-40)
\put( 7,21){\small${\bar m}$}%
\put(10,10){\line(1,0){9}}
\put(10,50){\line(1,0){9}}%
\put(11,18){\vector(0,-1){8}}
\put(11,30){\vector(0,01){20}}%
\put(10,-20){\line(1,0){9}}%
\put(11,-10){\vector(0,-1){10}}
\put(11,0){\vector(0,01){10}}%
\put( 6,-6){\small$q$}%
\begin{picture}(120,50)(-22,0)%
\put( 40, 10 ){\small $ \overline{{n}}_{{\bar m}}   $} \put(107, 45){\small
$\bar{n}_1\,  $ }
\multiput(0,-20)(0,5){6}{\usebox{\verp}}
\multiput(0,-20)(0,5){6}{\usebox{\gorp}}%
\multiput(5,-20)(0,5){6}{\usebox{\verp}}%
\multiput(0,10)(0,6){3}{\usebox{\ver}}
\multiput(0,10)(5,0){6}{\usebox{\gorp}}%
\multiput(0,10)(5,0){7}{\usebox{\verp}}%
\multiput(0,15)(5,0){6}{\usebox{\gorp}}%
\multiput(0,35)(5,0){15}{\usebox{\gorp}}%
\multiput(0,35)(5,0){16}{\usebox{\verp}}%
\multiput(0,40)(5,0){17}{\usebox{\gorp}}%
\multiput(0,40)(5,0){18}{\usebox{\verp}}%
\multiput(0,45)(5,0){20}{\usebox{\gorp}}%
\multiput(0,45)(5,0){21}{\usebox{\verp}}%
\multiput(0,50)(5,0){20}{\usebox{\gorp}}%
\multiput(3,30)(4,0){15}{\usebox{\toch}}%
\multiput(3,25)(4,0){14}{\usebox{\toch}}%
\multiput(3,20)(4,0){12}{\usebox{\toch}}
\end{picture}
\end{picture}
\eee
iff
 \be \label{2duMin}\!\!\!
\hei_1(\Y_0{})+\hei_1(\overline\Y_1{})=\hei_1(\overline\Y_0{})+\hei_1(\Y_1{}) =\K+1\,, \quad
\hei_k(\Y_0{})=\hei_k(\Y_1{}),\quad \hei_k(\overline\Y_0{})=\hei_k(\overline\Y_1{}) \,\quad  \forall k\ge2\,. \ee

\newcommand{\ind}{{k}}
Equations of motion in $\MM $ are most conveniently represented with the help
of the projecting tensor
 \be \label{DUALTM}
 \mathrm{E}_{\,\,i_1[h_1]\,, \quad\,\,\,\,\, i_2[h_2]\,, \,\,  \ldots,\,\,\, \,i_\ind [h_\ind ]\,;}
^{\ga_1[\K-\bar{h}_1+1]\,,\,\ga_2[ h_2 ]\,,\, \ldots,\, \ga_\ind [h_\ind ]\,;}
{\,\, }_{\,j_1[\bar{h}_1]\,,\quad \quad\,\,\, j_2[\bar{h}_2]\,, \,\,\,
\ldots,\,\, \,\, \,j_{\bar \ind }[\bar{h}_{\bar \ind }]}
^{\pa_1[\K-h_1+1]\,,\,\pa_2[\bar{h}_2]\,,\, \ldots,\pa_{\bar \ind }[\bar{h}_{\bar \ind }]}\,, \ee
that obeys  mutual tracelessness  condition (\ref{tr0M}) and has
symmetry  described by the two pairs of mutually traceless
 Young diagrams  \be\label{n1}\Y_0{}[h_1,h_2,h_3,\ldots,h_\ind ]\q
 \overline{\Y}_0[\bar{h}_1,\bar{h}_2 ,\ldots,\bar{h}_{\bar \ind }]\ee
in the color indices  and their two-column dual
\be\label{n11}\Y_1{}[\K-\bar{h}_1+1, h_2,\ldots,h_\ind ]\q
\overline{\Y}_1[\K-h_1+1,\bar{h}_2 ,\ldots,\bar{h}_{\bar \ind }]\, \ee
in the spinor ones.

The following   equation holds as a consequence of
rank-$\KM$ unfolded equations (\ref{minunK}):
 \bee \label{conseqrM}
\widehat{\mathrm{E}}_{\Y_0{},\overline{\Y}_0
}\C(y,\by|x)=0\q\rule{160pt}{0pt}\\  \label{conseqrMop}\rule{0pt}{14pt}
\widehat{\mathrm{E}}_{\Y_0{},\overline{\Y}_0 }:=
 \mathrm{E}_{\,\,i_1[h_1]\,, \quad\,\,\,\,\, i_2[h_2]\,, \,\,  \ldots,\,\,\, \,i_\ind [h_\ind ]\,;}
^{\ga_1[\K-\bar{h}_1+1]\,,\,\ga_2[ h_2 ]\,,\, \ldots,\, \ga_\ind [h_\ind ]\,;}
{\,\, }_{\,j_1[\bar{h}_1]\,,\quad \quad\,\,\,\,j_2[\bar{h}_2]\,, \,\,\,
\ldots,\,\, \,\, \,j_{\bar \ind }[\bar{h}_{\bar \ind }]}
^{\pa_1[\K-h_1+1]\,,\,\pa_2[\bar{h}_2]\,,\, \ldots,\pa_{\bar \ind }[\bar{h}_{\bar \ind }]} \qquad
\\ \nn
  \underbrace{\f{\ptl}{\ptl y_{i_1^1}^{\ga_1^1 }}
\ldots \f{\ptl}{\ptl  y_{i_1^{h_1}} ^{\ga_1^{h_1}}}}_{h_1}   \ldots
  \underbrace{\f{\ptl}{\ptl y_{i^1_\ind }^{\ga^1_\ind  }}
\ldots \f{\ptl}{\ptl  y_{i_\ind ^{h_\ind }} ^{\ga_\ind ^{h_\ind }}}}_{h_\ind }   \quad
  \underbrace{\f{\ptl}{\ptl \by_{j_1^1}^{\pa_1^1 }}
\ldots \f{\ptl}{\ptl  \by_{j_1^{\bar{h}_1}} ^{\pa_1^{\bar{h}_1}}
}}_{\bar{h}_1} \ldots
  \underbrace{\f{\ptl}{\ptl \by_{j^1_{\bar \ind }}^{\pa^1_{\bar \ind } }}
\ldots \f{\ptl}{\ptl  \by_{j_{\bar \ind }^{\bar{h}_{\bar \ind }}}
^{\pa_{\bar \ind }^{\bar{h}_{\bar \ind }}}}}_{\bar{h}_{\bar \ind }}\quad
\\\nn
 \underbrace{\f{\ptl}{\ptl x^{\ga_1^{h_1+1}}{} ^{ \,\pa_1^{{\bar{h}_1+1}} } }
\cdots
\f{\ptl}{\ptl x^{\ga_1^{^{\K+1-\bar{h}_1 }}}{} ^{ \,\pa_1^{{\K+1-h_1 }} }}
 }_{\K+1  -h_1-\bar{h}_1}%
 .  \qquad\qquad
\eee
Indeed, by virtue of  (\ref{minunK}), Eq.~(\ref{conseqrMop}) is
equivalent to
 \bee  \label{conseqrMopPr}
\widehat{\mathrm{E}}_{\Y_0{},\overline{\Y}_0 }=
 \mathrm{E}_{\,\,i_1[h_1]\,, \quad\,\,\,\,\, i_2[h_2]\,, \,\,  \ldots,\,\,\, \,i_{\bar \ind }[h_\ind ]\,;}
^{\ga_1[\K-\bar{h}_1+1]\,,\,\ga_2[ h_2 ]\,,\, \ldots,\, \ga_{\bar \ind }[h_\ind ]\,;}
{\,\, }_{\,j_1[\bar{h}_1]\,,\quad \quad\,\,\,\,j_2[\bar{h}_2]\,, \,\,\,
\ldots,\,\, \,\, \,j_n[\bar{h}_{\bar \ind }]}
^{\pa_1[\K-h_1+1]\,,\,\pa_2[\bar{h}_2]\,,\, \ldots,\pa_n[\bar{h}_{\bar \ind }]} \quad
\prod_{l=1}^{\K+1  -h_1-\bar{h}_1} \gd_{p_l\,q_l}\qquad
\\ \nn
  \underbrace{\f{\ptl}{\ptl y_{i_1^1}^{\ga_1^1 }}
\ldots \f{\ptl}{\ptl  y_{i_1^{h_1}} ^{\ga_1^{h_1}}}}_{h_1}   \ldots
  \underbrace{\f{\ptl}{\ptl y_{i^1_\ind }^{\ga^1_\ind  }}
\ldots \f{\ptl}{\ptl  y_{i_\ind ^{h_\ind }} ^{\ga_\ind ^{h_\ind }}}}_{h_\ind }   \quad
 \underbrace{\f{\ptl}{\ptl y_{p_1}^{\ga_1^{h_1+1}}}
\ldots \f{\ptl }{ \ptl  y_{p_N}^{\ga{}_{1}^{\K+1-\bar{h}_1 }}}
 }_{N=\K+1  -h_1-\bar{h}_1} \qquad\qquad
\\ \nn
  \underbrace{\f{\ptl}{\ptl \by_{j_1^1}^{\pa_1^1 }}
\ldots \f{\ptl}{\ptl  \by_{j_1^{\bar{h}_1}} ^{\pa_1^{\bar{h}_1}}
}}_{\bar{h}_1} \ldots
  \underbrace{\f{\ptl}{\ptl \by_{j^1_{\bar \ind }}^{\pa^1_{\bar \ind } }}
\ldots \f{\ptl}{\ptl  \by_{j_{\bar \ind }^{\bar{h}_{\bar \ind }}}
^{\pa_{\bar \ind }^{\bar{h}_{\bar \ind }}}}}_{\bar{h}_{\bar \ind }}\quad
\underbrace{\f{\ptl}{\ptl   \by_{q_1}^{{ \,\pa_1^{{\bar{h}_1+1}} } }}\ldots
\f{\ptl }{ \ptl \by_{q_{N}} ^{\pa{}_{1}^{\K+1-h_1 }}}
  }_{N=\K+1  -h_1-\bar{h}_1}%
 .  \qquad\qquad
  \eee
Since derivatives $\f{\ptl}{\ptl Y_{j}^{\pa }}$  commute while the indices in
columns of  YDs
(\ref{n1}) and (\ref{n11}) are antisymmetrized, the indices $p_1 \ldots
p_ {\K+1  -h_1-\bar{h}_1}$ are antisymmetrized with the indices $i_1[  h_1 ]$
while the indices $q_1 \ldots q_ {\K+1  -h_1-\bar{h}_1}$ are antisymmetrized
with the indices $j_1[ \bar{h}_1]$. This yields  zero by
virtue of   mutual tracelessness 
of tensor $ \mathrm{E}$ (\ref{DUALTM}) and {\it Lemma 1} on
p. \pageref{Lemma 1}.

Since   $\widehat{\mathrm{E}}  $   (\ref{conseqrMop}) is a homogeneous
 differential operator,
Eq.~(\ref{conseqrM}) is obeyed by  primary fields
(\ref{leadprimM}). The nontrivial part of the analysis is to show that
Eq.~(\ref{conseqrM}) yields the full list of dynamical equations. This follows
from the analysis of the cohomology groups $H^{0}(\gs_-^\KM{}^{Mnk})$ and
$H^{ 1}(\gs_-^\KM{}^{Mnk})$ of $\gs_-^\KM{}^{Mnk}$ (\ref{minunK}) for
arbitrary   $M=2K$ outlined in Section \ref{CohomMin}.

\subsection{$4d$  Minkowski space}
\label{Minkovsky40}

The dictionary between the tensor and  two-component spinor notations  is based
on \be A^{\ga\pb}= A^a \sigma_a^{\ga\pb}\,,\qquad \ee where
$\sigma_a^{\ga\pb}$ ($a = 0,1,2,3$)  are four Hermitian $2\times 2$ matrices.
Let us list the $4d$ Minkowski primary fields  and their field equations for
various $\KM$.

As follows from the analysis of \cite{Ann},  the $\KM=1$ primary fields are
 \be\label{fild41}       C (x)\,,\,\,  C (y|x)
\,, \overline{C} (\by|x)\,.
\ee
These have symmetry properties   described by the
following pairs of  Young diagrams \bee\label{H0mnk1Y}
 \begin{picture}(100,20)(30,235)%
\begin{picture}(100,20)(-24,0)%
\put(0,240) {$\quad \bullet$}%
\put(7,240){$ \quad,$} \put(-30,240){    $\Y_0{}  =$}
\end{picture}\end{picture}
\begin{picture}(100,20)(0,235)%
\begin{picture}(100,20)(-24,0)%
 \put(0,240) {$\bullet$}%
   \put(-60,240){    $ \qquad\overline{\Y}_0 =$}
  \put( 7,240){$;$}
\end{picture}
\end{picture}
\\ \nn 
 \begin{picture}(100,20)(30,235)%
\begin{picture}(100,20)(-24,0)%
\put(0,240) {$\quad\bullet$}%
\put( 7,240){$ \quad,$} \put(-30,240){    $\Y_0{}  =$}
\end{picture}\end{picture}
\begin{picture}(100,20)(0,235)%
\begin{picture}(100,20)(-24,0)%
 %
  \multiput(0,240)(5,0){6}{\usebox{\gorp}}%
\multiput(0,240)(5,0){7}{\usebox{\verp}}%
 \multiput(0,245)(5,0){6}{\usebox{\gorp}}%
  \put(-60,240){    $ \qquad\overline{\Y}_0 =$}
  \put( 33,240){$;$}
\end{picture}
\end{picture}
\\\nn
 \begin{picture}(100,20)(30,235)%
\begin{picture}(100,20)(-24,0)%
  \multiput(5,240)(5,0){8}{\usebox{\gorp}}%
\multiput(5,240)(5,0){9}{\usebox{\verp}}%
 \multiput(5,245)(5,0){8}{\usebox{\gorp}}%
 \put(-30,240){    $\Y_0{}  =$}
 \put( 43,240){$ \quad,$}
\end{picture}\end{picture}
\begin{picture}(100,20)(0,235)%
\begin{picture}(100,20)(-24,0)%
\put(0,240) {$\bullet$}%
\put(-60,240){    $\qquad\overline{\Y}_0 =$}
 \put( 7,240){$\,.$}
\end{picture}
\end{picture}
\eee  The    consequences of   (\ref{minunK}) for $\K=1$
 \bee\nn
\gvep^{\ga\gb}\gvep^{\pa\pb} \f{\p^2}{\p x^{\ga\pa}\p x^{\gb\pb}} {C}  (y,
\by|x)=0\q\qquad\\ \nn \gvep^{\ga\gb} \f{\p^2  }{\p x^{\ga\pa}\p y^{\gb }   }
{C}  (y, \by|x)=0\q \gvep^{\pa\pb} \f{\p^2  }{\p x^{\ga\pa} \p \by^{\pb } }
 {C}  (y, \by|x) =0
\eee impose  the equations on primaries (\ref{fild41})
 \bee\label{eqprim1M}
\gvep^{\ga\gb}\gvep^{\pa\pb} \f{\p^2}{\p x^{\ga\pa}\p x^{\gb\pb}} { C}  (
x)=0\q\qquad\\ \nn \gvep^{\ga\gb} \f{\p   }{\p x^{\ga\pa}\p y^\gb    } { C}
(y| x)=0\q \gvep^{\pa\pb} \f{\p   }{\p x^{\ga\pa} \p \by^\pb  }
  \overline{C} (\by| x) =0\,.
\eee
 The symmetry properties of the left-hand-sides of these
 equations
 are  described by
the following pairs of  Young diagrams  \bee\label{H1mnkY}
 \begin{picture}(100,20)(30,235)%
\begin{picture}(100,20)(-24,0)%
\multiput(0,240)(5,0){1}{\usebox{\gorp}}%
\multiput(0,240)(5,0){2}{\usebox{\verp}}%
\multiput(0,245)(5,0){1}{\usebox{\gorp}}%
\multiput(0,245)(5,0){2}{\usebox{\verp}}%
 \multiput(0,250)(5,0){1}{\usebox{\gorp}}%
  \put(-40,240){    $\Y_1{}  =$}
\end{picture}\end{picture}
\begin{picture}(100,20)(0,235)%
\begin{picture}(100,20)(-24,0)%
 \multiput(0,240)(5,0){1}{\usebox{\gorp}}%
\multiput(0,240)(5,0){2}{\usebox{\verp}}%
 \multiput(0,245)(5,0){1}{\usebox{\gorp}}%
\multiput(0,245)(5,0){2}{\usebox{\verp}}%
 \multiput(0,250)(5,0){1}{\usebox{\gorp}}%
  \put(-70,240){    $,\qquad\overline{\Y}_1 =$}
  \put( 10,240){$;$}
\end{picture}
\end{picture}
\\\nn 
\begin{picture}(100,20)(30,235)%
\begin{picture}(100,20)(-24,0)%
\multiput(0,245)(5,0){1}{\usebox{\gorp}}%
\multiput(0,245)(5,0){2}{\usebox{\verp}}%
\multiput(0,250)(5,0){1}{\usebox{\gorp}}%
 \put(-40,240){    $\Y_1 =$}
\end{picture}\end{picture}
\begin{picture}(100,20)(0,235)%
\begin{picture}(100,20)(-24,0)%
 %
 \multiput(0,240)(5,0){1}{\usebox{\gorp}}%
\multiput(0,240)(5,0){2}{\usebox{\verp}}%
%
\multiput(0,245)(5,0){6}{\usebox{\gorp}}%
\multiput(0,245)(5,0){7}{\usebox{\verp}}%
\multiput(0,250)(5,0){6}{\usebox{\gorp}}%
 \put(-70,240){    $,\qquad\overline{\Y}_1 =$}
 \put( 33,245){$ ;$}\end{picture}
\end{picture}\\\nn
 \begin{picture}(100,20)(30,235)%
\begin{picture}(100,20)(-24,0)%
\multiput(0,240)(5,0){1}{\usebox{\gorp}}%
\multiput(0,240)(5,0){2}{\usebox{\verp}}%
 \multiput(0,245)(5,0){8}{\usebox{\gorp}}%
\multiput(0,245)(5,0){9}{\usebox{\verp}}%
 \multiput(0,250)(5,0){8}{\usebox{\gorp}}%
 \put(-40,240){    $\Y_1{}  =$}
\end{picture}\end{picture}
\begin{picture}(100,20)(0,235)%
\begin{picture}(100,20)(-24,0)%
\multiput(0,245)(5,0){1}{\usebox{\gorp}}%
\multiput(0,245)(5,0){2}{\usebox{\verp}}%
 \multiput(0,250)(5,0){1}{\usebox{\gorp}}%
 \put(-70,240){    $,\qquad\overline{\Y}_1 =$}
 \put( 10,240){$.$}
\end{picture}
\end{picture}
\eee

 Consider the rank-two case. To obey (\ref{trlessmnk}),
  in the case of $M=4$, $\KM=2$
primary fields (\ref{leadprimM}) belong to the following list
  \be\label{fild412}          \quad C (y|x)
\,, \quad {C} (\by|x)\,,\quad C (y,\by|x) \,\, \ee
possessing   symmetries described by pairs of Young diagrams (\ref{halfdimK})
  \bee\label{H0mnk1Y+12}
  \begin{picture}(100,20)(30,235)%
\begin{picture}(100,20)(-10,0)%
  \multiput(5,245)(5,0){12}{\usebox{\gorp}}%
\multiput(5,245)(5,0){13}{\usebox{\verp}}%
 \multiput(5,250)(5,0){12}{\usebox{\gorp}}%
  \multiput(5,240)(5,0){7}{\usebox{\gorp}}%
\multiput(5,240)(5,0){8}{\usebox{\verp}}%
 \multiput(5,245)(5,0){7}{\usebox{\gorp}}%
 \put(-56,240){    $\Y_0{}  =$}
 \put( 63,240){$ \quad,$}
  \end{picture}\end{picture}
 \begin{picture}(100,20)(0,235)%
\begin{picture}(100,20)(-24,0)%
\put(20,240) {$\bullet$}%
\put(-60,240){    $\qquad\overline{\Y}_0 =$}
 \put( 27,240){$;$}
\end{picture}
\end{picture}\\
   \label{H0mnk1Y+1}
  \begin{picture}(100,25)(70,235)%
  \begin{picture}(100,20)(-24,0)
\put(20,240) {$\quad\bullet$}%
\put(27,240){$ \quad,$} \put(-30,240){    $\Y_0{}  =$}
\end{picture}\end{picture}
 \begin{picture}(100,25)(-40,235)%
\begin{picture}(100,20)(5,0)%
  \multiput(5,245)(5,0){12}{\usebox{\gorp}}%
\multiput(5,245)(5,0){13}{\usebox{\verp}}%
 \multiput(5,250)(5,0){12}{\usebox{\gorp}}%
  \multiput(5,240)(5,0){7}{\usebox{\gorp}}%
\multiput(5,240)(5,0){8}{\usebox{\verp}}%
 \multiput(5,245)(5,0){7}{\usebox{\gorp}}%
  \put(-74,240){    $ \qquad\overline{\Y}_0  =$}
  \put( 63,240){$ \quad;$}
 \end{picture}
\end{picture}
\\  \label{H0mnk1Y+13}
  \begin{picture}(100,25)(50,235)%
\begin{picture}(100,20)(-24,0)%
\multiput(5,240)(5,0){8}{\usebox{\gorp}}%
\multiput(5,240)(5,0){9}{\usebox{\verp}}%
 \multiput(5,245)(5,0){8}{\usebox{\gorp}}%
  \multiput(5,245)(5,0){1}{\usebox{\gorp}}%
  \put( 48,240){$ \,\,,$}
 \put(-50,240){    $\Y_0{}  =\quad$}
\end{picture}\end{picture}
 \begin{picture}(100,20)(-20,235)%
\begin{picture}(100,20)(-24,0)%
\multiput(0,240)(5,0){6}{\usebox{\gorp}}%
\multiput(0,240)(5,0){7}{\usebox{\verp}}%
 \multiput(0,245)(5,0){6}{\usebox{\gorp}}%
 \put(-80,240){    $ \qquad\overline{\Y}_0  =\quad$}
 \put( 33,240){$\,\,.$}\end{picture}
\end{picture}
\eee
Since primary fields  have to satisfy (\ref{minunK}),
to describe $C (y,\by|x)$ (\ref{fild412})
it is convenient to use the language of highest-weight modules. Solutions to (\ref{minunK}) form    $\mathfrak{u}({\KM})$-modules, where
  the algebra  $\mathfrak{u}({\KM}) $ spanned  by the generators
\be\label{glMM}
    \gp_{m}^{n}= y_m^{\ga}\frac{\ptl}{\ptl y ^{\ga}_n}-
  \by^{\pa}{}^n\frac{\ptl}{\ptl \by^{\pa}{}^m}\,\q
   [\gp_m^p ,\gp_q^j ]=
\gd^p_q     \gp_m^j-   \gd^j_m   \gp_q^p\, \ee
commutes with $\gs_-^\K{}^{Mnk}$ (\ref{Msigmar})\,.
Evidently,  ${\gp}^1_2 C(y_1, \by_2|x)=0$  and  $\Big({\gp}^2_1\Big)^k C(y_1, \by_2|x)$
 with  $0\le k\le max(n,\bar{n})$ form the full list of
 primary fields  with symmetry properties
(\ref{H0mnk1Y+13}).
Note, that these results were used in \cite{{Gelfond:2013xt}} to describe free
$4d$  conformal  primary currents.

Equations of motion are projected by tensors (\ref{DUALTM}).
Since $\ga_j$ and $\pa_j$ take just two values,
to be nonzero at $M=4$, $\KM=2$, these should have
$h_1=\bar{h}_1=1$ and  $h_j\le1$, $\bar{h}_j\le1$ in YDs (\ref{n1}),  (\ref{n11}).
Hence, the primary fields associated  with diagrams (\ref{H0mnk1Y+12}) and (\ref{H0mnk1Y+1})
are off-shell obeying no field equations,
while those  associated with diagrams (\ref{H0mnk1Y+13}) satisfy
 \bee\label{conseq24}
 \gvep^{\ga\gb}\gvep^{\pa\pb}
\f{\p^3 }{\p x^{\ga\pa}\p y_1^{\gb}\p \by_2^{\pb}} C ( y ,\by|x)  =0\q
\gvep^{\ga\gb}\gvep^{\pa\pb} \f{\p^3 }{\p x^{\ga\pa}\p y_2^{\gb}\p
\by_1^{\pb}} C( y ,\by|x)  =0\q\\ \nn \gvep^{\ga\gb}\gvep^{\pa\pb} \f{\p
}{\p x^{\ga\pa}}\left(
 \f{\p^2 }{\p y_1^{\gb}\p \by_1^{\pb}}-\f{\p^2 }{\p y_2^{\gb}\p \by_2^{\pb}}
\right)
 C( y ,\by|x) =0\,.\qquad
\eee
These equations have symmetries of
$ \begin{picture}(90,20)( -40,240)%
\begin{picture}(90,20)
\multiput(0,240)(5,0){1}{\usebox{\gorp}}%
\multiput(0,240)(5,0){2}{\usebox{\verp}}%
 \multiput(0,245)(5,0){8}{\usebox{\gorp}}%
\multiput(0,245)(5,0){9}{\usebox{\verp}}%
 \multiput(0,250)(5,0){8}{\usebox{\gorp}}%
 \put(-40,240){    $\Y_1{}  =$}
\end{picture}\end{picture}
$ and $\begin{picture}(70,20)(-40,240)%
\begin{picture}(70,20)
  \multiput(0,240)(5,0){1}{\usebox{\gorp}}%
\multiput(0,240)(5,0){2}{\usebox{\verp}}%
\multiput(0,245)(5,0){6}{\usebox{\gorp}}%
\multiput(0,245)(5,0){7}{\usebox{\verp}}%
\multiput(0,250)(5,0){6}{\usebox{\gorp}}%
 \put(-40,240){    $  \overline{\Y}_1 =$}
 \end{picture}
\end{picture}$
.

  For  $\KM\ge 3$,  dynamical fields  (\ref{leadprimM})
are described by  Young diagrams (\ref{halfdimK}) with at most two rows. Since
$H^1(\gs_-^\KM{}^{Mnk})$ is empty for  $\KM\ge 3$,  rank-$\KM\ge 3$
primary fields are off-shell.

\subsection{Higher $\gs_-^\KM{}^{Mnk}$-cohomology in $\M_M^{Mnk}$}
\label{ResHigMin}
Due to anticommutativity of differentials, the
differential forms $f(\xi^{\ga\pb})$ are described by pairs
of {\it mutually transposed} Young diagrams $\Y  $ and $\overline{\Y}$ related
by the reflection with respect to the diagonal, \ie
\be\label{mutinMnk}\len_i(\overline{\Y})=\hei_i(\Y) \q\qquad
 \len_j( {\Y})=\hei_j(\overline{\Y}) .\ee
 Such pairs of diagrams, whose role is analogous to that of almost
 symmetric YDs $\Y_A$  (\ref{exalsim}) for the $Sp(2M)$ invariant sytems,
 will be denoted as $\Y_{\aaa} \,,\,\,\overline{\Y}_{\aaa}$,
 describing symmetry  properties of
$f(\xi^{\ga\pb})$ with respect to  unprimed and primed indices,
respectively.

As sketched in Section \ref{Details}, $H^{p}(\gs_-^{\KM}{}^{Mnk} )$ is
characterized by a pair of mutually transposed   Young diagrams
$\Y_{\aaa} $ and $\overline{\Y}_{\aaa} $ such that $|\Y_\aaa|=|\overline{\Y}_\aaa| =p$ and a  pair of
Young diagrams $\Y_0$ and $ \overline{\Y}_0$ obeying the   condition
 $ {\hei}_1(\Y_{{0}})+\hei_1(\overline{\Y}_{{0}})\le \KM $.

To describe cohomology groups   we introduce
the following {\it infinite shift matrix} $
{ \Sm } _{n\,\,m}(\K| \Y_{{0}}, \overline{\Y}_{{0}})\,$
\bee\label{shiftSMin}
   \Sm_{n\,\,m}&=& \del_{m-n  } \qquad\!\! \mbox{ for }
 n< m \,,\quad n \ge1  \q      \\ \nn
\Sm_{n\,\,n} &=& \del_{0} \q\\ \nn
  \Sm_{m\,\,n}&=& \overline{\del}_{m-n}  \qquad \mbox{ for } n< m    \q
 \eee
where
 \bee\label{deltfCarMin} \del_j=\hei_j( {\Y}_{{0}})-\hei_{j+1}({\Y}_{{0}})
\q \overline{\del}_j=
\hei_j(\overline{\Y}_{{0}})-\hei_{j+1}(\overline{\Y}_{{0}})\,
\q \\\nn
{\del}_0=\KM- \hei_1( {\Y}_{{0}})-\hei_{1}(\overline{\Y}_{{0}})
.\qquad\qquad\qquad\qquad\eee
Pictorially
  \bee \label{nestSinfmink}
 \Sm  (\K| \Y_{{0}} , \overline{\Y}_{{0}})   =\left(
\beee{c c c c c c c c c   }
   {\del}_0 &{\del}_1 &{\del}_{2}&{\del}_3&{\del}_4&{\del}_5&{\del}_6& \ldots  \\
   {\bel}_1&{\del}_0&{\del}_1  &{\del}_2&{\del}_3  &{\del}_4&{\del}_5&   \ldots        \\
   {\bel}_ 2&{\bel}_ 1&{\del}_0 &{\del}_1 &{\del}_2  &{\del}_3&{\del}_4 & \ldots \\
   {\bel}_ 3&{\bel}_ 2&{\bel}_ 1  &{\del}_0&{\del}_1&{\del}_2&{\del}_3  &  \ldots    \\
   {\bel}_ 4&{\bel}_ 3&{\bel}_ 2  & {\bel}_ 1&{\del}_0 &{\del}_1 &{\del}_2   &\ldots    \\
   {\bel}_ 5&{\bel}_ 4&{\bel}_ 3  & {\bel}_ 2&{\bel}_ 1  &{\del}_0&{\del}_1& \ldots \\
   {\bel}_6&{\bel}_ 5&{\bel}_ 4  &{\bel}_ 3&{\bel}_ 2  & {\bel}_ 1&{\del}_0 &  \ldots \\
   \vdots  & \vdots & \vdots & \vdots & \vdots  & \vdots  &  \vdots& \ddots      \\
  \eeee\right)\,.
\eee

The  cohomology group $H^{p}(\gs_-^{\KM}{}^{Mnk} )$ consists
of
  the following    pairs of $\mathfrak{gl}_K\oplus \mathfrak{gl}_K$ diagrams
  \bee\label{otvetecMMin}
 \Ycoh &:&\quad \hei_j(\Ycoh)=\hei_j(\Y_{{0}})+\hei_j(\Y_\aaa)+\sum _i    \shm_{i\, j}\q\\ \nn
  \overline{{\Ycoh}}  &:&\quad
   \hei_j(\overline{\Ycoh} )=\hei_j(\overline{\Y}_{{0}})+\hei_j(\overline{\Y}_\aaa )
   +\sum _i     \shm_{j\,\,i} \,, \eee
where $\shm_{i\, j}$ results from the
  intersection of the {infinite  shift matrix} $\Sm(\K| \Y_{{0}}, \overline{\Y}_{{0}} )$
   (\ref{nestSinfmink})  with  $ \Y_\aaa$, \ie
  \be
  \shm_{n\,\,m}= \theta\big(\hei_m({\Y_\aaa})-n\big)
 \Sm_{n\,\,m} \,.
\ee

For example, for the     Young diagram
$\Y_\aaa[6,6,5,5,4,2,2,1 ]$   the
shift matrix is
  \be\!\!\!\label{MUTS1}\shm =\left( \beee{l l l l l l l l l }
   {\del}_0& {\del}_1 &{\del}_2&{\del}_3 &{\del}_4&{\del}_5&{\del}_6&{\del}_7\\
   \bel  _1&{\del}_0&{\del}_1&{\del}_2&{\del}_3&{\del}_4&{\del}_5&0  \\
   \bel  _2&\bel  _1&{\del}_0&\del_1 & {\del}_2 &0&0&0 \\
   \bel  _3&\bel  _2&\bel  _1&{\del}_0&\del_1 &0&0&  0     \\
   \bel  _4&\bel  _3&\bel  _2&\bel  _1&  0   &  0   &  0   &  0     \\
   \bel  _5&\bel  _4&0&0&  0   &  0   &  0   &  0     \\
      \eeee\right) .    \quad
\ee

\section{$\sigma_-$--cohomology analysis }

\label{Details} Our  analysis generalizes those of \cite{tens2}, where the
 rank-two case  was considered, and of \cite{Vasiliev:2009ck}, where conformal
field equations in Minkowski space of any dimension were obtained. The main
tool is the standard homotopy trick.

Let a linear operator $\Omega$ act in a linear space $V$ and satisfy
$\Omega^2=0$. By definition, $H (\Omega )= \ker \,\Omega  /\Imm \, \Omega $ is
the cohomology space. Let $\Omega^*$ be another nilpotent operator,
$(\Omega^*)^2=0$. Then the operator
\be \label{l1} \Delta=\{ \Omega,\Omega^*\, \} \ee
satisfies $ [\Omega , \Delta ] = [\Omega^*\,, \Delta ]=0\,. $ {}
From (\ref{l1}) it follows that
 $\Delta ({\ker \,\Omega})\subset { \Imm } \,\Omega$.
 Therefore
$H(\Omega)\subset \ker\, \Omega / \Delta ({\ker \,\Omega})$. Suppose now that $V$
is a Hilbert space in which $\Omega^*\,$ and $\Omega$ are conjugated. Then
$\Delta$ is positive semi-definite. If the operator $\Delta$
is also  quasifinite-dimensional, \ie $V=\sum \oplus V_\dda$ with
finite-dimensional subspaces $V_\dda$ such that $\Delta (V_\dda ) \subset
V_\dda $ and $V_\dda $ is orthogonal to $V_\ddb$ for $\dda \neq \ddb$, then
$\Delta$ can be diagonalized and it is easy to see that $\ker\, \Omega /
\Delta ({\ker \,\Omega})$ = $\ker \Delta  \cap \ker \Omega$. Therefore, in
this case,
\be {H(\Omega)}\subset \ker \,\Delta \cap  \ker\,\Omega\,. \ee

This formula is particularly useful for the practical analysis since,
to calculate   $H(\Omega)$, one
can use the  {\it Homotopy equation} \be\label{homotopycom}
\Delta  F=0. \ee

 \subsection{Homotopy equation in $\M_M$}
\label{Cohom}
 Defining
\be \f{\p}{\p \kk^{\ddb \ddd}} \kk^{\ddm \ddn} =\half \left(\gd^{\ddm}_{\ddb}
\gd^{\ddn}_{\ddd} +\gd^{\ddn}_{\ddb}\gd^{\ddm}_{\ddd}\right) - \kk^{\ddm \ddn
}\f{\p}{\p \kk^{\ddb \ddd}}\,, \ee we observe that the operators \be
\label{chigen} \gch^{\dda}_{\ddb}= 2 \kk^{\dda \ddd }\f{\p}{\p \kk^{\ddb
\ddd}}\,,\ee form $\mathfrak{gl}(M)$,
 obeying
\be\label{chichi} [\gch^{\dda}_{\ddb}\,,\,\gch^{\ddn}_{\ddm}]
 = \gd^{\ddn}_{\ddb}  \gch^{\dda}_{ \ddm }-
  \gd^{\dda}_{\ddm}\gch^{\ddb}_{ \ddn }
\,,\ee \be\label{chicom} [\gch^{\dda}_{\ddb}\,,\,\kk^{\ddm \ddn }]
 =   \gd^{\ddm}_{\ddb}\kk^{\dda \ddn }
+\gd^{\ddn}_{\ddb}\kk^{\dda \ddm }     \,,\ee
\be\label{chichi2} \gch^{\dda}_{\ddb}\gch_{\dda}^{\ddb}=  (M+1)\gch^{\dda}_{\dda}\,.
\ee

For
\be\label{Omega} \Omega{}^{\K}:=
\gs_-^\K = T_{AB }   \kk^{AB } \, \,
\ee with $\gs_-^\K$ (\ref{sigmar}) and  $T_{AB }$ (\ref{glM2}) we introduce
 \be\label{Omega*}
 \Omega^*{}^{\K}=
 T^{CD}
\frac{\ptl}{\ptl \kk^{CD}}\,\q \left(\Omega^*{}^{\K}\right)^2=0, \ee
\be \label{Deltar} \Delta^{\K}=\{
\Omega{}^{\K}  \,,  \Omega^*{}^{\K}
  \}\,.
\ee An important property of $\Delta^{\K}$ (\ref{Deltar}) is that it is
positive semi-definite because $ \Omega{}^{\K}$ and $\Omega^*{}^{\K}$ are conjugated
in the positive-definite  Fock space generated by $Y^A_i$ and $\xi^{AB}$
  as creation operators.

By virtue of (\ref{TT}), (\ref{chichi}) and (\ref{chicom}), $\Delta^{\K}$ can
be represented in the form \be \label{DeltachiTTT} \Delta^{\K} =
 (\TTT^\ddb_\dda\,+\gch^\ddb_\dda\,)
(\TTT^{\dda}_ {\ddb}+\gch^{\dda }_ {\ddb})+ \Big(T^{AB}  T_{AB} - \TTT^\ddb_\dda
\TTT^{\dda}_ {\ddb}    \Big) +(\K-(M+1)) \gch^{\dda }_ {\dda}\,, \ee
where \be \label{TglM}
\TTT^\dda_\ddb=:Y_j^{\dda}\,   \f{\p}{\p Y_j^\ddb} ={T}^\dda_\ddb-\half\K \gd ^\dda_\ddb 
\ee are generators of the $\mathfrak{gl} _M$ that act on      $Y^A_i$.

By virtue of   (\ref{glM})-(\ref{tt}) an elementary computation yields
\be
\label{schet} T^{AB}  T_{AB} - \TTT^\ddb_\dda \TTT^{\dda}_ {\ddb}   + (M+1-\K
)\TTT_A^A  =\half \gta_{mn}\gta^{mn} \,,
 \ee
 where $\gta_{mn}$ are $\mathfrak{o(\K)}$ generators (\ref{or}).
 Hence
  \be
\label{Deltachittt} \Delta^{\K}= \gn^\dda_\ddb \gn^{\dda }_ {\ddb}+ \half
\gta_{m}{}_{k} \gta^{m}{}^{k} - (M+1-\K )\gn^\dda_\dda \,,\qquad \ee where
 \be
 \gn^\dda_\ddb= \gch^\dda_\ddb+ \TTT^\dda_\ddb\,
 \ee
are generators of the $\mathfrak{gl}^{tot}_M$ that act on indices $A,B,\ldots$
carried by both  $Y^A_i$ and $\xi^{AB}$. The first and second terms on the
\rhs of (\ref{Deltachittt}) are the quadratic  Casimir operators of the
algebras $\mathfrak{gl}^{tot}_M$ and $\mathfrak{o}(\K)$, respectively.

The computation in terms of fermionic oscillator  realization of generators of the algebra  $\mathfrak{gl}_M$, that makes
the antisymmetrization manifest, yields
 (see  e.g.,\cite{Vasiliev:2009ck})
 \be\label{glMH}
\gn^\dda_\ddb \gn^{\dda }_ {\ddb} = -\sum_{ i } H_i(H_i-M-1-2(i-1))\,, \ee
where $i$ enumerates height-$H_i$ columns of the $\mathfrak{gl}_M$  YD $\YY[H_1, H_2,\ldots]$.

As mentioned above,   $\Y_0$ can be treated both as a
$\mathfrak{gl}(M)$   YD, and as a $\mathfrak{o}(\K)$ YD.
Recall that, in addition to the indices of a   $\mathfrak{o}(\K)$--traceless YD $\Y_0$,
the diagram  $\YY $ treated as an $\mathfrak{o}(\K)$ YD  can
contain indices  carried by  products of  $\delta^{ij}$.
Then the respective $\mathfrak{gl}(M)$ tensor   contains    products of
  the  ``tracefull" combinations (\ref{Kron})
   that, being symmetric in the indices $A,B$,  are  described by $\Y[1,1]$.
In addition, it can contain indices  carried  by the differentials $\xi^{AB}$ also described by $\Y[1,1]$.

  As a result
\be\label{yysubsety0}\YY[H_1, H_2,\ldots]\in\Y_0 [h_1, h_2, \ldots]\otimes\underbrace{  \Y[1,1]\otimes\ldots\otimes
\Y[1,1]}_{N}\ee for some $N$. Here the pairs of symmetric indices carried by $
U^{AB}$ (\ref{Kron})  and $\xi^{AB}$ are symmetrized and antisymmetrized,
respectively.

Analogous  computation for
 the orthogonal algebra yields
\be\label{ortMH} \gta_{m}{}_{k} \gta^{m}{}^{k}=2\sum_{ j }
h_i(h_i-\K-2(i-1))\,, \ee where $i$ enumerates height-$h_i$ columns of the
$\mathfrak{o}(\K)$--traceless  YD $\Y_0 [h_1, h_2,\ldots ]$.

Hence, Eq.~(\ref{DeltachiTTT}) yields
\be \label{DeltaKaz} \Delta^{\K}
  =
  -\sum_{ i } H_i(H_i-\K-2(i-1))
+  \sum_{j } h_j(h_j-\K-2(j-1))
  \,.\qquad \ee
 By virtue of Eqs.~(\ref{sumH})
this  yields
 \be
\label{Kazx-y1}
 \Delta^{\K}
=  2\,\,\chi^{\half(\K-1)}(\Y') - 2\,\,\chi^{\half(\K-1)}(\Y_0) \,.
\ee
Since $\chi^{a}$ is additive, from (\ref{Kazx-y1}) it follows that
 \be
\label{Kazx-y=}
   \,\,\chi^{\half(\K-1)}(\Y') -  \,\,\chi^{\half(\K-1)}(\Y_0)\,\equiv \,\chi^{\half(\K-1)}(\Y' \setminus  \Y_0)  \q
\ee where  summation in $\chi^{\half(\K-1)}(\Y' \setminus  \Y_0)$    is over the cells $S({x,y})$  of the supplement of $\Y_0$
 in  $ \Y' $ with the convention that
  $\Y_0$  is situated in the   upper left  corner of $\Y'$.

  Consider Homotopy  equation $\Delta^\K F=0$
(\ref{homotopycom})  on a tensor $F$ obeying symmetry properties of
 $\YY[H_1, H_2,\ldots] $ (\ref{yysubsety0}). By virtue of Eqs.~(\ref{sumH}),
 (\ref{DeltaKaz}) it  takes the form
 \be
\label{Kazx-y}
     \,\chi^{\half(\K-1)}(\Y' \setminus  \Y_0)  =0\,.
\ee

Note that since  equation (\ref{Kazx-y}) is independent  of $M$,
the assumption of Section \ref{allsigcohomology} that $M$ is large enough
to avoid the restriction on the maximal height of $\mathfrak{gl}_M$-Young
diagrams does not restrict the  generality.

 Analogously to \cite{Vasiliev:2009ck}, to find the complete set of solutions to
the homotopy equation one can use the positive semi-definiteness of $\Delta^{\K}$.
Indeed,  nontrivial cohomology
can appear at $\Delta^{\K}=0$ which is its minimal possible value.
Equation (\ref{chicaz}) implies that for a given set of cells of a
$\mathfrak{gl}(M)$-diagram, the minimal value of
$\chi^{\half(\K-1)}(\Y' \setminus  \Y_0) $ is reached when all cells are
situated maximally South-West, \ie possible values of $i$ and $j$ in
(\ref{chicaz}) are maximized and minimized, respectively. This minimization
will be referred to as {\it South-West (SW) principle}.

For a given
traceless YD $\Y_0 $ the minimization occurs with respect
 to the cells associated with the
  differentials $\xi^{AB}$ forming an  almost symmetric
 YD $\Y_{A}$ (\ref{exalsim})
 and   the tracefull blocks $U^{AB}$ (\ref{Kron}) forming a YD $\Y_\gd$ (\ref{Krond}). Any solution $\Y'$ of (\ref{Kazx-y}) is given by
  \be\label{SWcomp1}\Y'  \in SW \big(\Y_0\otimes\,\Y_\gd\, \otimes\, \Y_{A}\big)\,\ee
for an appropriate $\Y_\gd$, where
 $  SW\big(\Y_1\otimes\,\Y_2\otimes\,\ldots\big)$ denotes
{\it maximally South-West components ($\sim$ SW-diagrams)} of the tensor
product $\Y_1\otimes\,\Y_2\otimes\,\ldots$.
For given $\Y_0$ and $\Y_A$,  the problem is to find  $\Y_\gd$ (\ref{Krond}) such that
the maximally South-West component of the tensor product $\Y_0\otimes\,\Y_\gd\, \otimes\, \Y_{A}$ solves (\ref{Kazx-y}).

The   details of the analysis of the Homotopy equation
are given in Appendix  A.
 However, for  the final results  presented in Section \ref{allsigcohomology} it
 is easy to
check that $\Y' $ (\ref{otvetec})  solves   homotopy equation (\ref{Kazx-y}).
Indeed,  substitution of $\Y'$ (\ref{otvetec}) into (\ref{Kazx-y}) yields using (\ref{sumH})
\be\label{sumnoshift}
\chi^{\half(\K-1)}(\Y' \setminus  \Y_0) =\sum_{j\le\la}\,\,\sum _{k=1 }^{\hh_j }  (\sh_{k\,j}+1) \Big(j-
h_j -\!\sum_{n<k} (\sh_{n\,j}+1) - \half(\sh_{k\,j}+2)
+\half(\K-1)\Big) \,. \ee
 For $ \Y_A = \Yn{}^{nest} \{a_1,\ldots a_\la\}$ (\ref{nestedhooks}),
using the nested structure of the nested shift matrix (\ref{nestS1}) along with
Eq.~(\ref{shiftinfin}), the \lhs of (\ref{sumnoshift}) can be  rewritten in the form
\be\label{chiproof1}\chi^{\half(\K-1)}(\Y' \setminus  \Y_0) =
 \sum_{k \le\la}
       \sum_{m  = k }^{a_k}
                     (\sh_{k\, m }+1) \Big\{A_{mk}+B_{mk}   \Big\}
\,,\,\, \ee
where\be\label{exprcurly1}A_{mk}=  \K  -h_k -h_{m+1}
                     -  \sh_{m\,k  }  - \sum_{n< m } \sh_{n\, k }
                       -\sum_{n< k} \sh_{m\, n },\ee
   \be\label{exprcurly2}B_{mk}=  m+k-1 -1 - \sum_{n< m } 1
                       -\sum_{n< k} 1.  \ee
One can see that $A_{mk}=0$ for all $  k,m$ by virtue of (\ref{shiftS}) while $B_{mk}$ also vanishes.

\subsection{Sketch of the $\sigma_-$--cohomology analysis in $\M_M^{Mnk}$}
\label{CohomMin}

Here we consider  $\sigma_-$--cohomology   in the generalised Minkowski space $\M_M^{Mnk}$  with $ M=2K$, \ie $H(\gs_-^\KM{}^{Mnk})$ for $\gs_-^\KM{}^{Mnk}$
(\ref{Msigmar}).

 The operators
 \be \label{chigenM}\phi^{\ga}_{\gb}=
  \kk^{\ga \pga }\f{\p}{\p \kk^{\gb \pga}}\q
  \bhi^{\pa}_{\pb}=
  \kk^{\ga \pa }\f{\p}{\p \kk^{\ga \pb}}\,,\qquad
\ee form  $\mathfrak{gl}_{K}({\mathbb C})$ while the
operators
 \bee\label{uMM}
T^{\ga\pb}   = y_j^{\ga} y^{\pb}{}^j\,,\quad
T_{\ga \pb}   = \frac{\p }{\p y_j^{\ga}}\frac{\p }{\p y^{\pb}{}^j} \,,\quad
 T^\ga_\gb= \half \left\{ y_j^{\ga}\,,\frac{\p }{\p  y_j^{\gb}}\right\}  \,,\quad
  T^\pa_\pb= \half \left\{ \by^{\pa}_j\,,\frac{\p }{\p  \by^{\pb}_j}\right\}\quad
  \eee
form   $\mathfrak{u}(K\,,K)$.

 Setting
 \be\label{OmegaM}
\Omega{}^{\KM}_{Mnk}=\gs_-^\KM{}^{Mnk}=
 T_{\ga\pb }
  \kk^{\ga\pb } \q
 \Omega^*{}^{\KM}_{Mnk}=
 T^{\ga\pb}
\frac{\ptl}{\ptl \kk^{\ga\pb}}\,, \ee
we obtain using  (\ref{uMM}), (\ref{glMM})
   \bee\label{DeltaCazM}
\!\!\Delta^{\KM}_{Mnk} = \half  \Big(   \gn^\ga_\gb \gn^{\gb }_ {\ga}+
 \bar{\gn}^\pa_\pb \bar{\gn}^{\pb }_ {\pa}
-  \gp_m^p  \gp^m_p
 + ( \KM-K ) ({\gn}^\ga_\ga+\bar{\gn}^\pa_\pa)
\Big)\q
 \eee
where $$\gn^\ga_\gb= \phi^\ga_\gb+ \TTT^\ga_\gb\q\bar{\gn}^\pa_\pb=
\phi^\pa_\pb+ \TTT^\pa_\pb
 \q \TTT^\ga_\gb =y_j^{\ga}\, \frac{\p }{\p  y_j^{\gb}}\,, \quad
  \TTT^\pa_\pb= \by^{\pa}_j\, \frac{\p }{\p  \by^{\pb}_j}
  $$ are $\mathfrak{gl}_{K}({\mathbb C})$  generators while  $$    \gp_{m}^{n}= y_m^{\ga}\frac{\ptl}{\ptl y ^{\ga}_n}-
  \by^{\pa}{}^n\frac{\ptl}{\ptl \by^{\pa}{}^m}\,
$$   are   $\mathfrak{u}(\KM) $-generators (\ref{glMM}).
The sum of the  first two  terms and the third one
on the \rhs of (\ref{DeltaCazM}) are the quadratic
Casimir operators of $\mathfrak{gl}_{K}({\mathbb C})$ and $\mathfrak{u}(\KM)$,
respectively.

  The $\mathfrak{u}(\KM) $-generators $\gp_{m}^{n}$ (\ref{glMM})
     commute with $\mathfrak{u}(K,K)$ (\ref{uMM}). Hence
$\mathfrak{u}(\KM) $ acts on solutions of the higher-rank equations which
 form $\mathfrak{u}(\KM) $-modules. Irreducible  $\mathfrak{u}(\KM) $-modules
 satisfy the
mutual  tracelessness conditions (\ref{trrM}) equivalent to \be\label{trrMP}
  T_{\ga\,\pb}  P(y,\by,\gx|x) =0\,.
\ee

An elementary  computation in terms of fermionic oscillator realization of
$\mathfrak{u}(\KM)$, that makes the  anti-symmetrization manifest, yields with the help of
mutual tracelessness condition (\ref{trrMP})
\be \gp_m^p  \gp^m_p=-
 \sum_{ i } h_i(h_i -\K-2 (i-1) -1 )-
 \sum_{ j} \bar{h}_j(\bar{h}_j -\K-2(j-1)-1  ) ,
\ee where $h_i$ and $\overline{h}_i$ are heights of the respective columns of $ {\Y}_0
[{h}_1 \ldots,  {h}_{{0}}]$ and $\overline{\Y}_0 [\bar{h}_1, \ldots,
\bar{h}_k]$.
Using   (\ref{glMH})   for the   Casimir operators
$\gn^\ga_\gb \gn^{\gb }_{\ga}$ and $ \bar{\gn}^\pa_\pb \bar{\gn}^{\pb }_
{\pa}$ ,    Eq.~(\ref{DeltaCazM}) yields \bee
\label{DeltaKazM} \Delta^{\K}_{Mnk} =
 \half\Big(
-\sum_{ m } H_m(H_m -2(m-1) ) -\sum_{ n } \overline{H}_n(\overline{H}_n
-2(n-1) )\qquad\qquad
\\ \nn +\sum_{ i } h_i(h_i -2(i-1) ) +\sum_{ j }
\bar{h}_j(\bar{h}_j -2(j-1)) +  (1+\KM )\sum_{ p }\left(H_p+ \overline{H}_p-
h_p -\bar{h}_p\right) \Big) \,,\qquad \eee
where $m$ enumerates columns of
the $\mathfrak{gl}_{K}({\mathbb C})$ YD  $\Yodm[H_1 , H_2\ldots]$ and $n$
enumerates columns of the $\mathfrak{gl}_{K}({\mathbb C})$
YD $\overline{\Y}'[\overline{H}_1, \overline{H}_2 \ldots]$\,.

 Analogously to  Section \ref{Cohom}, in addition to indices
  associated with the   given mutually traceless $\Y_0$ and $\overline{\Y}_0$,  the diagrams
  $\YY$ and $\YYY$ contain indices
  of the differentials $\xi^{\ga\pb}$ and of the
  $\mathfrak{u}(\K)$ invariant tracefull combinations
  \be \label{KronMM}
  u^\ga{}^\pb \,=\gd^{ij} y^\ga_i\by^\pb _j.
  \ee
   As a result $\YY\otimes \YYY$  belongs to
$\Y_0 \otimes \overline{\Y}_0\underbrace{\otimes  \Y[1]\ldots\otimes  \Y[1]}_{N}
\underbrace{\otimes  \overline{\Y} [1]\ldots\otimes  \overline{\Y}[1]}_{N}$
for some $N\ge1$. Hence $ H_i =h_i+s_i $, $ \overline{H}_i =\bar{h}_i+\bar s_i $
  with some $s_i\ge0$, $\bar s_i\ge0$, $\sum s_i=\sum \bar s_i=N $\,.

By virtue of Eqs.~(\ref{DeltaKazM}) and (\ref{sumH}) the equation
$\Delta^{\KM}_{Mnk} =0$ yields
\be  \label{Kazx-yM}
\chi^{\half\K }(  {\Y}'\setminus   {\Y}_0)
+\chi^{\half\K }( \overline{\Y}'\setminus  \overline{\Y}_0)
 =0  \q
\ee
where, as before,   summation  in $\chi^{\half\K }(\Y' \setminus  \Y_0)$
is over the cells $S({x,y})$  of the supplement of $\Y_0$
 in  $ \Y' $ with the convention, that
  $\Y_0$  is situated in the upper left  corner of $\Y'$,
  and analogously for the conjugated YDs.

This form of the homotopy equation allows us to make sure that any pair
of Young diagrams $ \Ycoh$,  $\overline{\Y}' $ of the
form (\ref{otvetecMMin})  indeed solves (\ref{Kazx-yM}), therefore
describing the $\gs_{}^{Mnk}$--cohomology.
Using the positive semi-definiteness of $\Delta^{\KM}_{Mnk}$ and the South-West
principle along the lines of Section \ref{Cohom} and Appendix A, it can be shown that the
list of solutions of Section \ref{ResHigMin} is complete.

\section{Conclusion}
\label{Conclusions}   Results of
this paper raise a number of interesting problems. One of the most interesting is to study
 conserved charges generated by the constructed currents.
A single conserved current is expected to generate many different charges
upon integration with different global symmetry parameters $\eta$. (For
example, a traceless stress tensor generates the full conformal algebra being integrated
with the parameters of translations, Lorentz rotations, dilatations and
special conformal transformations.) See, e.g., \cite{{Gelfond:2013xt}} for
the analysis of this issue for rank-two conserved currents. It is therefore
important to find the full space of the corresponding symmetry parameters
$\eta$ leading to independent charges. An interesting peculiarity of this
analysis is that, as shown in \cite{gelcur} for the case of rank two, to
obtain non-zero charges in  Minkowski subspace of $\M_M$ it is necessary to
consider parameters $\eta$ that are singular in some of coordinates in $\M_M$
transversal to Minkowski space. It is therefore necessary to find what is an
appropriate singularity of $\eta$ in the general case of any rank that gives
rise to non-zero conserved charges.

Another peculiarity is that, being multilinear in the dynamical fields, the
charges resulting from the  proposed currents cannot be represented as
integrals in usual Minkowski space, requiring integration over a larger space
like $\M_M$ or its  product with the twistor space. Nevertheless, being
nonlocal from the perspective of Minkowski space, the charges are
well-defined and should form some algebra. An  interesting question is what
is this algebra and,  specifically, what is its relation to the
multiparticle algebra proposed recently in \cite{Vasiliev:2012tv} and
\cite{{Gelfond:2013xt}} where it was shown, in particular, that the usual
bilinear (\i.e., rank-two) currents give rise to a set of charges that forms
the higher-spin algebra.

A very interesting possible application of the obtained results
can  be related  to the analysis of
multi-particle amplitudes  in QFT. Indeed, the higher-rank fields
considered in this paper can be interpreted as being associated with
the asymptotic states in the scattering processes which
are on shell, \ie obey free field equations. The idea is to associate the
constructed conserved charges with the amplitudes. Of course, this will not
allow us to determine amplitudes exactly since for this it is necessary
to determine the parameters $\eta$ in (\ref{Jeta}) which remain arbitrary
functions of the respective twistor variables
in our approach and have to be determined by the dynamics of a nonlinear
model in question. However, this will determine the amplitude  kinematics.

The reason why we believe that the amplitudes should be associated with the
higher-rank conserved charges is that,
being represented by integrals of on-shell closed forms,
 in this case they will only be determined
by certain singularities independent of local variations of the integration
contour. This interpretation  is not only very similar to the what happens in the
Grassmanian computations (see e.g. \cite{Arkani-Hamed:2014bca,Bork:2016xfn} and references therein)
but can open a unique opportunity
for establishing explicit relation between the space-time computational schemes
and those in the twistor space. To this end the constructed conserved charges
should be extended to differential forms in the correspondence space unifying
space-time with the twistor space. This program was initiated in \cite{gelcur}
for general tensor products.  We anticipate that this construction
allows an extension to  irreducible conserved charges considered in this paper,
integrated over the full correspondence space with appropriate singular parameters
analogously to the construction of Minkowski currents
from those in $\M_4$ presented in \cite{gelcur}. We leave this exciting problem for the future.

Finally, the analysis of higher cohomology performed in this paper may have
applications to the construction of equations of motion of higher gauge theories
associated with higher-rank fields. The latter  are related  to multiparticle
states in the original field theoretic model. Such higher gauge theories are
likely to be related to string-like higher-spin gauge theories.

\section*{Acknowledgments}

We are grateful to Nathan Berkovits for hospitality at ICTP-SAIFR Sao-Paulo
where a part of this work was done.
Also we thank the organizers
and participants of winter 2014
 Kavli Institute for Theoretical Physics in Santa Barbara
program ``New Methods in Nonperturbative Quantum Field Theory",
where this work was continued, for creation of friendly and
productive atmosphere and stimulating conversations.
The authors are grateful to Leonid Bork for the stimulating discussion
and Yegor Goncharov for pointing out a typo in the draft. The first
version of this work  was supported in part  by RFBR Grant No 11-02-00814-a.
The extension of the original version  by the evaluation of
higher $\sigma_-^\K$- cohomologies in $\M_M$ and $\M_M^{Mnk}$   presented
in Sections \ref{allsigcohomology} and \ref{ResHigMin}  and in Appendix A
was supported    by the Russian Science Foundation grant 14-42-00047.

\newcounter{appendix}
\setcounter{appendix}{1}
\renewcommand{\theequation}{\Alph{appendix}.\arabic{equation}}
\addtocounter{section}{1} \setcounter{equation}{0}
 \renewcommand{\thesection}{\Alph{appendix}.}
  \addtocounter{section}{1}
\addcontentsline{toc}{section}{\,\,\,\,\,\,\,Appendix A.  {Details of analysis of
Homotopy equation in $\M_M$}}
\section*{Appendix A.  Details of analysis of
Homotopy equation in $\M_M$}\label{Appendix}

   For any pair of  Young diagrams
     $\Y_{1,2}$,   $\Y_1  \oSW\Y_2= \Y_2  \oSW\Y_1$   is defined  as
   \be\label{OSW}
  \Y_1  \oSW\Y_2:\quad \hei_j(\Y_1\oSW\Y_2)=\hei_j(\Y_1 )+\hei_j( \Y_2)
\quad \forall j\,. \ee Evidently,   $\Y_1  \oSW\Y_2$  is a Young
diagram. From definition (\ref{chicaz1S}) it follows that \be
\label{chisw} \chi^a (\Y_1  \oSW\Y_2) = \chi^a(\Y_1)  +\chi^a(\Y_2)
- \sum_j h_j (\Y_1) h_j (\Y_2)\,. \ee

To solve homotopy equation (\ref{Kazx-y})
  we observe that at sufficiently large  $M$  a solution $\Y'$ (\ref{SWcomp1})
 to (\ref{Kazx-y}) has the form
 \be
\label{Y''oSWYA} \Yoda =\Y''\oSW \Y_A
\ee
with some \be\label{Y''inYdel}  \Y''\equiv\Yod\in  \Y_0\otimes\,\Y_\gd\, .\ee
 Indeed, since  there are no constraints on the heights of components of
 $\Y_1{}\otimes\Y_2$ at sufficiently large $M$ for given $\mathfrak{gl}_M$-diagrams $\Y_1{} $ and $\Y_{2 } $,
   $ \Y_1  \oSW\Y_2$
  is nonzero and maximally South-West. Note that a similar argument cannot
  be applied to the construction of $\Y''$ since $ \Y_0 \oSW\,\Y_\gd$
  can  be zero by {\it Lemma 1} (see p.\pageref{Lemma 1}).

Hence, Eq.~(\ref{Kazx-y}) takes the form
\be \label{DeltaKazSW0}\Delta^{\K}=2\chi^{\half(\K-1)}\big((\Y''\oSW \Y_A )\big/  \Y_0\big) =0.
\ee

From (\ref{OSW}) and (\ref{chisw}) it follows that
for any YD $\Y$ and almost symmetric YD  $\Y_A$
\be\label{chisw2}
 \chi^{\half(\K-1)}(\Y\oSW \Y_A) = \chi^{\half(\K-1)}(\Y)+\half
\sum_{i=1}^{\len_1(\Y_A)} \hei_i(\Y_A)(\K-2\hei_i(\Y ))
\,.\quad
\ee
As a result, using additivity of   $\chi^{a}$ (\ref{chicaz}),
  condition  (\ref{DeltaKazSW0}) amounts to
\be \label{DeltaKazSW} \chi^{\half(\K-1)}(\Y'' \big  )-\chi^{\half(\K-1)}(  \Y_0)+\half
 \sum_i^{\len_1(\Y_A)} \hh_i(\K-2\Hh_i)=0
\,\q  \ee
where
\be\label{defhs}\hh_i=\hei_i(\Y_A)\q
\Hh_i=\hei_i(\Y'')\,.\ee

Since $\Delta^{\K}$ is positive semi-definite,
 Eq.~(\ref{chisw2}) implies that, for given  $ \Y_0$ and $\Y''$,  a solution
 to (\ref{DeltaKazSW}) demands $\Y_A$ to minimize the last term, \ie
 if  $ {\Y}_A$  solves
(\ref{DeltaKazSW}) then
\be\label{semipostil}
 \sum_j^{\len_1(\widetilde{\Y}_A)} \hei_j(\widetilde{\Y}_A)(\K-2\Hh_j)\ge
 \sum_i^{\len_1(\Y_A)} \hei_i(\Y_A)(\K-2\Hh_i)\,\ee
for any almost symmetric $\widetilde{\Y}_A$.   Consideration of almost symmetric  diagrams
$\widetilde{\Y}_A $ resulting from  varying    $\Y_A$
by  a pair of   cells severely   restricts  $\Hh_i$  allowed
 by (\ref{semipostil}). Analogous trick applied to the Kronecker YD $\Y_\gd$  makes it possible to solve (\ref{DeltaKazSW}).

For instance, consider $ \Yn^{nest} \{\underbrace{10,9,8}_{3}\,,  \underbrace{4,3 }_{2}  \,\}
 $ that consists of two block hooks.
 \sbox{\gorp}{\linethickness{.250mm}\line(1,0){10}}
\sbox{\verp}{\linethickness{.250mm}\line(0,1){10}}
 \bee\label{nested122}
  \begin{picture}(200,130)(-40,-30)
\multiput(-20,-10)( 0,10){11}{\usebox{\gorp}}%
\multiput(-10,-10)( 0,10){11}{\usebox{\gorp}}
\put(-20,-10){\linethickness{.250mm}\line(0,1){100}}%
\put(-10,-10){\linethickness{.250mm}\line(0,1){100}}%
\multiput(0,70)(10,0){10}{\usebox{\verp}}
\multiput(0,80)(10,0){10}{\usebox{\verp}}
\put(0,90) {\linethickness{.250mm}\line(1,0){90}}
\put(0,80){\linethickness{.250mm}\line(1,0){90}}
\multiput(0,-10)(10,0){01}{\usebox{\gorp}}
\multiput(0,00)(10,0){01}{\usebox{\gorp}}
\multiput(0,10)(10,0){01}{\usebox{\gorp}}
\multiput(0,20)(10,0){03}{\usebox{\gorp}}
\multiput(0,30)(10,0){03}{\usebox{\gorp}}
\multiput(0,70)(10,0){09}{\usebox{\gorp}}
\multiput(0,60)(10,0){09}{\usebox{\gorp}}
\multiput(0,50)(10,0){06}{\usebox{\gorp}}
\multiput(0,40)(10,0){06}{\usebox{\gorp}}
\multiput(0,60)(10,0){10}{\usebox{\verp}}
\multiput(0,50)(10,0){07}{\usebox{\verp}}
\multiput(0,40)(10,0){07}{\usebox{\verp}}
\multiput(0,30)(10,0){04}{\usebox{\verp}}
\multiput(0,20)(10,0){04}{\usebox{\verp}}
\multiput(0,10)(10,0){02}{\usebox{\verp}}
\multiput(0,00)(10,0){02}{\usebox{\verp}}
\multiput(0,-10)(10,0){02}{\usebox{\verp}}
{   %
 \put(37.3,30){  $\blacksquare$} %
 \put(27.5, 30){  $\blacksquare$}
%
} \put(-60 , 20){$\Yn^{nest}_{(3+)}
=\rule{140pt}{0pt} $}
  \end{picture}
  \quad
  \begin{picture}(80,130)(-40,-30)
\multiput(-20,-10)( 0,10){11}{\usebox{\gorp}}%
\multiput(-10,-10)( 0,10){11}{\usebox{\gorp}}
\put(-20,-10){\linethickness{.250mm}\line(0,1){100}}%
\put(-10,-10){\linethickness{.250mm}\line(0,1){100}}%
\multiput(0,70)(10,0){10}{\usebox{\verp}}
\multiput(0,80)(10,0){10}{\usebox{\verp}}
\put(0,90) {\linethickness{.250mm}\line(1,0){90}}
\put(0,80){\linethickness{.250mm}\line(1,0){90}}
\multiput(0,-10)(10,0){01}{\usebox{\gorp}}
\multiput(0,00)(10,0){01}{\usebox{\gorp}}
\multiput(0,10)(10,0){01}{\usebox{\gorp}}
\multiput(0,20)(10,0){03}{\usebox{\gorp}}
\multiput(0,30)(10,0){03}{\usebox{\gorp}}
\multiput(0,70)(10,0){09}{\usebox{\gorp}}
\multiput(0,60)(10,0){09}{\usebox{\gorp}}
\multiput(0,50)(10,0){06}{\usebox{\gorp}}
\multiput(0,40)(10,0){06}{\usebox{\gorp}}
\multiput(0,60)(10,0){10}{\usebox{\verp}}
\multiput(0,50)(10,0){07}{\usebox{\verp}}
\multiput(0,40)(10,0){07}{\usebox{\verp}}
\multiput(0,30)(10,0){04}{\usebox{\verp}}
\multiput(0,20)(10,0){04}{\usebox{\verp}}
\multiput(0,10)(10,0){02}{\usebox{\verp}}
\multiput(0,00)(10,0){02}{\usebox{\verp}}
\multiput(0,-10)(10,0){02}{\usebox{\verp}}
{   %
 \put(-2.5,-8){  $\times$}%
\put(77.5, 62){  $\times$}%
} \put(-60 , 20){$\Yn^{nest}_{(1-)}
=\rule{140pt}{0pt} $}
  \end{picture}
\eee
 The  black cells of the left YD  exemplify an addition of a pair  to form  the
 third  block hook,
while the crossed cells of the right YD can be removed from the first block hook to
produce another almost symmetric diagram.

Another useful example is a rectangular block.
\bee\label{nested123}
 \begin{picture}(180,70)
\put(-40, 40){$
\Yn^{nest}_{(2+)}
= $}
 \multiput(0,10)(10,0){07}{\usebox{\gorp}}
\multiput(0,20)(10,0){07}{\usebox{\gorp}}
\multiput(0,30)(10,0){07}{\usebox{\gorp}}
\multiput(0,70)(10,0){07}{\usebox{\gorp}}
\multiput(0,60)(10,0){07}{\usebox{\gorp}}
\multiput(0,50)(10,0){07}{\usebox{\gorp}}
\multiput(0,40)(10,0){07}{\usebox{\gorp}}
\multiput(0,60)(10,0){08}{\usebox{\verp}}
\multiput(0,50)(10,0){08}{\usebox{\verp}}
\multiput(0,40)(10,0){08}{\usebox{\verp}}
\multiput(0,30)(10,0){08}{\usebox{\verp}}
\multiput(0,20)(10,0){08}{\usebox{\verp}}
\multiput(0,10)(10,0){08}{\usebox{\verp}}
  \put( 68,62){  $\blacksquare$} %
\put(   -04,0){  $\blacksquare\,\,\qquad$}
  \end{picture}
\quad\begin{picture}(80,70)
\put(-40, 40){$
\Yn^{nest}_{(1-)}
= $}
\multiput(0,10)(10,0){07}{\usebox{\gorp}}
\multiput(0,20)(10,0){07}{\usebox{\gorp}}
\multiput(0,30)(10,0){07}{\usebox{\gorp}}
\multiput(0,70)(10,0){07}{\usebox{\gorp}}
\multiput(0,60)(10,0){07}{\usebox{\gorp}}
\multiput(0,50)(10,0){07}{\usebox{\gorp}}
\multiput(0,40)(10,0){07}{\usebox{\gorp}}
\multiput(0,60)(10,0){08}{\usebox{\verp}}
\multiput(0,50)(10,0){08}{\usebox{\verp}}
\multiput(0,40)(10,0){08}{\usebox{\verp}}
\multiput(0,30)(10,0){08}{\usebox{\verp}}
\multiput(0,20)(10,0){08}{\usebox{\verp}}
\multiput(0,10)(10,0){08}{\usebox{\verp}}
 \put(47,10){  $\times$} %
\put(57,10){  $\times$}
   \end{picture}
\eee
 Here the two black cells form the only pair that can be added,
while the  two crossed cells form the only pair  that can be removed.

Generally, in the language of  nested-block hook diagrams
(\ref{nested'blocs})
 $$  \Yn{}^{nest} \{\underbrace{\ath_1  ,\ath_1-1 ,\ldots ,\ath_1-n_1+1}_{n_1}\,,\,\ldots\,,\,
 \underbrace{\ath_{\ppi}\,,\ath_{\ppi}-1 \,,\ldots ,\,\ath_{\ppi}-n_{\ppi}+1}_{n_{\ppi}}  \,\}\,
 ,$$    one can either add two cells to a $(n_1+\ldots+n_{k-1}+1)$-th hook
 \be \label{nested'blocs+}
\!\! \!\Yn{}^{nest}_{ k+}  \{\underbrace{\ath_1,\ath_1-1 , \ldots ,\ath_1-n_1+1}_{n_1},\ldots,
\underbrace{\ath_{k }+1}_{1},
 \underbrace{ \ath_k-1 \,,\ldots ,\,\ath_k-n_k+1}_{n_k-1},\ldots,
 \underbrace{\ath_{\ppi},\ldots ,\ath_{\ppi}-n_{\ppi}+1}_{n_{\ppi}}\},\quad \ee
or  remove two cells from   any $(n_1+\ldots+n_{k})$-th hook
 \be \label{nested'blocs-}
\!\! \Yn{}^{nest}_{ k-}   \{\underbrace{\ath_1,\ath_1-1 , \ldots ,\ath_1-n_1+1}_{n_1}\,,\ldots,
  \underbrace{\ath_{k } \,,\ldots ,\,\ath_{k }-n_{k }+2}_{n_{k }-1} \,,\,
\underbrace{\ath_{k }-n_k}_{1},
\ldots,
 \underbrace{\ath_{\ppi},\ldots ,\,\ath_{\ppi}-n_{\ppi}+1}_{n_{\ppi}}\}\quad \ee
  for any $k=1,\ldots, \ppi$  where $\ppi$ is the number of block hooks.
   If the $\ppi$-th   block hook  is not rectangular, \ie
\,$\ath_{\ppi}\ne n_{\ppi} $\,,  then two cells can also be added to the
 $(n_1+\ldots+n_{\ppi}+1)$-th zero hook as in   (\ref{nested122})
\be\label{nblast+}
\Yn{}^{nest}_{ {(p+1)}+}   \{\underbrace{\ath_1, \ldots ,\ath_1-n_1+1}_{n_1}\,,\ldots,
  \underbrace{\ath_{\ppi},\ldots ,\,\ath_{\ppi}-n_{\ppi}+1}_{n_{\ppi}}\,,\,
  \underbrace{ 1}_{1}\}\,.\ee

Substitution
of $\widetilde{\Y}_A=\Yn{}^{nest}_{k+}  $ (\ref{nested'blocs+}) into Eq.~(\ref{semipostil})
    for  $k\le \ppi$ yields
\be\label{raznost1}
  \Hh_{N_{k-1}+1}+ \Hh_{\ath_k+N_{k-1}+2} \le \K\,,
\ee
where      \be\label{Ns} N_s=n_1+\ldots+n_s\q N_0=0 .\ee
A useful consequence of these inequalities  is \be\label{last}
   \Hh_{\la+1}+ \Hh_{\la+2} \le\K \,,\quad \la=\sharp(\Y_A)\, =N_\ppi.
\ee
(Recall, that $ \sharp(\Y_A)$ is the number of the nested
hooks in $ \Y_A \, $, while $\ppi$ is the    number of the nested block
hooks.)
Indeed, for a non-rectangular last block hook, \ie
at   $\ath_\ppi\ne n_\ppi $,    substitution
of $\widetilde{\Y}_A=\Yn{}^{nest}_{( \ppi+1\,+)} $ (\ref{nblast+}) into Eq.~(\ref{semipostil}) yields
(\ref{last}).
At $\ath_\ppi= n_\ppi $   (\ref{last}) follows from
(\ref{raznost1}) since in this case  $ \ath_\ppi+N_{\ppi-1} =N_\ppi  =\sharp(\Y_A)$ and
$\Hh_{N_{\ppi}+1}=\Hh_{N_{\ppi-1}+1}.$

Analogously, for $\widetilde{\Y}_A=\Yn{}^{nest}_{(k-)} $ (\ref{nested'blocs-}), Eq.~(\ref{semipostil}) yields
 \be\label{raznost2}
    \Hh_{\ath_k+N_{k-1}+1 }+\Hh_{N_k} \ge \K\quad\forall\, k\le \ppi\,.
\ee
 In particular, using that  $ {N_\ppi}=\sharp(\Y_A)$ and    $\ath_k+N_{k-1}\ge N_\ppi$, (\ref{raznost2}) yields
   \be\label{last-} \Hh_{\la} + \Hh_{\la+1} \ge\K \,,\quad \la=\sharp(\Y_A). \qquad\ee

 Now let us show that the addition  of cells to    $\Y''$
on the left from the $(\la+1)$-th column
decreases  $\chi^{\half(\K-1)}$. Let
   \be
 \label{Yod+k} \Y {}{}_{ n_1,n_2,...n_k \, + }\,:=\,\,
\Y {}  \cup \S(   \hei_{n_1}(\Y) +1,n_1)\cup\ldots\cup \S(   \hei_{n_k}(\Y) +1,n_k)  \ee
for different $n_i$.
 We add here $k$ cells into $n_1,n_2,...n_k$-th   columns of $\Y$ at the condition
 that the resulting diagram exists.

Let \be\label{oneadd}\mathbf{Q}_{{j_1,\ldots,j_k\,+}}:=
  \chi^{\half(\K-1)}\Big(\big(\Y''{}_{j_1,\ldots,j_k\,+}  \big)\oSW \Y_A\Big)
  -\chi^{\half(\K-1)}\Big( \Y'' \oSW \Y_A\Big)     \,.
 \ee
Properties of $\chi^a$ (\ref{chicaz}) and Eq.~(\ref{OSW})
imply that $\mathbf{Q}_{{j_1,\ldots,j_k\,+}}=\sum_{i=1}^k \mathbf{Q}_{j_i+}$ and
\be\label{Q-1}
\mathbf{Q}_{j+}
=  (j-1 -\hh_j) + \half(\K  - 2\Hh_{j} )-\half
  , \qquad
\ee
where $\hh_j=\hei_j(\Y_A).$
One can see that
$(j-1 -\hh_j) \le0$ for $ j\le\la+1 $
because  $\hh_{j}\ge\hh_{\la+1}$ while $\hh_{\la+1}=\la$ (cf (\ref{nesthn})).
Hence,   taking into account Eq.~(\ref{last-}),
   \be\label{twoadd}
 \mathbf{Q}_{i,j+}= \chi^{\half(\K-1)}\Big( \Ydelnol{}_{i,j\,+}  \oSW \Y_A\Big)
   -\chi^{\half(\K-1)}\Big( \Ydelnol{} \oSW \Y_A\Big) < 0  \,\quad \forall \,\, j< i\le \la+1.\qquad
 \ee

 The following statements that hold by virtue of {\it Lemma 1} and Young properties of
 tensor products will be used below. Any nonzero component
of the tensor product $\Y\in \Y_0\otimes\,\Y_\gd$    has to
satisfy
  \be\label{konTm}
 \hei_{2k}(\Y)+\hei_{2k-1}(\Y)\le \K+\hei_{2k}(\Y_\gd) \,\quad\forall k \,,
 \ee
 \be  \label{younten1} \max \big(\hei_{ i}(\Y_0 )\,,\,  \hei_{ i}(   \Y_\gd)\big)\le
 \hei_{ i}(\Y  )
   \quad \forall i \, .
   \ee
   Eqs.~(\ref{chisw}), (\ref{last}), (\ref{last-})  and (\ref{twoadd})    yield the  following\\

\noindent
{\it Lemma~2}\label{Lemma 2}
\\
Given $\Y_0$ and $\Y_A\ne\bullet$ any solution $\Y'$  of the homotopy equation
\be\Y' =\Y''\oSW \Y_A \,,\qquad  \Y'' \in  \Y_0\otimes\,\Y_\gd\,\ee
obeys   the  conditions
 \be\label{konTm=}
 \Hh _{2k} +\Hh_{2k-1} = \K+d_k \,\quad\forall k\le
 \ld : =\left[\half(\la+1)\right]\,\q\la= \sharp(\Y_A)  \,,\quad
 \ee
where $d_k=\hei_{2k}(\Y_\gd)$, $\Hh_n=\hei_{n}(\Ydelnol{})$ and $[x]$ is the integer part
of $x$, $[x]\leq x$.

We prove {\it Lemma~2} by induction.
 To this end we observe
that if inequalities (\ref{konTm=}) hold for $\forall k\le\mathbf{j}$
with some $\mathbf{j} \le \ld   $ then from the property that  heights of
columns of a YD do not increase from the left to the right it follows that
if $d_l=d_{ i} \mbox{ for some } l<  i\le\mathbf{j}$ then
 \be \label{assd}
\Hh_{2l }=\ldots= \Hh _{2i-1}
=  \Hh_{ 2i }
\ee
and hence, if $\Hh _{2l-1}>
 \Hh_{2l}   \mbox{ for some }  l\le\mathbf{j}$, then
\be
  \label{assdineq1}
   d_{\,l-1}>d_{\,l }\,\,\mbox{ if } l\leq\mathbf{j} \,\,\mbox{ and}\quad d_{\,l }>d_{\,l+1}
   \,\,\mbox{ if } l<\mathbf{j}
 \ee
 and, if    $\Hh _{2l-2}>
 \Hh_{2l-1}   \mbox{ for some }  l\le\mathbf{j}$, then
  \be\label{assdineq2}
d_{\,l-1}> d_{\,l }\, .
\ee
Here and after we use the convention $d_0:=\K.$

Firstly let us prove  (\ref{konTm=}) at $k=1$.
Suppose that this is not  true, \ie
 \be\label{d1ineq}\Hh _{1} +\Hh_{2 } < \K+d_1.\ee
From  (\ref{younten1}), (\ref{d1ineq}) it follows that $d_1<\K$.
This demands  $\Hh_1<\K$ since otherwise
 $ \Hh_{2 } <  d_1<\K$  in contradiction with (\ref{younten1}). Therefore $\K\ge\Hh_1+1\ge\Hh_2+1$ and $\K\ge d_1+1$. Hence YD
    $  \Ydelnol{}{}_{ 1,2+}$ and $ \Y_\gd{}_{ 1,2+}$ (\ref{Yod+k}) are nonzero.
 Since by virtue of (\ref{d1ineq}) $$\Hh_1+\Hh_2+2\le\! \K+(d_1+1)\,,$$
         \ie conditions (\ref{konTm}) hold,  $ \Ydelnol{}{}_{ 1,2+}$ is a nonzero component of  $ \Y_0\otimes  \Y_\gd{}_{ 1,2+}$.
Substitution of $ \Ydelnol{}{}_{ 1,2+}$ into  (\ref{twoadd}) yields by virtue of
(\ref{semipostil}) that
 $\chi^{\half(\K-1)}\Big( \Ydelnol{} \oSW \Y_A\Big)> 0 $ in contradiction  with
the assumption of  {\it Lemma~2}. This proves  (\ref{konTm=}) at $k=1$.

 Now suppose  that   (\ref{konTm=})   holds  for all
$k \le\mathbf{j}$ with some   $\mathbf{j}\ge1$ but does not hold for $k= \mathbf{j}+1$, \ie taking into account {\it Lemma~1},
\be\label{djineq}\Hh _{ 2\mathbf{j}+1} +\Hh_{2 \mathbf{j}+2 } < \K+d_ {\mathbf{j}+1},\ee
\be\label{djieq}\Hh _{ 2k+1} +\Hh_{2k+2 } = \K+d_ {k+1}\qquad \forall k<\mathbf{j}\,.\ee
If  $ \Hh _{2\mathbf{j}}=\K$   then  as for the case of $k=1$
Eqs.~(\ref{younten1}), (\ref{djineq})
      yield $d_\mathbf{j}=\K$,  $ \Hh_{2\mathbf{j}+1}<\K$
   and $ d_{\mathbf{j}+1}<\K$. Therefore,  substitution of  $ \Ydelnol{}{}_{(2\mathbf{j}+ 1),(2\mathbf{j}+2)+}$
   into  (\ref{twoadd}) leads  by virtue of
(\ref{semipostil})       to the contradiction  with
the assumption of  {\it Lemma~2}.
       Hence it remains to consider the case of   $\Hh _{2\mathbf{j} }<\K$, $d_\mathbf{j}<\K$.

By  Eq.~(\ref{assd}), assumption (\ref{djineq}) implies that at least
one of the  equalities
$\Hh_{2\mathbf{j}-1 }=\Hh_{2\mathbf{j} }=\Hh_{2\mathbf{j}+1}=\Hh_{2\mathbf{j}+2}$
 fails. Consider different cases.

  Let $\Hh_{2\mathbf{j} }>\Hh_{2\mathbf{j}+1}$.
 If  $d_\mathbf{j}>d_{\mathbf{j}+1}$ then the YD
    $ \Ydelnol{}{}_{ 2\mathbf{j}+1,2\mathbf{j}+2 \,+}$ and
 $  \Y_\gd{}_{ 2\mathbf{j}+1,2\mathbf{j}+2\, +}$ have an allowed shape and
 are nonzero since conditions (\ref{konTm}) hold for them  by virtue of (\ref{djineq}).
    Hence     $ \Ydelnol{}{}_{ 2\mathbf{j}+1,2\mathbf{j}+2 \,+}\in \Y_0\otimes
     \Y_\gd{}_{ 2\mathbf{j}+1,2\mathbf{j}+2\, +}$
   and (\ref{twoadd})  yields  $\chi^{\half(\K-1)}\Big( \Ydelnol{} \oSW \Y_A\Big)> 0 $
   in contradiction with the assumption of {\it Lemma~2}.

   If  $d_\mathbf{j}=d_{\mathbf{j}+1}$ then
   by virtue of (\ref{assdineq1}),  (\ref{assdineq2}) there exist some
          $ n\le \mathbf{j}$ and $m\in[2n-1,2n] $ such that
          $\Ydelnol{}{}_{m,2\mathbf{j}+1 \,+}$ and $\Y_\gd{}_{ 2n-1,2n \, +}$
          are nonzero YD satisfying (\ref{konTm}) by virtue of (\ref{djineq}) and  (\ref{djieq}).
          Hence $\Ydelnol{}{}_{m,2\mathbf{j}+1 \,+}\in \Y_0\otimes
           \Y_\gd{}_{ 2n-1,2n \, +}$
          in  contradiction with the assumption of  {\it Lemma~2}.

  This allows us to set $\Hh_{2\mathbf{j} }=\Hh_{2\mathbf{j}+1}$.

If $\Hh_{2\mathbf{j}-1 }>\Hh_{2\mathbf{j}}$,  (\ref{assdineq1}) implies
    $d_{\mathbf{j}-1}>d_{\mathbf{j} }$ and there exist     nonzero YD $\Ydelnol{}{}_{2\mathbf{j},2\mathbf{j}+1 \,+}$
    and $ \Y_\gd{}_{2\mathbf{j} -1,2\mathbf{j} \, +}$
     satisfying (\ref{konTm}) by virtue of (\ref{djineq}) and  (\ref{djieq}).
 Then   $\Ydelnol{}{}_{2\mathbf{j},2\mathbf{j}+1 \,+}\in \Y_0\otimes  \Y_\gd{}_{2\mathbf{j} -1,2\mathbf{j} \, +}$
 which by (\ref{twoadd})  is in
     contradiction with  the assumption of {\it Lemma~2}.

If $\Hh_{2\mathbf{j}-1 }=\Hh_{2\mathbf{j}}$,  then the only remaining option is
   $\Hh_{2\mathbf{j}+1}> \Hh_{2\mathbf{j}+2} $. Hence
    by  (\ref{assdineq1}),  (\ref{assdineq2}) there exist some
          $ n< \mathbf{j}$ and $m\in[2n-1,2n] $
          such that
          $\Ydelnol{}{}_{m,2\mathbf{j}+2 \,+}$ and $\Y_\gd{}_{ 2n-1,2n \, +}$
          are nonzero YD satisfying (\ref{konTm}) by virtue of (\ref{djineq}) and  (\ref{djieq}).
Then   $\Ydelnol{}{}_{m,2\mathbf{j}+2 \,+}\in \Y_0\otimes  \Y_\gd{}_{ 2n-1,2n \, +}  $
 in  contradiction with  the assumption of {\it Lemma~2} thus completing the proof.
       $\square$
\\

It is convenient to introduce notations
\be\label{Htilde}
\Hr_j=\K-\Hh_j\qquad \forall \,\, j\le  {2\ld} \,.
\ee Substitution of $\Hr_j$ (\ref{Htilde}) into   (\ref{raznost1}), (\ref{last}) and
(\ref{raznost2})
yields   inequalities on $\Hr_j$ with $j\le2\ld$ and $\Hh_k$ with $k>2\ld$.
In particular,   (\ref{raznost1}) and  (\ref{raznost2}) yield, respectively,
\be\label{raznost1Hr}
  \Hr_{N_{k-1}+1}\ge \Hh_{\ath_k+N_{k-1}+2}  \,,\quad k\le \ppi\,,
\ee
\be\label{raznost2Hr}
  \Hr_{N_k}\le \Hh_{\ath_k+N_{k-1}+1 } \,,\quad k\le \ppi\,.
  \ee
Since by construction
\be \label{neravHH}
\Hr_{k+1}\ge  \Hr_{k} \quad \forall \, k<{2\ld} \q
\Hh_{j}\ge  \Hh_{j+1} \quad \forall \, j\ge{2\ld}+1\,,\ee
Eqs.~(\ref{raznost1Hr})-(\ref{neravHH}) give the following chain of inequalities
\be\label{order}\ls
\ldots\ge \Hh_{\ath_k+N_{k-1}+1 }\ge \Hr_{N_k}\ge\ldots\ge\Hr_{N_{k-1}+1}\ge \Hh_{\ath_k+N_{k-1}+2}\ge\ldots
\ge \Hh_{\ath_{k+1}+N_{k}+1 }\ge\Hr_{N_{k+1}}\ge\ldots,
\ee
where the starting point   depends on the oddness
of $\la\equiv N_\ppi$ and on whether $a_{\ppi}-n_{\ppi}$ is zero or not.
There are two reasons for this. Firstly there is a difference
between diagrams with rectangular ($a_{\ppi}=n_{\ppi}$) and not
rectangular ($a_{\ppi}>n_{\ppi}$) inner block hooks
that gives different numbers of inequalities on $\Hh$ in these two cases.
Secondly, $\ {2\ld }= {N_\ppi+1}$ for odd $\la$ and  $   {2\ld }= {N_\ppi }$
for even  $\la$,
that leads to different
relations between $\Hr_j$ and $\Hh_{2\ld+i}$.

Namely for even $\la$ and   $a_{\ppi}>n_{\ppi} $
\be\label{ordere1}\ls
    \Hh_{ N_{p}+1}\ge\ldots
\ge \Hh_{\ath_{p}+N_{p-1}+1 }\ge\Hr_{N_{p}}\ge\ldots \ge\Hr_{N_{p-1}+1}\ge
\Hh_{\ath_p+N_{p-1}+2}\ge\ldots\,.
\ee
For odd $\la$ and   $a_{\ppi}>n_{\ppi} $
\be\label{ordere2}\ls
  \Hr_{N_{p}+1}\ge  \Hh_{ N_{p}+2}\ge\ldots
\ge \Hh_{\ath_{p}+N_{p-1}+1 }\ge
\Hr_{N_{p}}\ge\ldots \ge\Hr_{N_{p-1}+1}\ge
\Hh_{\ath_p+N_{p-1}+2}\ge\ldots\,.
\ee
For even $\la$ and   $a_{\ppi}=n_{\ppi} $
\be\label{ordere3}\ls
    \Hh_{ N_{p}+1}\ge  \Hr_{N_{p}}\ge \ldots\ge\Hr_{N_{p-1}+1}\ge
\Hh_{\ath_p+N_{p-1}+2}\ge\ldots\,.
\ee
For odd $\la$ and   $a_{\ppi}=n_{\ppi} $
\be\label{ordere4}\ls
  \Hr_{N_{p}+1}\ge \ldots
\ge \Hr_{N_{p-1}+1  } \ge \Hh_{ N_{p}+2}\ge\ldots
\ge \Hh_{\ath_{p}+N_{p-1}+1 }\ge
\Hr_{N_{p-1}}\ge\ldots \,.
\ee

To prove  {\it Theorem} (see p.\pageref{ Theorem}) we first prove \\

{\it Lemma 3} \label{Lemma 3}

Given $\Y_0$ and $\Y_A\ne\bullet$, any solution $\Y'$  of the homotopy equation
\be\Y' =\Y''\oSW \Y_A \,,\qquad  \Y'' \in  \Y_0\otimes\,\Y_\gd\,,\ee
obeys   relations (\ref{otvetec1}) of {\it Theorem}.

Firstly let us assume that
\be\label{len2m} \len_1(Y_\gd)=2\ld.  \ee
(Recall that $\ld  =\left[\half(\la+1)\right]$, $\la= \sharp(\Y_A) $ (\ref{Ydi}).)
  Since   $\Y''[\Hh_1,\ldots]\in \Y_0\otimes Y_\gd$ ,
\be\label{sumHi}\sum_{i }  \Hh_i= \sum_{j }   \hei_j(\Y_0)+  \sum_{k }\hei_k(\Y_\gd)\q
     \, \ee
     where summation is over all columns of each diagram. Since by virtue of Eqs.~(\ref{konTm=}) and (\ref{Htilde})
\be\label{d+H}
\qquad\Hh_{2i-1}=d_{i}+\Hr_{2i }\q
\Hh_{2i}=d_{i}+\Hr_{2i-1 }\q i\le  {\ld}\,,
  \ee
  the
    substitution of (\ref{d+H}) into (\ref{sumHi})    yields
    using (\ref{len2m})
\be  \label{yountend=} \sum_{i =1}^{{2\ld}} \Hr_i
  +\sum_{i >{2\ld}} \Hh_i= \sum_{j}   h_{j}\q h_{j}=\hei_j(\Y_0)
     \,.
\ee

 Let
\be\label{tensY0d}
T^{B_1{[h_1]},B_2{[h_2]} \ldots }\q D^{C_1{[d_1]}, \ldots,C_{{2\ld}}{[d_{\ld}]}}
 \ee
be tensors with symmetry properties of given    traceless YD
 $\Y_0[h_1,h_2,\ldots]$ and \\Kronecker YD $\Y_\gd[d_1,d_1,d_2,d_2,\ldots]$ (\ref{Ydi}), respectively, while
 \be\label{tensY''}
F_{A_1{[\Hh_1]},A_2{[\Hh_2]},\ldots}
\ee
be a tensor with symmetry of $\Ydelnol{}[\Hh_1,\Hh_2,\ldots] =\Yod\in \Y_0\otimes  \Y_\gd $.
From (\ref{len2m}) and  {\it Lemma~2} it follows that $\Y_\gd$ in
(\ref{Y''oSWYA}) is  determined by (\ref{konTm=}).
Since  $\Ydelnol{}$ is a component of the tensor product $ \Y_0\otimes\Y_\gd$,
one can contract all indices of $D$ (\ref{tensY0d}) with indices of $F$ (\ref{tensY''})
obtaining by virtue of (\ref{d+H}) the  tensor
\be\label{FY0}
\F_{A_1{[\Hr_2]};A_2{[\Hr_1]};\ldots;A_{{2\ld}-1}{[\Hr_{2\ld}]};A_{{2\ld}}{[\Hr_{{2\ld}-1}]};
A_{{2\ld}+1}{[\Hh_{{2\ld}+1}]},A_{{2\ld}+2}{[\Hh_{{2\ld}+2}]},\ldots}\,\,,\ee
which is antisymmetric in all indices $A_j[.]$ but   not  necessarily has  properties
of a YD. The numbers of  antisymmetrized indices $A_j[.]$ are called 'heights'
of  $\F$.
In accordance with (\ref{yountend=}) all indices of  $\F$ (\ref{FY0})
have to be contracted with all indices of  $T$ (\ref{tensY0d}).
Hence the maximal height  of   $\F$ cannot be greater than $h_1=\hei_1(\Y_0)$.
On the other hand, the SW-principle demands the maximal height   of   $\F$
 be exactly equal to $h_1$. Analogously, the subleading height cannot be greater than $h_2=\hei_2(\Y_0)$, {\it etc.}
Hence the SW-principle demands  ordered heights  of   $\F$
 be  equal to $h_i$. Since the ordering of heights of $\F$ is determined by
inequalities (\ref{order})-(\ref{ordere4}) this expresses the heights of $\F$ via $h_i$.
 Straightforward computation gives
$\Hh_j$    in the form (\ref{otvetec1}). This completes the proof at the condition
that (\ref{len2m}) holds.

To show that condition (\ref{len2m})  holds true it is enough to show that
   existence of such a  solution $ {\Y}'[ {\Hh}_{1},  {\Hh}_{2 },\ldots]$
   of Eq.~(\ref{Kazx-y}) that
  \be
\label{Y{}''oSWYA>}\qquad  {\Y}' = {\Y}{}''\oSW \Y_A\q  {\Y}{}''
  \in  \Y_0\otimes\, {\Y}_\gd\,, \quad {\Y}_\gd= {\Y} _\gd[ {d}_1, {d}_1,\ldots, {d}_K, {d}_K] \,,\ee
 \be\label{Ydi>}
       {d}_{k}=  {\Hh}_{2k}+ {\Hh}_{2k-1}-\K\,
   \quad \forall \,\, k\le  {\ld}\,,\quad K\ge\ld+1 \,, \quad {d}_{K}>0\,, \quad
   {\Hh}_{j}=\hei_j( {\Y}{}'')
    \ee
  contradicts the assumption that $ {\Y}'$ solves Eq.~(\ref{Kazx-y}).

Since \be {\Y} _\gd \in {\Y} ^-_\gd  \otimes \Y_\gd[1,1] \q
{\Y} ^-_\gd:={\Y} ^-_\gd [ {d}_1, {d}_1,\ldots, {d}_K-1, {d}_K-1]\ee
then
${\Y}{}'' \in  \Y_0\otimes\, {\Y}_\gd \in \Y_0\otimes {\Y} ^-_\gd  \otimes \Y_\gd[1,1]$
 and by associativity of the tensor
product
   \be\label{YD''min} {{\Y}}{}''\in {{\Y}}{}''{}^-\otimes \Y_\gd[1,1] \q {{\Y}}{}''{}^-
 \in  \Y_0\otimes\, {\Y} ^-_\gd \,.
\ee
 By construction,
 ${{\Y}}{}''={{\Y}}{}''{}^-\cup \S(  \Hh_{j_1}  ,j_1)\cup
\S(   \Hh_{j_2}   ,j_2) $ for some $j_1\ne j_2 $ since $\Y_\gd[1,1]$ is symmetric.
For definiteness we set $j_1<j_2$.
  Additivity of $\chi^a$ (\ref{chicaz})
  and Eq.~(\ref{OSW})
imply that
\be\label{two--}
   \chi^{\half(\K-1)}\Big( \Ydelnol{} \oSW \Y_A\Big)-\chi^{\half(\K-1)}\Big( \Ydelnol{}^-   \oSW \Y_A\Big)
    =  j_1 -1 -\hh_{j_1}+j_2  -\hh_{j_2}  +   \K  -   \Hh_{{j_1}}-\Hh_{{j_2}}
             \,,
 \ee
where $\hh_j=\hei_j(\Y_A).$
  Consider different cases.

Let   $ j_1\ge\la+1 $. Then, by properties of almost symmetric YD,
$j_1 -1 -\hh_{j_1} \ge 0$
because  $\hh_{j_1}\le\hh_{\la+1}$ while $\hh_{\la+1}=\la$ (cf Eq.~(\ref{nesthn})).
Analogously, $j_2  -\hh_{j_2}>0$ since $j_2>j_1 $. By virtue of (\ref{last})
$\K  -   \Hh_{{j_1}}-\Hh_{{j_2}} \ge0$.
Hence  Eq.~(\ref{two--}) yields
 $\chi^{\half(\K-1)}\Big( \Ydelnol{} \oSW \Y_A\Big)>\chi^{\half(\K-1)}\Big( \Ydelnol{}^-   \oSW \Y_A\Big)$.
  By (\ref{semipostil})  this    contradicts
  the assumption that $ {\Y}'$ solves Homotopy condition.

Let     $j_1 \le 2\ld $. 
  Since $\Y{}''$ satisfies (\ref{konTm=}), this is equivalent to
\bee\label{djieq5}\hei _{ 2l-1}(\Y {}''{}^-) +\hei_{2l  }(\Y {}''{}^-) = \K+d_ {l }-1\qquad
 \mbox{for } l=\left[\half (j_1+1)\right]\,,\\ \label{djieq6}\hei _{ 2k-1}(\Y {}''{}^-) +\hei_{2k  }(\Y {}''{}^-) = \K+d_ {k }\qquad
  \mbox{for }  k\le
 \ld, k\ne l.\eee  It can be shown along the lines of the proof of {\it Lemma 2} that  there
 exist YD  $\Y {}''{}^-{}{}_{m ,j_1  \,+}$
   with some $m\leq 2l $ and some 
 Kronecker YD  $\widetilde{\Y}_\gd\supset {\Y}_\gd^- $ such that
 $\Y {}''{}^-{}{}_{m ,j_1  \,+}\in \Y_0\otimes \widetilde{\Y}_\gd$.
 By construction $\Y {}''{}^-{}{}_{m ,j_1  \,+}\prec\Y{}''$ in contradiction
 with
the assumption that $ {\Y}'$ (\ref{Y{}''oSWYA>}) solves Homotopy condition. This
  completes the proof of {\it Lemma 3 } $\square$

\bigskip

 To complete the proof of {\it Theorem} (see p.\pageref{ Theorem})
 it remains to prove

 {\it Lemma 4} \label{Lemma 5}

Differential forms obeying symmetry properties of YD $\Y'  $ (\ref{otvetec})
 are  $\gs_-^\K$-closed but not $\gs_-^\K$--exact.

   Indeed, consider a tensor  $P(\gx,Y )$ with symmetry properties of YD
    $ {\Y}' = (\Y_0\otimes\, {\Y}_\gd\,)\oSW \Y_A
     $  (\ref{otvetec}).
 The action of  $\gs_-^\K = \dis{   \gd_{ij}
  \kk^{AB
  }\frac{\p^2 }{\p Y{}_i{}^{A}\p Y{}_j{}^{B}} \,}$
 (\ref{sigmar}) on   $P(\gx,Y)$ (\ref{F})
 yields zero by virtue of one of the following three mechanisms
 \\$\bullet\quad$  antisymmetrization of the indices of $\gx^{AB}$
 in $\gs_-$  with  those of the almost symmetric YD $\Y_A$ gives zero allowing no
 proper contraction of the resulting almost symmetric YDs within the representation
 of     $P(\gx,Y|X)$  (\ref{F}).
 \\
 $\bullet\quad$  contraction of the indices of the Kronecker symbol with
  those of a traceless diagram,
\\$\bullet\quad$   antisymmetrization of  the  lower case Latin indices of the
Kronecker symbol with
the indices of $\Y_\gd$ (\ref{Ydi}) gives zero by virtue of {\it Lemma 1}.

To show that $\ker \Delta$  (\ref{l1}) does not contain $\gs_-^\K$--exact
elements, suppose that
$  \Delta  a = 0$ , $ a= \gs_-^\K b$
for some $b$. Since   $ [ \gs_-^\K   , \Delta ]=0$, the expansion of $b$ in eigenvectors
of $\Delta$ can only
contain those with zero eigenvalues, \ie  $\Delta b=0$.
Since it is shown that every $b$ obeying this condition  is $\gs_-^\K$-closed,  every exact $a\in
\ker\Delta$ is zero $\square$

\end{document}